\documentclass[fleqn,usenatbib]{mnras}

\usepackage{newtxtext,newtxmath}

\usepackage[T1]{fontenc}
\usepackage{ae,aecompl}

\usepackage{graphicx}	
\usepackage{amsmath}	
\title[DES Y3: cosmology from cosmic shear in harmonic space]{Dark Energy Survey Year 3 results: cosmological constraints from the analysis of cosmic shear in harmonic space}

\author[Doux, C. et al.]{
\parbox{\textwidth}{
\Large
C.~Doux,$^{1}$\thanks{E-mail: cdoux@sas.upenn.edu}
B.~Jain,$^{1}$
D.~Zeurcher,$^{2}$
J.~Lee,$^{1}$
X.~Fang,$^{3,4}$
R.~Rosenfeld,$^{5,6}$
A.~Amon,$^{7}$
H.~Camacho,$^{8,6}$
A.~Choi,$^{9}$
L.~F.~Secco,$^{10}$
% ...
\begin{center} (DES Collaboration) \end{center}
}
}

\date{Accepted XXX. Received YYY; in original form ZZZ}

\pubyear{2022}

\usepackage{xspace}
\usepackage[dvipsnames]{xcolor}

\usepackage{physics}
\usepackage{siunitx}
\DeclareSIUnit \h {\ensuremath{\mathit{h}}}
\DeclareSIUnit \parsec {pc}
\DeclareSIUnit \deg {deg}
\usepackage{bm}
\usepackage{url}
\usepackage{cleveref}
\usepackage{comment}
\usepackage{ulem}
\usepackage{paralist}

\newcommand{\mcal}{\textsc{Metacalibration}\xspace}
\newcommand{\healpy}{\textsc{healpy}\xspace}
\newcommand{\healpix}{\textsc{HealPix}\xspace}
\newcommand{\namaster}{\textsc{NaMaster}\xspace}
\newcommand{\cosmosis}{\textsc{CosmoSIS}\xspace}
\newcommand{\polychord}{\textsc{PolyChord}\xspace}
\newcommand{\cosmolike}{\textsc{CosmoLike}\xspace}
\newcommand{\camb}{\textsc{CAMB}\xspace}
\newcommand{\halofit}{\textsc{HaloFit}\xspace}
\newcommand{\hmcode}{\textsc{HMCode}\xspace}
\newcommand{\gadget}{\textsc{Gadget}\xspace}
\newcommand{\calclens}{\textsc{CalcLens}\xspace}
\newcommand{\pkd}{\textsc{PKDGrav3}\xspace}
\newcommand{\dgv}{\textsc{DarkGridV1}\xspace}
\newcommand{\sompz}{\textsc{Sompz}\xspace}
\newcommand{\hyperrank}{\textsc{HyperRank}\xspace}
\newcommand{\multirank}{\textsc{MultiRank}\xspace}
\newcommand{\planck}{\textit{Planck}\xspace}

\newcommand{\cell}{$C_\ell$\xspace}
\newcommand{\cl}{C_\ell}
\newcommand{\xip}{\xi_{+}}
\newcommand{\xim}{\xi_{-}}
\newcommand{\xipm}{\xi_{\pm}}
\newcommand{\fsky}{f_{\rm sky}}
\newcommand{\nside}{N_{\rm side}}
\newcommand{\kmax}{k_{\rm max}}
\newcommand{\lmin}{\ell_{\rm min}}
\newcommand{\lmax}{\ell_{\rm max}}

\newcommand{\lcdm}{{$\Lambda$CDM}\xspace}
\newcommand{\wcdm}{{$w$CDM}\xspace}
\newcommand{\Om}{\Omega_{\rm m}}
\newcommand{\Ob}{\Omega_{\rm b}}
\newcommand{\Onu}{\Omega_{\nu}}
\newcommand{\As}{A_{\rm s}}
\newcommand{\ns}{n_{\rm s}}
\newcommand{\Ata}{A_{\rm TA}}
\newcommand{\ata}{\alpha_{\rm TA}}
\newcommand{\Att}{A_{\rm TT}}
\newcommand{\att}{\alpha_{\rm TT}}
\newcommand{\bta}{b_{\rm TA}}
\newcommand{\Ahm}{A_{\rm HM}}
\newcommand{\etahm}{\eta_{\rm HM}}

\DeclareMathOperator{\cov}{cov}

\newcommand{\ie}{{i.e.}\xspace}
\newcommand{\eg}{{e.g.}\xspace}
\newcommand{\vs}{\textit{vs}\xspace}

\usepackage{eso-pic}

\AddToShipoutPictureBG*{%
  \AtPageUpperLeft{%
    \hspace{0.75\paperwidth}%
    \raisebox{-3.5\baselineskip}{%
      \makebox[0pt][l]{\textnormal{DES-2015-0048}}
}}}%

\AddToShipoutPictureBG*{%
  \AtPageUpperLeft{%
    \hspace{0.75\paperwidth}%
    \raisebox{-4.5\baselineskip}{%
      \makebox[0pt][l]{\textnormal{FERMILAB-PUB-22-042-PPD-SCD}}
}}}%

\begin{document}
\label{firstpage}
\pagerange{\pageref{firstpage}--\pageref{lastpage}}
\maketitle

\begin{abstract}
We present cosmological constraints from the analysis of angular power spectra of cosmic shear maps based on data from the first three years of observations by the Dark Energy Survey (DES~Y3). The shape catalog contains ellipticity measurements for over 100~million galaxies within a footprint of 4143~square degrees. Our measurements are based on the pseudo-\cell method and offer a view complementary to that of the two-point correlation functions in real space, as the two estimators are known to compress and select Gaussian information in different ways, due to scale cuts. They may also be differently affected by systematic effects and theoretical uncertainties, such as baryons and intrinsic alignments (IA), making this analysis an important cross-check.
In the context of \lcdm, and using the same fiducial model as in the DES Y3 real space analysis, we find ${S_8 \equiv \sigma_8 \sqrt{\Om/0.3} = 0.793^{+0.038}_{-0.025}}$, which further improves to ${S_8 = 0.784\pm 0.026 }$ when including shear ratios.
This constraint is within expected statistical fluctuations from the real space analysis, and in agreement with DES~Y3 analyses of non-Gaussian statistics, but favors a slightly higher value of $S_8$, which reduces the tension with the \planck cosmic microwave background 2018 results from \SI{2.3}{$\sigma$} {in the real space analysis} to \SI{1.5}{$\sigma$} in this work. We explore less conservative IA models than the one adopted in our fiducial analysis, finding no clear preference for a more complex model. We also include small scales, using an increased Fourier mode cut-off up to $\kmax=\SI{5}{\h\per\mega\parsec}$, which allows to constrain baryonic feedback while leaving cosmological constraints essentially unchanged. Finally, we present an approximate reconstruction of the linear matter power spectrum at present time, which is found to be about 20\% lower than predicted by \planck 2018, as reflected by the \SI{1.5}{$\sigma$} lower $S_8$ value.

\end{abstract}

\begin{keywords}
gravitational lensing: weak --  cosmological parameters -- large-scale structure of Universe.
\end{keywords}



\section{Introduction}
\label{sec:introduction}

Gravitational lensing by the large-scale structure coherently distorts the apparent shapes of distant galaxies. The measured effect, \textit{cosmic shear}, is sensitive to both the geometry of the Universe and the growth of structure, making it, in principle, a powerful tool for probing the origin of the accelerated expansion of the Universe and, consequently, the nature of dark energy. After the first detections two decades ago \citep{2000Natur.405..143W,2000astro.ph..3338K,2000A&A...358...30V,2000MNRAS.318..625B}, methodological advances in measurement algorithms were permitted by newly collected data, \eg from the {Deep Lens Survey \citep[DLS,][]{2002SPIE.4836...73W, 2013ApJ...765...74J,2016ApJ...824...77J}, the COSMOS survey \citep{2007ApJS..172....1S}, the Canada-France-Hawaii Telescope Legacy Survey \citep[CFHTLS,][]{2006A&A...452...51S} and Canada-France-Hawaii Telescope Lensing Survey \citep[CFHTLenS,][]{2017MNRAS.465.2033J}} and the Sloan Digital Sky Survey \citep[SDSS,][]{2014MNRAS.440.1322H}.
These were fostered by community challenges \citep[see, \eg,][]{2006MNRAS.368.1323H,2007MNRAS.376...13M,2009AnApS...3....6B,2012MNRAS.423.3163K,2014ApJS..212....5M}.
Ongoing surveys, such as the Dark Energy Survey\footnote{\url{https://www.darkenergysurvey.org/}} \citep[DES,][]{2005IJMPA..20.3121F}, the ESO Kilo-Degree Survey\footnote{\url{http://kids.strw.leidenuniv.nl/}} \citep[KiDS,][]{2013Msngr.154...44D,2015MNRAS.454.3500K}, and the Hyper Suprime-Cam Subaru Strategic Program\footnote{\url{https://hsc.mtk.nao.ac.jp/ssp/}} \citep[HSC,][]{2018PASJ...70S...4A,2018PASJ...70S...8A}, have produced data sets capable of achieving cosmological constraints that are competitive with cosmic microwave background observations on the amplitude of structure, $\sigma_8$, and the density of matter, $\Om$, through the parameter combination $S_8\equiv\sigma_8\sqrt{\Om/0.3}$ \citep{2018PhRvD..98d3528T,2019PASJ...71...43H,y3-3x2ptkp,2020PASJ...72...16H,2020A&A...641A...6P,2021A&A...645A.104A}.
These surveys are paving the way for the next generation of surveys, namely the Vera Rubin Observatory Legacy Survey of Space and Time\footnote{\url{https://www.lsst.org/}} \citep[LSST,][]{2019ApJ...873..111I}, the ESA satellite Euclid\footnote{\url{https://sci.esa.int/web/euclid}} \citep{2012SPIE.8442E..0TL}, and NASA's Nancy Grace Roman Space Telescope\footnote{\url{https://roman.gsfc.nasa.gov/}} \citep{2019arXiv190205569A}, which will improve upon current observations in quality, area, depth and spectral coverage, in the hope of {better determining the nature} of dark energy. However, the level of precision needed to fully exploit the cosmological information contained in these future observations pushes the community to dissect every component of the analysis framework, from data collection to inference of cosmological parameters.

The two-point statistics of the cosmic shear field are most commonly used to extract cosmological information. While it is well known that the shear or convergence fields are, to some extent, non-Gaussian \citep{2006Natur.440.1137S,2011PhRvD..84d3529Y}, \ie that there is information in higher-order statistics (\eg in peaks, \citealt{2010MNRAS.402.1049D,2018MNRAS.474..712M,2020arXiv201202777H,2021JCAP...01..028Z,2021MNRAS.501..954J}, or three-point functions, \citealt{2003MNRAS.340..580T,2014MNRAS.441.2725F}), the two-point functions remain the primary source of information, as they can be predicted by numerical integration of analytical models \citep{2015A&C....12...45Z,2017MNRAS.465.2033J,y3-generalmethods,2019ApJS..242....2C} {and efficiently measured \citep{2015ascl.soft08007J}}. The shear two-point function can be characterized by its two components, $\xip(\theta)$ and $\xim(\theta)$, as a function of angular separation $\theta$, or by its Fourier (or harmonic) counterpart, the shear angular power spectrum, \cell, as a function of multipole $\ell$ (with an approximate mapping ${\ell\sim\pi/\theta}$). Both have been measured on recent data from the DES (DES Year~1, \citealt{2018PhRvD..98d3528T,2020arXiv201009717N,2021arXiv211107203C}, and DES Year~3, \citealt*{y3-cosmicshear1,y3-cosmicshear2}), KiDS (KiDS-450, \citealt{2017MNRAS.465.1454H,2017MNRAS.471.4412K}, and KiDS-1000, \citealt{2021A&A...645A.104A,2021arXiv211006947L}) and HSC \citep{2019PASJ...71...43H,2020PASJ...72...16H}.

While, in principle, the two statistics summarize the same information, practical considerations require discarding some of the measurements for cosmological analyses via scale cuts. As a consequence, the information retained by the two statistics differs in practice, which introduces some statistical variance in cosmological constraints, {on top of potential differences due to differential systematic effects}. Indeed, constraints reported for the analyses of cosmic shear with KiDS-450 data showed a difference {between the real- and harmonic-space analyses} of $\Delta{S_8}=0.094$ \citep{2017MNRAS.465.1454H,2017MNRAS.471.4412K}, and that of HSC~Year~1 data a difference of $\Delta{\sigma_8}=0.28$ \citep{2019PASJ...71...43H,2020PASJ...72...16H,2022arXiv220112698H}, both corresponding to about $2\sigma$ discrepancies (see also \cref{fig:1d_all}, discussed below).
{More recently, the comparison between three different estimators presented for KiDS-1000 data, on the other hand, showed excellent agreement \citep{2021A&A...645A.104A}, including a newly developed pseudo-$\cl$ estimator in \citet{2021arXiv211006947L}.}
In a preparatory study \citep{2020arXiv201106469D}, we quantified this effect for DES~Y3 by means of simulations and showed
\begin{inparaenum}[(i)]
    \item that the difference on the $S_8$ parameter is expected to fluctuate by about $\sigma(\Delta{S_8})\sim\num{0.02}$ for typical scale cuts, and
    \item that the observed difference is the result of the interplay between scale cuts and systematic effects, and how these impact each statistic.
\end{inparaenum}

In this work, we present measurements of (tomographic) cosmic shear power spectra measured from data based on the first three years of observations by the Dark Energy Survey (DES~Y3), which we use to infer cosmological constraints on the \lcdm model. We then extend our analysis and vary scale cuts to derive constraints on intrinsic alignments and baryonic feedback at small scales, the two largest astrophysical sources of uncertainty on cosmic shear studies {\citep{2018MNRAS.480.3962C,2018ARA&A..56..393M,y3-cosmicshear2}}. Finally, we study the consistency of these constraints with those inferred from other DES~Y3 weak lensing analyses, using two-point functions \citep*{y3-cosmicshear1,y3-cosmicshear2} and non-Gaussian statistics \citep{2022MNRAS.tmp..151Z,2021arXiv211010141G}. 

The paper is organized as follows: \cref{sec:data} presents DES~Y3 data; \cref{sec:methods} introduces the formalism relevant to the estimation of cosmic shear power spectra and the cosmological model, including systematic effects, intrinsic alignments and baryonic feedback; \cref{sec:validation} highlights the different tests we performed to validate both the measurement and modeling pipelines, some of which rely on simulations (Gaussian, $N$-body and hydrodynamical); \cref{sec:blinding} details the three-step blinding procedure we adopted in this work; \cref{sec:res} presents our main results, \ie cosmological constraints inferred from the analysis of DES~Y3 cosmic shear power spectra, and compares them to other weak lensing studies; and finally \cref{sec:conclusion} summarizes our results.

\section{Dark Energy Survey Year 3 data}
\label{sec:data}

\begin{figure*}
    \centering
    \includegraphics[scale=0.65, trim={0 5mm 0 5mm}, clip]{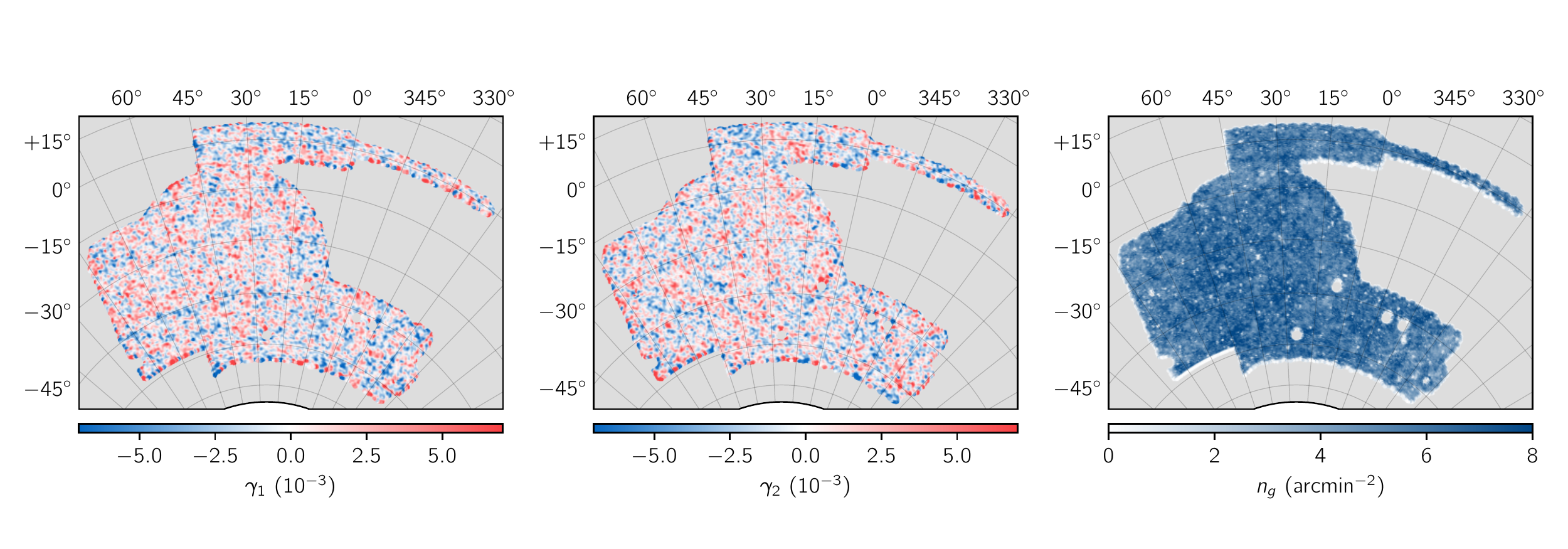}
    \caption{Maps of the two shear components, $\gamma_1$ and $\gamma_2$, and density, $n_g$, of the full DES~Y3 weak lensing catalog.}
    \label{fig:maps_e1_e2_density}
\end{figure*}

The Dark Energy Survey \citet[DES,][]{2005astro.ph.10346T} is a photometric imaging survey that covers around 5000 square degrees of the southern hemisphere in five optical and near-infrared bands ($grizY$). Its observations were carried out at the Cerro Tololo Inter-American Observatory (CTIO) in Chile, using the 570-megapixel DECam camera mounted on the Blanco telescope \citep{2015AJ....150..150F}, during a six-year campaign (2013-2019).  This work is based on data collected during the first three years (Y3) of observations, in particular the DES~Y3 weak lensing shape catalog presented in \citet*{y3-shapecatalog}, which is a subsample of the Y3 Gold catalogue \citep*{y3-gold}, and the inferred redshift distributions presented in \citet*{y3-sompz}.

\subsection{Shape catalog}

{Galaxy shape calibration biases are usually parameterized in terms of multiplicative and additive components.
The DES~Y3 shape measurements} are based on the \mcal algorithm, which allows to self-calibrate most shear multiplicative biases, including selection effects, by measuring the response of the shape measurement pipeline to an artificial shear~\citep{2017ApJ...841...24S,2017arXiv170202600H}. The residual multiplicative biases, at the \numrange{2}{3}\% level, are dominated by shear-dependent detection and blending effects, and the correction was measured on a suite of realistic, DES-Y3-like image simulations presented in \citet*{y3-imagesims}.

The shape catalog was validated by a series of (null) tests presented in \citet*{y3-shapecatalog} and found to be robust to both multiplicative and additive biases. The fiducial DES~Y3 catalog used here comprises ellipticity measurements for \num{100204026} galaxies, with inverse-variance weights based on signal-to-noise ratio and size. The effective area of the sample is \SI{4143}{\deg\squared} \citep*[see][for details]{y3-gold}, corresponding to an effective density of ${\bar{n}=\SI{5.59}{gal\per arcmin\squared}}$. \Cref{fig:maps_e1_e2_density} shows the two ellipticity components and the density of the entire sample.
{We will construct similar maps for each of the four tomographic bin (see next section) and use them to measure cosmic shear power spectra.}

\subsection{Redshift distributions}

\begin{figure}
    \centering
    \includegraphics[scale=0.65]{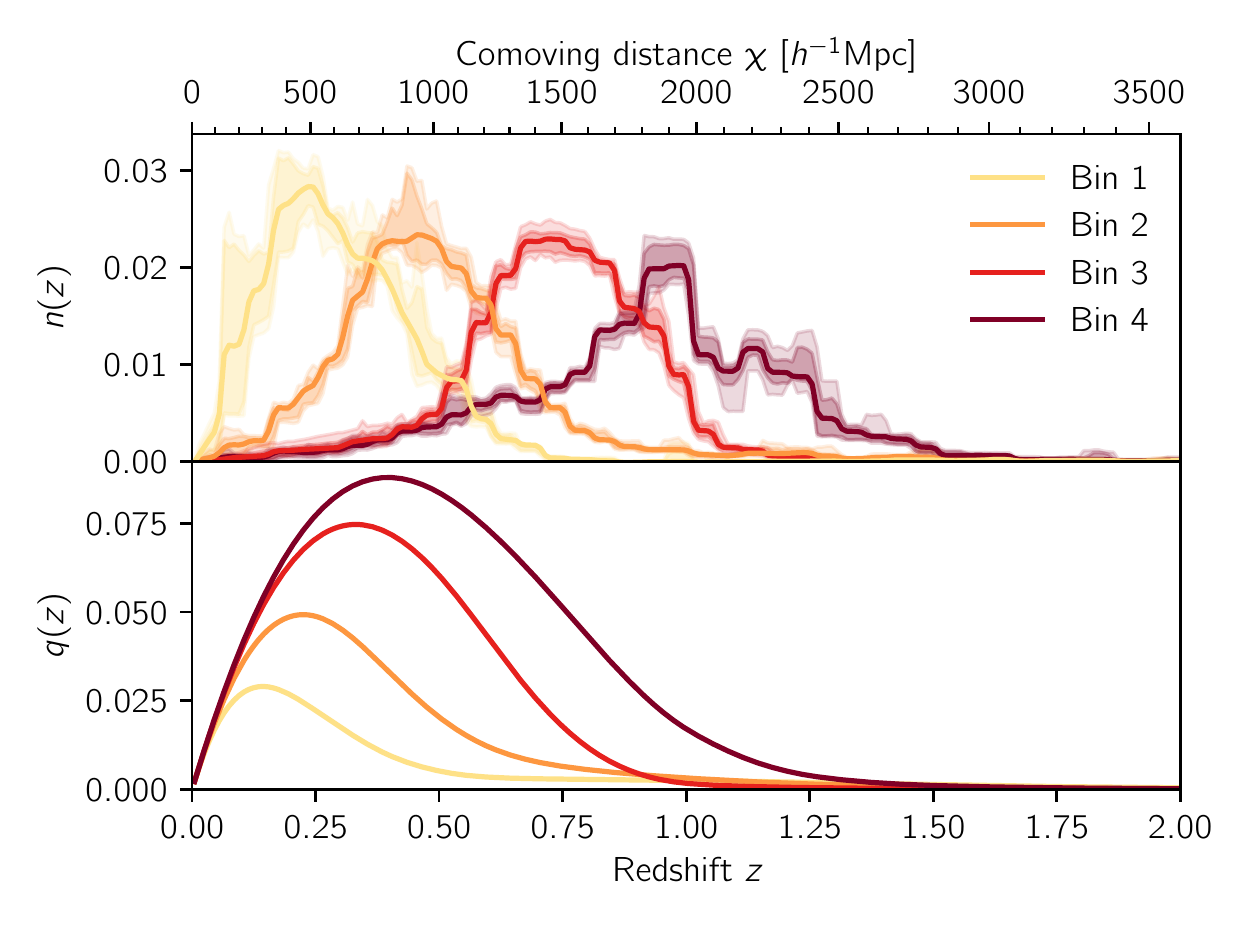}
    \caption{Redshift distributions (top) and corresponding lensing efficiency functions (bottom) for the four tomographic bins. The upper panel shows the mean (solid lines), $\pm1\sigma$ and $\pm2\sigma$ (light bands) percentiles of the ensemble of redshift distributions \citep*{y3-sompz,y3-sourcewz}.}
    \label{fig:dndz_efficiency}
\end{figure}

The DES~Y3 shape catalogue was further divided into four tomographic bins, based on photometric redshifts inferred with the \sompz algorithm \citep[phenotypic redshifts with self-organizing maps,][]{2019MNRAS.489..820B}. {The DES~Y3 implementation is detailed in \citet*{y3-sompz} and connects DES wide-field photometry to
\begin{inparaenum}
    \item deep-field observations \citep*{y3-deepfields}, using image injection with the Balrog software \citep*{y3-balrog}, and to
    \item external spectroscopic and high-quality photometric samples, to calibrate redshifts.
\end{inparaenum}
This Bayesian framework allows to consistently sample the posterior distribution of the four redshift distributions,
while propagating calibration and sample uncertainties. 
Given an ensemble of realizations, uncertainties can be marginalized-over during sampling by means of the \hyperrank method \citep*{y3-hyperrank}. The initial ensemble that was generated for DES~Y3 was subsequently} filtered using constraints on redshifts from cross-correlations with spectroscopic samples, as detailed in \citet*{y3-sourcewz}. The residual uncertainty on the mean redshift of each tomographic bin is of order $\sigma_{\expval{z}}\sim0.01$. Redshift distributions are shown in the upper panel of \cref{fig:dndz_efficiency}, where, for each bin, the ensemble mean is represented by a solid line, and the ensemble dispersion is represented by the light bands. The lensing efficiency functions corresponding to the mean distributions at the fiducial cosmology are shown in the lower panel. 

\section{Methods}
\label{sec:methods}

In this work, we aim at extracting cosmological constraints from the measurements of the angular auto- and cross-power spectra of the tomographic cosmic shear fields inferred from DES~Y3 data. This section describes the estimation of angular spectra from data and the multivariate Gaussian likelihood model, including theoretical predictions for power spectra and their covariance matrix.

\subsection{Angular power spectrum measurements}
\label{sec:pseudocl}

Cosmic shear is represented by a spin-2 field $\bm{\gamma}\equiv(\gamma_1,\gamma_2)$ on the sphere that describes, to linear order, the distortions of the ellipticities of background galaxies. 
A pixelized representation of the cosmic shear field can therefore be obtained by computing the weighted average of the observed ellipticities ${\vb{e}\equiv(e_1,e_2)}$ of galaxies within pixels on the sphere. For each pixel $p$ at angular position $\bm{\theta}_p$, we thus compute 
\begin{equation}
    \hat{\bm{\gamma}}\qty(\bm{\theta}_p) = \frac{ \sum_{i \in p} w_i \vb{e}_i}{ \sum_{i \in p} w_i},
    \label{eq:gmap}
\end{equation}
where the sums run over galaxies, indexed by $i$ and with {inverse-variance} weight $w_i$, that fall into pixel~$p$. The two components of the shear field estimated from the full DES~Y3 weak lensing sample are represented in the left and middle panel of \cref{fig:maps_e1_e2_density}. For the cosmological analysis, we compute maps of the two components of the shear field for each tomographic bin using the \healpy software \citep{2005ApJ...622..759G,2019JOSS....4.1298Z} with a resolution of $\nside=1024$, following the same procedure.
{Note that, prior to \cref{eq:gmap}, observed ellipticities were corrected for additive and multiplicative biases by subtracting the (weighted) mean ellipticity (as done in \citealt*{y3-shapecatalog}) and dividing by the \mcal response, both of which were computed for each bin.}

We now turn to the estimation of shear power spectra.
For full-sky observations, the true shear field for redshift bin $a$, ${\bm{\gamma}^a\equiv(\gamma_1^a,\gamma_2^a)}$, can be decomposed on the basis of spherical harmonics as
\begin{equation}
    \qty(\gamma_1^a \pm i \gamma_2^a)(\bm{\theta}) = - \sum_{\ell m} \qty[E_{\ell m}^a \pm i B_{\ell m}^a] {\,}_{\pm 2}Y_{\ell m}(\bm{\theta}),
    \label{eq:mapmaking}
\end{equation}
where ${}_{s}Y_{\ell m}$ are the spin-weighted spherical harmonics \citep{2011MNRAS.412...65H}. Here, we have used the decomposition of the field into $E$- and $B$-modes, \ie its curl-free and divergence-free components. The shear power spectra are then defined by the covariance matrix of the spherical harmonic coefficients,
\begin{align}
    \expval{E_{\ell m}^a E_{\ell^\prime m^\prime}^{b *}} &= C_\ell^{EE} \qty(\bm{\gamma}^a, \bm{\gamma}^b) \delta_{\ell\ell^\prime} \delta_{mm^\prime}, \label{eq:defclEE} \\
    \expval{E_{\ell m}^a B_{\ell^\prime m^\prime}^{b *}} &= C_\ell^{EB} \qty(\bm{\gamma}^a, \bm{\gamma}^b) \delta_{\ell\ell^\prime} \delta_{mm^\prime}, \label{eq:defclEB} \\
    \expval{B_{\ell m}^a B_{\ell^\prime m^\prime}^{b *}} &= C_\ell^{BB} \qty(\bm{\gamma}^a, \bm{\gamma}^b) \delta_{\ell\ell^\prime} \delta_{mm^\prime}, \label{eq:defclBB}
\end{align}
which can be estimated by
\begin{align}
    \hat{C}_\ell^{EE} \qty(\bm{\gamma}^a, \bm{\gamma}^b) &= \frac{1}{2\ell+1} \sum_{m} E_{\ell m}^a E_{\ell m}^{b *}, \label{eq:Emode}\\
    \hat{C}_\ell^{EB} \qty(\bm{\gamma}^a, \bm{\gamma}^b) &= \frac{1}{2\ell+1} \sum_{m} E_{\ell m}^a B_{\ell m}^{b *}, \label{eq:EBmode}\\
    \hat{C}_\ell^{BB} \qty(\bm{\gamma}^a, \bm{\gamma}^b) &= \frac{1}{2\ell+1} \sum_{m} B_{\ell m}^a B_{\ell m}^{b *}\label{eq:Bmode}.
\end{align}

Gravitational lensing, to first order, does not create $B$-modes, therefore the cosmological signal is contained within $E$-mode power spectra, and $B$-modes can be used to detect potential systematic effects in the data, such as contamination by the point spread function (PSF, see \cref{sec:bmodes_and_psf,app:psf}). However, a number of effects may generate small $B$-modes power spectra (small in comparison to to $E$-mode spectra), including second-order lensing effects (\eg \citealt{2010A&A...523A..28K}), clustering of source galaxies \citep{2002A&A...389..729S}, and intrinsic alignments, as is the case with the model used in our fiducial analysis \citep[TATT, including tidal alignment and tidal torquing mechanisms, from][see \cref{sec:ia_th}]{2019PhRvD.100j3506B}. Therefore, we preserve both components of the field and introduce the vector notation 
\begin{equation}
    \vb{C}_\ell^{ab} \equiv \mqty[C_\ell^{EE}\qty( \bm{\gamma}^a, \bm{\gamma}^b ) \\ C_\ell^{EB}\qty( \bm{\gamma}^a, \bm{\gamma}^b ) \\ C_\ell^{BB}\qty( \bm{\gamma}^a, \bm{\gamma}^b )]
\end{equation}
to denote the vectors made of the two components of the shear power spectra.

{The formalism introduced so far is valid for a full-sky observations.}
In practice, however, the cosmic shear field is only sampled within the survey footprint, at the positions of galaxies. This induces a complicated sky window function, or mask, that correlates different multipoles and biases the estimators defined in \cref{eq:Emode,eq:Bmode}.
We therefore estimate angular power spectra with the so-called pseudo-\cell or MASTER formalism \citep{2002ApJ...567....2H} using the \namaster software \citep{2019MNRAS.484.4127A} to correct for the effect of the mask. We provide a summary of the method here and refer the reader to \citet{2011MNRAS.412...65H} for the development of the pseudo-\cell formalism for cosmic shear,
to \citet{2019MNRAS.484.4127A} for the \namaster implementation and to \citet{2020arXiv201009717N} and \citet{2021arXiv211107203C}  for recent applications of the pseudo-\cell formalism with \namaster to DES~Y1 and  HSC  cosmic shear data.

\begin{figure}
    \centering
    \includegraphics[scale=0.65]{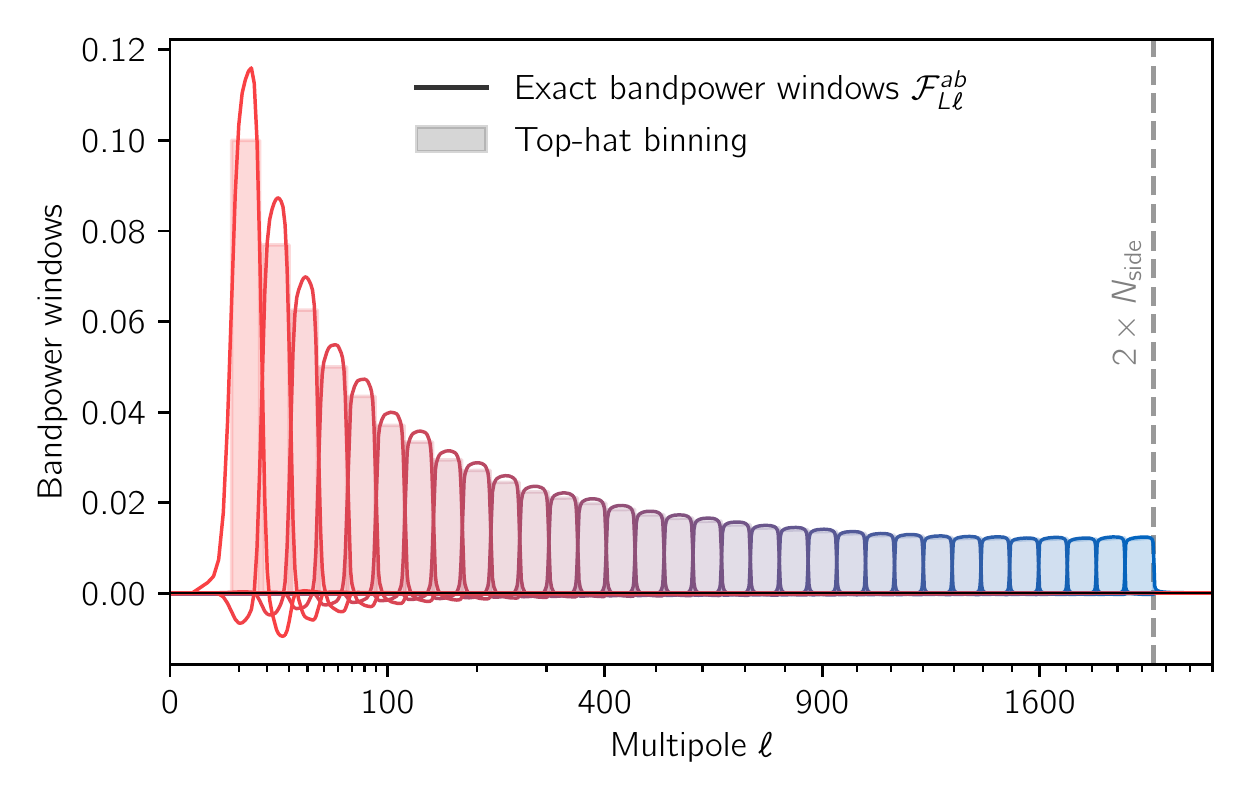}
    \caption{Bandpower window functions $\mathcal{F}_{L\ell}^{ab}$ from \cref{eq:bpws}. Each curve corresponds to one of the 32 bandpowers $L$ from $\lmin=8$ to ${\lmax=2\nside=2048}$, which are equally spaced on a square-root scale throughout this work. The naive binning function is shown by the filled histogram behind.}
    \label{fig:bpws}
\end{figure}

Let $W^a(\bm{\theta})$ be the mask for the shear field in bin $a$, which is zero outside the survey footprint, and let us define the masked shear field ${\tilde{\bm{\gamma}}^a(\bm{\theta}) \equiv W^a(\bm{\theta}) \bm{\gamma}^a(\bm{\theta})}$. Then the cross-power spectrum of the masked fields, \ie the pseudo-spectrum of the fields, has an expectation value given by
\begin{equation}
    \expval{\tilde{\vb{C}}_\ell^{ab}} = \sum_{\ell^\prime} \mathbfss{M}_{\ell \ell^\prime}^{ab} \vb{C}_\ell^{ab},
    \label{eq:clcoupling}
\end{equation}
where $\mathbfss{M}_{\ell \ell^\prime}^{ab}$ is the mode-coupling (or mixing) matrix of the masks, computed analytically from their spherical harmonic coefficients (see, \eg, \citealt{2019MNRAS.484.4127A} for formul\ae). This matrix describes how the mask correlates different multipoles, otherwise independent for full-sky observations, as well as leakages between $E$- and $B$-modes.  While this equation may not be directly inverted due to the loss of information pertaining to masking, one can define an estimator for the binned power spectrum, defined as
\begin{equation}
    \vb{C}_L^{ab} \equiv \sum_{\ell \in L} \omega_L^\ell \vb{C}_\ell^{ab},
    \label{eq:clbinning}
\end{equation}
where $\omega_L^\ell$ is a set of weights defined for multipoles $\ell$ in bandpower $L$ and normalized such that ${\sum_{\ell \in L} \omega_L^\ell=1}$. We also define the mean multipole of each bin as ${\bar{L}\equiv{\sum_{\ell\in L} \omega_L^{\ell} \ell}}$. The binned pseudo-spectrum $\tilde{\vb{C}}_{L}^{ab}$ is similarly defined from the unbinned pseudo-power spectrum $\tilde{\vb{C}}_{\ell}^{ab}$. The estimator for the binned power spectrum is then given by
\begin{equation}
    \hat{\vb{C}}_L^{ab} = \sum_{L^\prime} \qty(\mathbfss{M}^{ab})_{LL^\prime}^{-1} \tilde{\vb{C}}_{L^\prime}^{ab},
    \label{eq:cldecoupling}
\end{equation}
where the binned coupling matrix is
\begin{equation}
    \mathbfss{M}^{ab}_{LL^\prime} \equiv \sum_{\ell \in L} \sum_{\ell^\prime \in L^\prime} \omega_L^\ell \mathbfss{M}_{\ell \ell^\prime}^{ab}.
\end{equation}
The successive operations of masking, binning and decoupling described by \cref{eq:clcoupling,eq:clbinning,eq:cldecoupling} are generally not permutable, such that the expectation value of the estimator in \cref{eq:cldecoupling} can differ from a naive binning of the theoretical prediction for $\vb{C}_\ell^{ab}$, as in \cref{eq:clbinning}. Instead, the estimated shear power spectra must be compared to
\begin{equation}
    \expval{\hat{\vb{C}}_L^{ab}} = \sum_{\ell} \mathcal{F}_{L\ell}^{ab} \vb{C}_L^{ab}
    \label{eq:bpws_cl}
\end{equation}
where the bandpower windows $\mathcal{F}_{L\ell}^{ab}$ are given by
\begin{equation}
    \mathcal{F}_{L\ell}^{ab} = \sum_{L^\prime} \qty(\mathbfss{M}^{ab})_{LL^\prime}^{-1} \sum_{\ell^\prime \in L^\prime} \omega_{L^\prime}^{\ell^\prime} \mathbfss{M}_{\ell^\prime \ell}^{ab}.
    \label{eq:bpws}
\end{equation}
Throughout this work, we adopt an equal-weight binning scheme {(\ie $\omega_\ell=1$ if ${\ell\in L}$, $0$ otherwise)} with 32 square-root-spaced bins defined between multipoles ${\ell_{\rm min}=8}$ and ${\ell_{\rm max}=2048}$ (shown by the colored bars in \cref{fig:bpws}). This choice ensures a good balance of signal-to-noise ratio across bandpowers $L$ while remaining flexible for scale cuts at both low and high multipoles, \ie large and small scales (in comparison to linear and logarithmic bins that are too coarse for low and high multipoles respectively).
We use weighted galaxy count maps as masks, using the weights computed by the \mcal algorithm. This is a close approximation to inverse-variance masks since the \mcal weights are themselves inverse-variance weights of ellipticity measurements \citep[see][]{2020arXiv201009717N}.
The exact bandpower windows $\mathcal{F}_{L\ell}^{ab}$ for these binning and masking schemes are compared to the naive binning {(\ie top-hat)} windows in \cref{fig:bpws}.
{In particular, we observe that the exact windows extend beyond the top-hat ones, with some negative terms, especially for small multipoles below $\ell\lesssim200$.}

We compute tomographic cosmic shear power spectra with \namaster, given our binning and masking schemes, from the shear maps computed from \cref{eq:mapmaking}. These include a shape-noise component due to the intrinsic ellipticities of galaxies, which contributes an additive noise bias to the estimated auto-power spectra (whereas cross-spectra do not receive such contributions). For each tomographic bin, the noise power spectrum $N_\ell^a$ is flat for full-sky observations, and can be approximated by $N_\ell^a\approx\sigma_{e,a}^2/\bar{n}^a$, where $\sigma_{e,a}^2$ is the standard deviation of single-component (measured) ellipticity and $\bar{n}^a$ is the galaxy density in redshift bin $a$. We follow \citet{2020arXiv201009717N} and estimate the binned noise pseudo-power spectrum, which is constant, by
\begin{equation}
    \tilde{\vb{N}}_L = \Omega_{\rm pix} \expval{\sum_{i \in p} w_i^2 \frac{e_{1,i}^2+e_{2,i}^2}{2} }_{p},
    \label{eq:noise_pcl}
\end{equation}
where $\Omega_{\rm pix}$ is the pixel area in steradians (about \SI{11.8}{arcmin\squared} for ${\nside=1024}$), and the expectation value is computed for all pixels, including those outside the survey footprint (where the value is zero).
The binned noise power spectrum can then be computed with \cref{eq:cldecoupling} and subtracted from the estimated spectra. 
We note that this analytical estimation coincides with the expectation value of the auto-power spectra measured after applying random rotations to galaxies. Random rotations preserve the density of galaxies and the ellipticity distribution of the catalog and therefore properties of shape-noise {(including its potential spatial variations)}, while canceling any spatial correlation (that is, both in the $E$- and $B$-modes). We also applied this procedure and verified that the result agrees with the analytical estimation, which has the advantage of being noiseless and is therefore preferred for our measurements.

We do not apply any purification of $E$- and $B$-modes
{\citep{2001PhRvD..65b3505L,2006PhRvD..74h3002S,2009PhRvD..79l3515G,2019MNRAS.484.4127A}} since the $B$-mode signal is largely subdominant {and does not contain cosmological information, to first order}. Moreover, this would require an apodization of the mask, that is speckled with empty pixels due to fluctuations in the density of source galaxies and small vetoed areas, and thus significantly decrease the effective survey area.

Finally, we correct for the effect of the pixelization of the shear fields into \healpix maps. As noted in \citet{2020arXiv201009717N}, it depends on the density of galaxies, at fixed resolution: at low density, each pixel contains at most one galaxy and the map is sampling the shear field itself (but has many empty pixels), whereas at higher density, we are estimating the average of the shear field within each pixel. Here, for a resolution of $\nside=1024$, we find that pixels with at least one galaxy contain on average \num{17.2} to \num{17.5} galaxies for all four tomographic bins, meaning that we are indeed sampling the averaged shear field (although a small fraction of pixels, especially on the footprint edges, have only one galaxy).
This is then corrected for by dividing the pseudo-spectra $\tilde{\vb{C}}_\ell^{ab}$ by the (squared) \healpix pixel window function $F_\ell^2${, or equivalently, assigning weights $w_{L}^{\ell}=1/F_\ell^2$ for ${\ell\in L}$ for measurements (except for theoretical predictions). We test the effect of the resolution parameter in \cref{app:ic_cont}, and verify that it has negligible impact on cosmological constraints.} In \cref{sec:validation}, we validate these hypotheses and the measurement pipeline with Gaussian and $N$-body simulations.

\begin{figure*}
    \centering
    \includegraphics[scale=0.65]{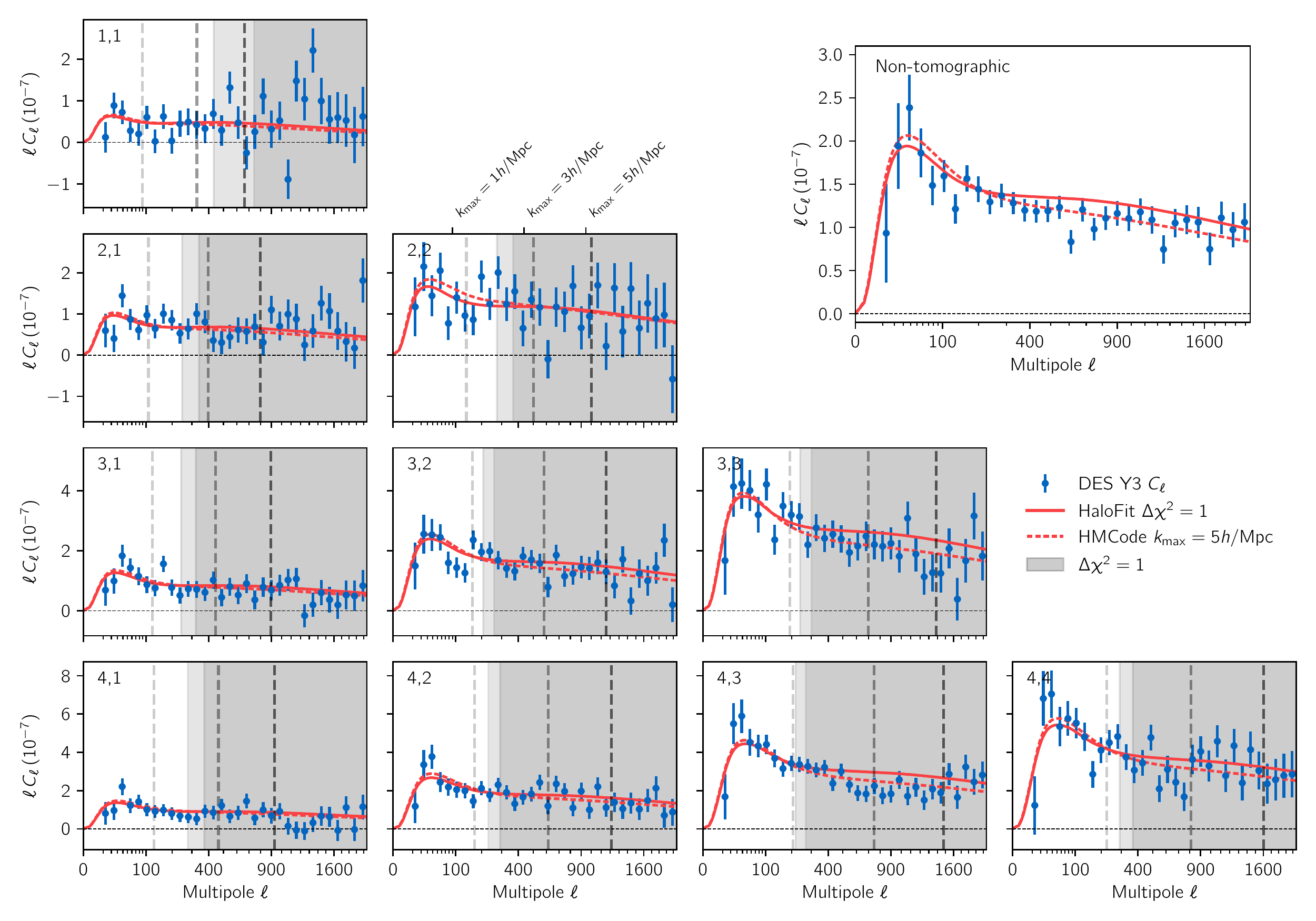}
    \caption{Cosmic shear power spectra measured from DES~Y3 data. Each panel {in the lower left triangle} corresponds to a redshift bin pair indicated in the upper left corner. The measured $E$-mode component of the binned, noise-bias corrected power spectra is shown in blue with error bars from an analytic covariance matrix (see \cref{sec:like_cov}). The gray shaded regions show scales that are not used in the fiducial analysis ($\Delta\chi^2=1$) where the effect of baryons is neglected, {with extra points removed when combining with shear ratios shown in light gray (see \cref{sec:scalecuts_fid})}. The corresponding best-fit model within \lcdm, discussed in \cref{sec:res_lcdm}, is represented by red {solid lines}. The gray dashed lines show the scale cuts corresponding to $\kmax=$\SIlist{1;3;5}{\h\per\mega\parsec} (see also \cref{sec:scalecuts_kmax}), and the corresponding best-fit model using \hmcode and $\kmax=\SI{5}{\h\per\mega\parsec}$, discussed in \cref{sec:res_baryons}, is represented by red dashed lines. The upper right panel shows the measured non-tomographic shear power spectrum of DES~Y3 data in blue, along with the theory expectation corresponding to the best-fit of the tomographic analysis, in red. For readability, all measurements and errors bars are scaled by the mean multipole $\bar{L}$ of each bandpower $L$, \ie the data points show $\bar{L}\hat{\vb{C}}_L^{EE}$ and are compared to theoretical predictions of $\ell C_\ell$.}
    \label{fig:clobs}
\end{figure*}

The estimated shear power spectra for DES Y3 data are shown in \cref{fig:clobs}, along with the best-fit model {for our fiducial \lcdm results}.

\subsection{Modeling}
\label{sec:modeling}

In this section, we describe the theoretical model for the observed shear power spectra, including systematic uncertainties.

\subsubsection{Theoretical background}

Gravitational lensing deflects photons from {straight} trajectories and the deflection angle can be written as the gradient (on the sphere) of the lensing potential $\psi(\bm{\theta})$. In the Born approximation, the lensing potential up to comoving distance $\chi$ is given by the projection of the three-dimensional Newtonian gravitational potential~$\Psi$ along the line of sight, such that
\begin{equation}
    \psi(\bm{\theta}, \chi) = 2 \int_0^\chi \dd{\chi'} \frac{\chi-\chi'}{\chi\chi'} \Psi\qty(\chi'\bm{\theta},\chi'),
    \label{eq:lens_pot}
\end{equation}
where we assumed a flat Universe \citep{2010CQGra..27w3001B}. The Jacobian of the deflection angle can further be decomposed into its trace and trace-less parts, defining the spin-0 convergence field, $\kappa$, and the spin-2 shear field, $\bm{\gamma}$. Both fields can therefore be expressed in terms of second-order derivatives of the lensing potential. In the spherical harmonics representation, we have
\begin{align}
    \kappa &= \frac{1}{4} \qty(\eth \bar{\eth} + \bar{\eth} \eth) \psi = \frac{1}{2} \nabla_{\bm{\theta}}^2 \psi, \label{eq:kappa_psi} \\
    \bm{\gamma} &= \gamma_1 + i\gamma_2 = \frac{1}{2} \eth \eth \psi \label{eq:gamma_psi},
\end{align}
where $\eth$ and $\bar{\eth}$ are the raising and lowering operators of the spin-weighted spherical harmonics, ${}_{s}Y_{\ell m}$ {(see \citealt{2005PhRvD..72b3516C} for details and, \eg, \citealt{2018MNRAS.475.3165C} for an application to curved-sky lensing mass maps)}. The Newtonian potential is related to the matter overdensity field $\delta$ via the Poisson equation,
\begin{equation}
    \nabla^2\Psi= \frac{3 \Om H_0^2}{2ac^2} \delta,
\end{equation}
where $\Om$ is the matter density parameter, $H_0$ is the Hubble constant today and $a=1/(1+z)$ is the scale factor. 
Combining \cref{eq:lens_pot,eq:kappa_psi}, we obtain
\begin{equation}
    \kappa(\bm{\theta}, \chi) = \frac{3 \Om H_0^2}{2c^2} \int_0^\chi \frac{\dd{\chi'}}{a(\chi')} \frac{\chi-\chi'}{\chi\chi'} \delta\qty(\chi'\bm{\theta},\chi'),
\end{equation}
where we have added the radial component of the Laplacian of the potential, $\nabla_{\chi}^2\Psi$, that vanishes in the integration.

For a sample of galaxies, the observable convergence and shear fields are integrated over comoving distance and weighted by their redshift distribution $n_a(\chi)$, where $a$ denotes the bin index.
In the Limber approximation \citep{1953ApJ...117..134L,1992ApJ...388..272K,1998ApJ...498...26K,2008PhRvD..78l3506L}, the convergence cross-power spectrum for bins $a$ and $b$ is
\begin{equation}
    C^{\kappa_a \kappa_b}_\ell = \int \dd{\chi} \frac{q_a(\chi) q_b(\chi)}{\chi^2} P_{\rm NL}\qty(k=\frac{\ell+1/2}{\chi}, z(\chi)),
    \label{eq:limber}
\end{equation}
where the lensing efficiency is given by
\begin{equation}
    q_a(\chi) = \frac{3 \Om H_0^2}{2c^2} \frac{\chi}{a(\chi)} \int_{\chi}^{\chi_H} \dd{\chi'} n_a(\chi') \frac{\chi-\chi'}{\chi'},
\end{equation}
where $\chi_H$ is the distance to the horizon (effectively, the comoving distance where the redshift distributions vanish). The lensing efficiency functions for DES~Y3 galaxies are shown in the lower panel of \cref{fig:dndz_efficiency}. Given \cref{eq:kappa_psi,eq:gamma_psi}, the cosmic shear $E$-mode power spectrum is given by 
\begin{equation}
    C^{ab}_\ell = T_\ell C^{\kappa_a \kappa_b}_\ell,
\end{equation}
where the prefactor, $T_\ell = (\ell+2)(\ell+1)\ell(\ell-1) / (\ell+1/2)^4$, is often replaced by 1, an excellent approximation for $\ell \gtrsim 10$  \citet[see][for a complete discussion]{2017MNRAS.469.2737K,2017MNRAS.472.2126K}. We verified that these two approximations are correct, given our binning scheme, with an error of at most 0.2\% on the largest scales considered.

\subsubsection{Non-linear power spectrum}
\label{sec:pnl}

\begin{figure*}
    \centering
    \includegraphics[scale=0.65]{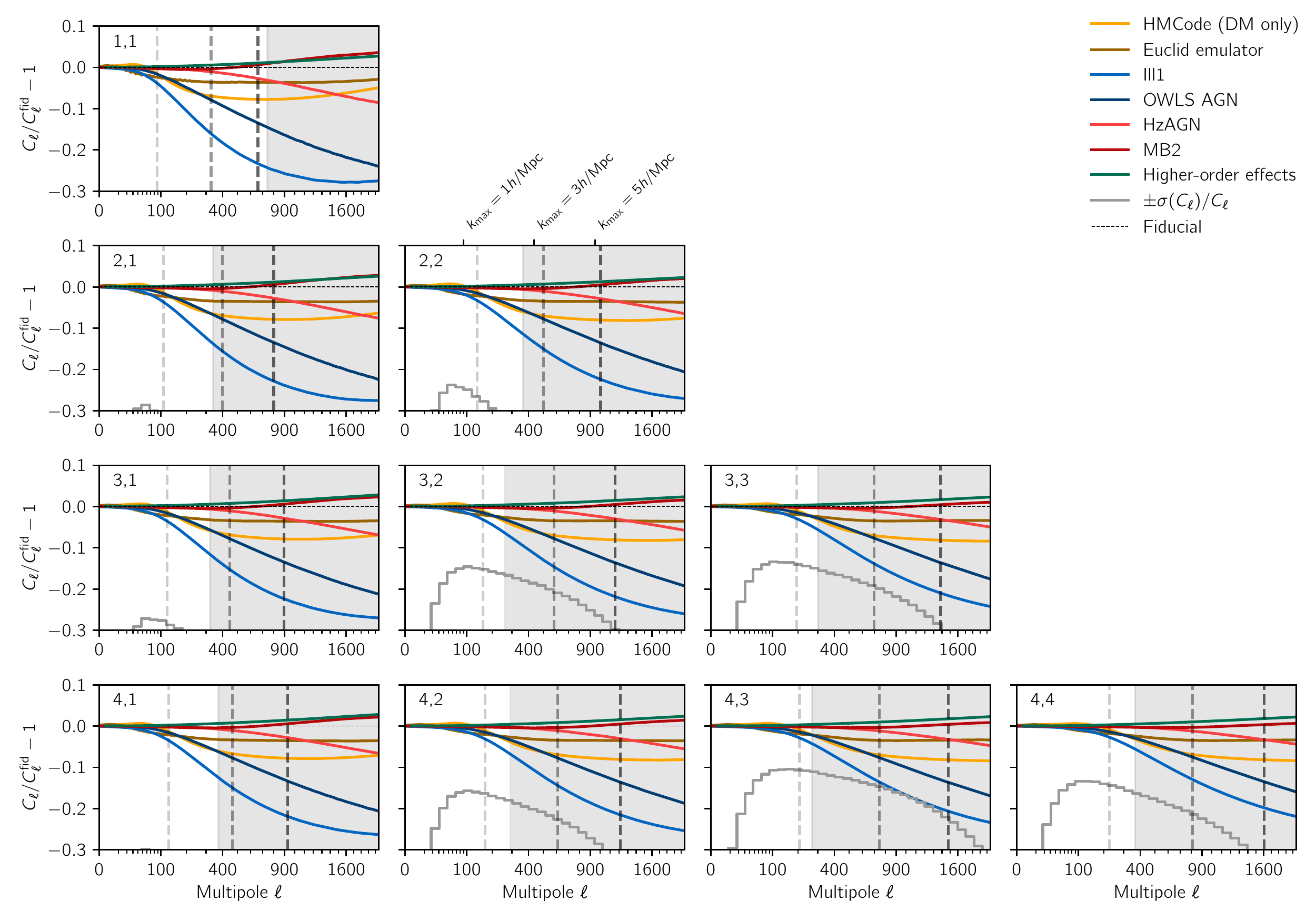}
    \caption{Residual shear power spectra with respect to the fiducial power spectra, $C_\ell^{\rm fid}$. The orange (\hmcode) and brown (\textsc{Euclid Emulator}) curves show residuals for alternative prescription of the non-linear power spectrum (see \cref{sec:pnl}). The blue and red curves show the effect of baryons as predicted by} four hydrodynamical simulations (Illustris, OWLS AGN, Horizon AGN and MassiveBlack~II). Higher-order lensing effects computed with \cosmolike are also shown, in green, to be small. The error bars are shown by the gray step-wise lines which represent $\pm\sigma(C_\ell)/C_\ell$ on the same scale (only $-\sigma(C_\ell)/C_\ell$ is visible). The gray shaded regions show scales that are not used in the fiducial analysis where the effect of baryons is neglected. The gray dashed lines show the scale cuts corresponding to $\kmax=$\SIlist{1;3;5}{\h\per\mega\parsec} (see \cref{sec:scalecuts_kmax}).
    \label{fig:clres}
\end{figure*}

Following the general methodology of the DES~Y3 large-scale structure analysis set in \citet*{y3-generalmethods}, for our fiducial model we compute the non-linear matter power spectrum~$P_{\rm NL}(k,z)$ using the Boltzmann code \camb \citep{2000ApJ...538..473L,2012JCAP...04..027H} with the \halofit extension to non-linear scales \citep{2003MNRAS.341.1311S}, with updates to dark energy and massive neutrinos from \citet{2012ApJ...761..152T}.
\halofit is reported to be accurate at the 5\% level for $k \leq \SI{1}{\h\per\mega\parsec}$, when compared to $N$-nody simulations, and degrading for smaller scales.
However, \citet*{y3-generalmethods} showed that DES Y3 cosmic shear is insensitive to varying the prescription to model the small-scale power spectrum by substituting \halofit for \hmcode (with dark matter only), the \textsc{Euclid Emulator}, or the \textsc{Mira-Titan Emulator} \citep{2015MNRAS.454.1958M,2019MNRAS.484.5509E,2017ApJ...847...50L}. {We show a comparison of some of these prescriptions in \cref{fig:clres}} and we verify {the robustness of our fiducial choice in} in \cref{sec:modeling_validation_synth}.

\subsubsection{Intrinsic alignments}
\label{sec:ia_th}

Galaxies are extended objects and therefore subject to tidal forces. Their intrinsic shapes, or ellipticities, are consequently not fully random but rather tend to align with the tidal field of the gravitational potential and therefore each other \citep{2004PhRvD..70f3526H,2007NJPh....9..444B}. As a consequence, the shear power spectrum estimated from galaxies receives additional contributions from the correlation of intrinsic shapes, $C_{\ell,{\rm I}{\rm I}}^{ab}$, and the cross-correlations of intrinsic shapes with the cosmological shear field, $C_{\ell,{\rm GI}}^{ab}$ and $C_{\ell, {\rm IG}}^{ab}$, such that the theoretical spectrum of the observed signal reads ${C_\ell^{ab}+C_{\ell,{\rm GI}}^{ab}+C_{\ell, {\rm IG}}^{ab}+C_{\ell,{\rm I}{\rm I}}^{ab}}$.

In this work, we follow the DES Y3 analysis of cosmic shear in real space \citep*{y3-generalmethods,y3-cosmicshear1,y3-cosmicshear2} and use the so-called TATT framework \citep{2019PhRvD.100j3506B} as our fiducial choice to model these extra terms stemming from intrinsic alignments (IA). This model unified tidal alignment (TA) with tidal torquing (TT) mechanisms, proposed by \citet{2001MNRAS.320L...7C,2001ApJ...559..552C,2002MNRAS.332..788M}, thanks to a perturbative expansion of the intrinsic galaxy shape field in the density and tidal fields, up to second order in the tidal field. We refer the reader to \citet*{y3-cosmicshear2} for full details of the implementation and a justification of this choice. The TA and TT contributions are each modulated by an amplitude (respectively $\Ata$ and $\Att$) and a redshift-dependence parameter (respectively $\ata$ and $\att$), with an additional linear bias $\bta$ of sources contributing to the TA signal. The non-linear alignment model \citep[NLA,][]{2004PhRvD..70f3526H,2007NJPh....9..444B}, commonly used in cosmic shear analyses \citep{2018PhRvD..98d3528T,2021A&A...645A.104A,2020PASJ...72...16H,2019PASJ...71...43H} is contained in the TATT framework and corresponds to the case $\Att=\bta=0$.

The TATT model also predicts a small, but non-zero $B$-mode power spectrum, when $\bta\neq0$ or $\Att\neq0$. In the main parts of the analysis, the $B$-mode spectrum is not used for cosmological analysis. Instead, it is demonstrated in \cref{sec:bmodes} that DES~Y3 data is consistent with no $B$-modes, rejecting the hypothesis of a strong contamination of the signal by systematic effects that would source $B$-modes, such as leakage from the PSF, measured in \cref{sec:psf} and \cref{app:psf}. This test thereby also excludes a detectable contribution of the IA $B$-mode signal, with the unlikely caveat that systematic effects and IA may cancel each other.
In addition, the PSF test allows us to predict the contamination of $B$-mode spectra, which is found to be subdominant, by an order of magnitude, to the TATT-predicted $B$-mode signal for $\Att=1$, which is well within current $E$-mode constraints.
Therefore, we will extend the cosmological analysis in \cref{sec:res_IA} and include $B$-mode measurements to improve constraints on the TATT parameters. To do so, we employ the same pseudo-\cell formalism and extend the mode-coupling matrices in \cref{eq:clcoupling,eq:bpws_cl} to account for the $B$-mode component. Note that \namaster computes both $E$ and $B$ components of the mixing matrices as well as the cross-terms accounting for leakages between the two components. The fiducial analysis simply discards those terms, as $B$-to-$E$ mode leakage is found to be negligible. However, $E$-to-$B$ mode leakage is found to significantly contribute to the $B$-mode signal, in comparison to the TATT-predicted $B$-mode signal ({they are of comparable magnitude} for $\Att$ of order unity). Therefore, the extended analysis including $B$-mode measurements uses consistent modeling of multipole coupling and $E$/$B$-mode leakage. The covariance matrix for the $B$-mode measurement as well as the cross-covariance between $E$- and $B$-mode measurements are computed from a set of \num{10000} Gaussian simulations based on DES~Y3 data, as detailed in \cref{sec:sims_gaussian}.

\subsubsection{Effects of baryons}
\label{sec:baryons}

Astrophysical, baryonic processes redistribute matter within dark-matter halos and modify the matter power spectrum at small scales \citep{2018MNRAS.480.3962C,2019JCAP...03..020S,2020JCAP...04..019S,2021MNRAS.502.6010H}. Feedback mechanisms from active galactic nuclei and supernov{\ae} heat up their environment and suppress clustering in the range $k\sim$\SIrange{1}{10}{\h\per\mega\parsec}, while cooling mechanisms enhance clustering on smaller scales. The complex physics involved in these mechanisms has been modeled in multiple hydrodynamical simulations \citep{2011MNRAS.415.3649V,2014MNRAS.444.1453D,2014MNRAS.444.1518V,2015MNRAS.450.1349K}. However the absolute and relative amplitudes of the various effects remain poorly understood and constitute a major source of uncertainty, at the level of tens of percent, on the matter power spectrum at scales $k\gtrsim\SI{5}{\h\per\mega\parsec}$, and on the shear power spectrum at multipoles as low as $\ell\gtrsim100$, as shown on \cref{fig:clres} \citep[see also][]{2019MNRAS.488.1652H}.

Our fiducial analysis follows the DES~Y3 analysis and discards scales that are strongly affected by baryonic effects, as detailed in \cref{sec:scalecuts_fid}. In general, the impact of baryons on the shear power spectrum can be computed by rescaling the matter power spectrum,
\begin{equation}
    P_{\rm NL}(k,z) \rightarrow P_{\rm NL}(k,z) \frac{P_{\rm hydro}(k,z)}{P_{\rm DM}(k,z)},
    \label{eq:baryon_Pk_ratio}
\end{equation}
where $P_{\rm hydro}(k,z)$ and $P_{\rm DM}(k,z)$ are the matter power spectra measured from hydrodynamical simulations, respectively with and without the effects of baryons. {In particular, we will use four simulations, selected to provide a diverse range of scenarios: Illustris \citep{2014MNRAS.444.1518V}, OWLS AGN \citep{2011MNRAS.415.3649V}, Horizon AGN \citep{2014MNRAS.444.1453D} and MassiveBlack II \citep{2015MNRAS.450.1349K}. We will use this approach to evaluate the impact of baryons, shown in \cref{fig:clres}, and determine our fiducial set of scale cuts, in \cref{sec:scalecuts_fid}.}

We will later extend our analysis to smaller scales, which requires {to model and marginalize over} baryonic effects. To do so, we will use \hmcode\footnote{\url{https://github.com/alexander-mead/HMcode}} \citep{2015MNRAS.454.1958M}, instead of \halofit, to simultaneously model the effects of non-linearities and baryonic feedback on the matter power spectrum. This adds one or two extra parameters, namely the minimum halo concentration $\Ahm$ and the halo bloating parameter $\etahm$, which were shown to approximately follow the linear relation $\etahm=1.03-0.11\Ahm$ for various simulations \citep[see][]{2015MNRAS.454.1958M}. {Although \citet{2021MNRAS.502.1401M} recently presented an updated version of \hmcode with improved treatment of baryon-acoustic oscillation damping and massive neutrinos, we will only consider the 2015 version of the code, which was available at the onset of this work. We note that \citet{2021arXiv210904458T} found only a small impact of \hmcode versions on cosmological constraints derived from cosmic shear and Sunyaev-Zeldovich effect cross-correlations.}

\subsubsection{Shear and redshift uncertainties}
\label{sec:dz_mz}

We include uncertainties on the shear calibration and redshift distributions following the DES Y3 real-space analysis \citep*{y3-generalmethods,y3-cosmicshear1,y3-cosmicshear2}.

In our fiducial model, uncertainties in redshift distributions are captured by allowing overall translations of the fiducial redshift distributions, shown in \cref{fig:dndz_efficiency}, such that
\begin{equation}
    n_a(z) \rightarrow n_a(z+\Delta z_a).
    \label{eq:deltaz}
\end{equation}
We parametrize the residual uncertainty in the shear calibration following a standard procedure which amounts to an overall rescaling of the shear signal in each redshift bin, such that
\begin{equation}
    C_\ell^{ab} \rightarrow (1+m_a)(1+m_b) C_\ell^{ab}.
    \label{eq:shearm}
\end{equation}
The four shear biases, $m_a$, are assumed to be redshift-independent within each bin.
Both of these choices are approximations to the more sophisticated approaches developed over the course of the DES~Y3 analysis.

For redshift uncertainties, the \sompz method provides a ensemble of redshift distributions encapsulating the full uncertainty \citep*{y3-sompz}, and not just that of the mean redshift. However, it was shown in \citet*{y3-hyperrank} and \citet*{y3-cosmicshear1} that the simpler parametrization of \cref{eq:deltaz} is sufficient for DES~Y3, {which we test in \cref{app:ic_cont}}.
For shear calibration, a new approach was developed alongside the image simulations presented in \citet*{y3-imagesims}.
In short, it was shown that the redshift distribution of a sample, $n(z)$, corresponds to the response of the shear estimated from this sample to a cosmological shear signal, as a function of the redshift of the signal. In the presence of galaxy blending, the response is modified, which may be captured by an effective redshift distribution, $n_\gamma(z)$, normalized to $1+m$. Realistic simulations, that used the same pipelines as DES Y3 data for co-addition, detection and shear measurements, allowed to jointly estimate residual uncertainties in shear and redshift biases. These results were subsequently mapped onto the standard parametrization of \cref{eq:deltaz,eq:shearm}, thus defining the priors over these parameters, as detailed in \cref{tab:params}. Extensive testing demonstrated that our fiducial approach is sufficiently accurate given the statistical uncertainties in DES~Y3 \citep*[see][for details]{y3-imagesims,y3-hyperrank,y3-cosmicshear1}.

\subsubsection{Higher-order shear}
\label{sec:higher_order}

Our modeling ignores higher-order contributions to the shear signal due to the magnification and clustering of the galaxy sample as well as the fact we can only access the reduced shear, given by $\gamma/(1-\kappa)$. These contributions are computed in \citet*{y3-generalmethods,y3-cosmicshear2} and found to be below 5\% for the scales used in this analysis, as shown by the orange curves in \cref{fig:clres}. We verified that they have a negligible impact on cosmological constraints for DES~Y3.

\subsection{Likelihood and covariance}
\label{sec:like_cov}

We assume cosmic shear spectrum measurements follow a multivariate Gaussian distribution with fixed covariance. The theoretical predictions detailed in the previous section are convolved with the bandpower windows, following \cref{eq:bpws_cl,eq:bpws}.

\begin{figure*}
    \centering
    \includegraphics[scale=0.65]{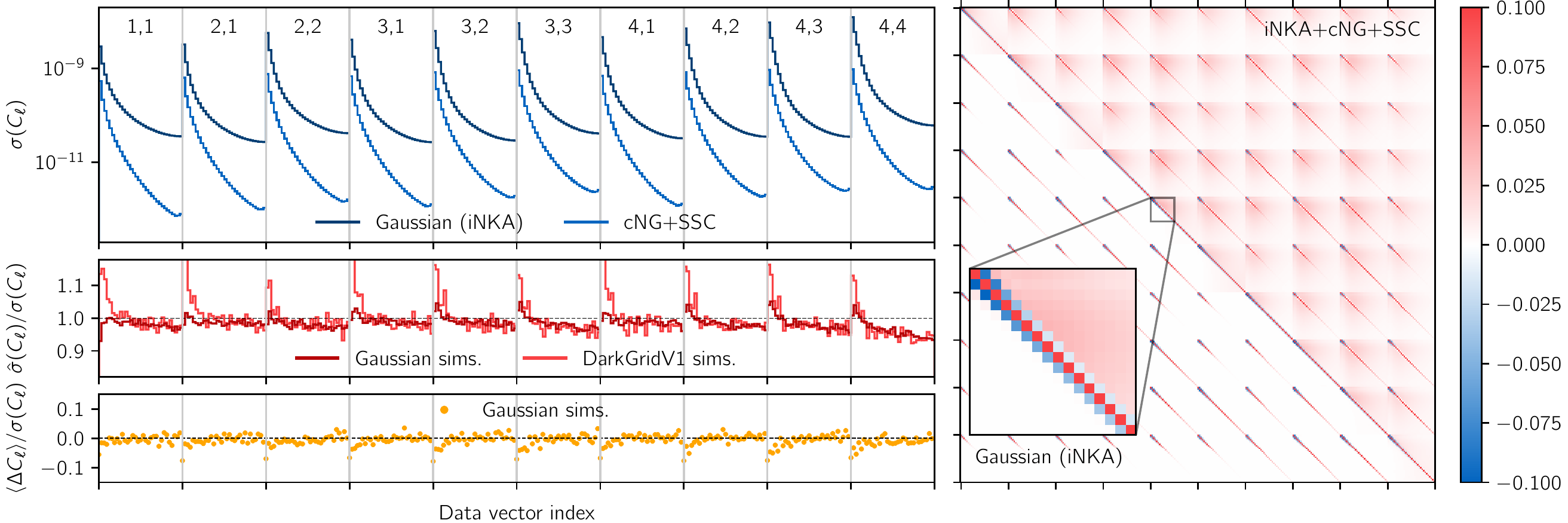}
    \caption{Features and validation of the analytical covariance matrix used in this work, computed with \namaster and \cosmolike. \textit{Upper left}: error bars given by the square-root of the diagonal of the Gaussian (dark blue) and non-Gaussian (light blue) contributions to the covariance matrix. \textit{Middle left}: comparison of the error bars computed from Gaussian simulations (dark red) and \dgv simulations (light red) with the fiducial error bars. \textit{Lower left}: residuals of the pseudo-\cell measurements from the Gaussian simulations with respect to the input (binned) spectra. In all left panels, the horizontal axis corresponds to indices of the components stacked data vector. {The corresponding redshift bin pairs are indicated at the top of the upper panel, with each block corresponding to multipoles in the range \numrange{8}{2048}.} \textit{Right}: correlation matrix, with only the Gaussian contribution in the lower triangle, and both Gaussian and non-Gaussian contributions in the upper triangle (note the normalization in the range \numrange{-0.1}{+0.1}).}
    \label{fig:cov}
\end{figure*}

The covariance of $E$-mode shear power spectra is computed analytically as follows. It is decomposed as a sum of Gaussian and non-Gaussian contributions from the shear field. The Gaussian contribution is computed with \namaster using the improved narrow-kernel approximation (iNKA) estimator developed in \citet{2019JCAP...11..043G} and optimized by \citet{2020arXiv201009717N}. This estimator correctly accounts for mode-mixing pertaining to masking and binning, consistently with the pseudo-\cell framework presented in \cref{sec:pseudocl}.
It requires the mode-coupled pseudo-\cell spectra, computed from the theoretical full-sky spectra convolved by the mixing matrix from \cref{eq:clcoupling}, and including noise bias for auto-spectra, computed from the data with \cref{eq:noise_pcl}. These are then rescaled by the product of masks over all pixels \citet[for details, see][]{2020arXiv201009717N}.

The non-Gaussian contribution is the sum of two terms: the connected four-point covariance (cNG) arising from the shear field trispectrum, and the so-called super-sample covariance (SSC), accounting for correlations of multipoles used in the analysis with super-survey modes. Both non-Gaussian terms are computed using the \cosmolike software \citep{2014MNRAS.440.1379E,2017MNRAS.470.2100K}, with formulae derived in \citet{2009MNRAS.395.2065T,2014PhRvD..90l3523S}. These analytic expressions do not account for the exact survey geometry and only apply a scaling by the fraction of observed sky, $\fsky$. Therefore, we interpolate these computations at all pairs of integer-valued multipoles and use the bandpower windows from \cref{eq:bpws} to obtain an approximation of the non-Gaussian covariance terms for the binned power spectrum estimator described in the previous section. The non-Gaussian terms (cNG+SSC) are subdominant with respect to the Gaussian contribution (see the upper left panel of \cref{fig:cov}) and this represents a good approximation to the extra covariance of different multipoles (\ie off-diagonal terms), which becomes non-negligible only on the smallest scales.

\Cref{fig:cov} illustrates properties of the fiducial covariance matrix, {computed as explained above}. First, as can be seen on the left panel, the non-Gaussian terms are largely subdominant in the computation of the error bars. Then, the right panel, showing the correlation matrix, reveals that multipole bins are largely uncorrelated in the Gaussian covariance, and only correlated at the 10\% level at most due to the non-Gaussian contributions. Adjacent multipole bins are actually slightly anti-correlated due to mode coupling and decoupling, at the 6\% level for the lowest bins to below 1\% for the highest bins.

The covariance matrix of $B$-mode shear power spectra and the cross-covariance between $E$- and $B$-mode power spectra are computed from Gaussian simulations, presented in \cref{sec:sims_gaussian}, as the original NKA estimator was found to be unreliable for these spectra in \citet{2019JCAP...11..043G}.

\subsection{Parameters and priors}

For our fiducial analysis, we vary six parameters of the \lcdm model, namely the total matter density parameter $\Om$, the baryon density parameter $\Ob$, the Hubble parameter $h$ (where ${H_0=\SI{100}{\h \kilo\meter\per\second\per\mega\parsec}}$), the amplitude of primordial curvature power spectrum $\As$ and the spectral index $\ns$, and the neutrino physical density parameter $\Onu h^2$.

We also vary the five parameters of the intrinsic alignments model, TATT. When restricting to the NLA model, we fix ${\Att=\att=\bta=0}$. Our validation tests are carried out assuming the TATT model, but using the NLA best-fit values from \citet{2019MNRAS.489.5453S} based on DES Year~1 data, since this work found no strong preference for the more complex model.

In addition to the cosmological and astrophysical parameters described above, our analysis includes two nuisance parameters per redshift bin to account for uncertainties in shape calibration ($m_a$) and redshift distributions (${\Delta z_a}$), as described in \cref{sec:dz_mz}.

The full list of parameters for the baseline \lcdm model with their priors is shown in \cref{tab:params}.
Throughout this paper we assume the \planck 2018 \citep{2020A&A...641A...6P} best-fit cosmology derived from TT,TE,EE+lowE+lensing+BAO data as our fiducial parameter values.

In addition, we will consider alternative models that require extra varied parameters:
\begin{itemize}
    \item When using \hmcode to model small scales, we vary either $\Ahm$ only \citep[using the relationship between $\Ahm$ and $\etahm$ suggested in][]{2015MNRAS.454.1958M}, or both $\Ahm$ and $\etahm$ parameters, applying uniform priors {$\Ahm\sim\mathcal{U}(0,10)$ and $\etahm\sim\mathcal{U}(0,2)$}.
    \item When constraining the \wcdm model, we vary the dark energy equation-of-state $w$, with a uniform prior in the range $[-2,-1/3]$.
\end{itemize}

Finally, we will, in some cases, include independent (geometric) information from measurements of ratios of galaxy-galaxy lensing two-point functions at small scales, as presented in \citet*{y3-shearratio}. Given an independent lens sample \citet*[here, \textsc{MagLim}, presented in][]{y3-2x2maglimforecast}, the ratios of tangential shear signals for two redshift bins of the source sample around the same galaxies from a common redshift bin of the lens sample depend largely on distances to these samples. Shear ratios (SR) can therefore be used to constrain uncertainties in the redshift distributions. We only exploit small-scale measurements, corresponding to scales of approximately \SIrange{2}{6}{\per\h\mega\parsec}, or $\lmin\sim$\numrange{360}{1200} for redshift bins \numrange{1}{4}, that are largely independent from the scales we use in this analysis (see \cref{fig:clobs} and \cref{sec:scalecuts}). In these cases, we incorporate shear ratios at the likelihood level, using a Gaussian likelihood. The modeling of shear ratios necessitates extra parameters, namely the clustering biases and redshift distribution uncertainties for each of the three lens bins used here. Details about the shear-ratio likelihood and priors can be found in \citet*{y3-shearratio}.

\begin{table}
    \centering
    \begin{tabular}{l l l}
        Parameter & Symbol & Prior \\
        \hline
        Total matter density            & $\Om$ & $\mathcal{U}(0.1,0.9)$ \\
        Baryon density                  & $\Ob$ & $\mathcal{U}(0.03,0.07)$\\
        Hubble parameter                & $h$ & $\mathcal{U}(0.55,0.91)$ \\
        Primordial spectrum amplitude   & $\As\times10^9$ & $\mathcal{U}(0.5,5)$ \\
        Spectral index                  & $\ns$ & $\mathcal{U}(0.87,1.07)$ \\
        Physical neutrino density       & $\Onu h^2$ & $\mathcal{U}(0.0006, 0.00644)$ \\
        \hline
        IA amplitude (TA)   & $\Ata$ & $\mathcal{U}(-5,5)$ \\
        IA redshift dependence (TA) & $\ata$ & $\mathcal{U}(-5,5)$ \\
        IA amplitude (TT)   & $\Att$ & $\mathcal{U}(-5,5)$ \\
        IA redshift dependence (TT) & $\att$ & $\mathcal{U}(-5,5)$ \\
        IA linear bias (TA) & $\bta$ & $\mathcal{U}(0,2)$\\
        \hline
        Photo-$z$ shift in bin 1 & $\Delta z_1$ & $\mathcal{N}(0, 0.018)$ \\
        Photo-$z$ shift in bin 2 & $\Delta z_2$ & $\mathcal{N}(0, 0.015)$ \\
        Photo-$z$ shift in bin 3 & $\Delta z_3$ & $\mathcal{N}(0, 0.011)$ \\
        Photo-$z$ shift in bin 4 & $\Delta z_4$ & $\mathcal{N}(0, 0.017)$ \\
        \hline
        Shear bias in bin 1 & $m_1$ & $\mathcal{N}(-0.0063,0.0091)$  \\
        Shear bias in bin 2 & $m_2$ & $\mathcal{N}(-0.0198,0.0078)$ \\
        Shear bias in bin 3 & $m_3$ & $\mathcal{N}(-0.0241,0.0076)$ \\
        Shear bias in bin 4 & $m_4$ & $\mathcal{N}(-0.0369,0.0076)$ \\
        \hline
    \end{tabular}
    \caption{Cosmological and nuisance parameters in the baseline \lcdm model. Uniform distributions in the range $[a,b]$ are denoted $\mathcal{U}(a,b)$ and Gaussian distributions with mean $\mu$ and standard deviation $\sigma$ are denoted $\mathcal{N}(\mu,\sigma)$.}
    \label{tab:params}
\end{table}

\subsection{Scale cuts}
\label{sec:scalecuts}

\subsubsection{Fiducial scale cuts ($\Delta\chi^2$)}
\label{sec:scalecuts_fid}

As stated in \cref{sec:baryons}, baryonic feedback is a major source of uncertainty on the matter power spectrum at small scales. Therefore, we follow the DES~Y3 methodology presented in \citet*{y3-generalmethods,y3-cosmicshear2} and remove multipole bins that are significantly affected by baryonic effects.

\begin{figure}
    \centering
    \includegraphics[scale=0.65]{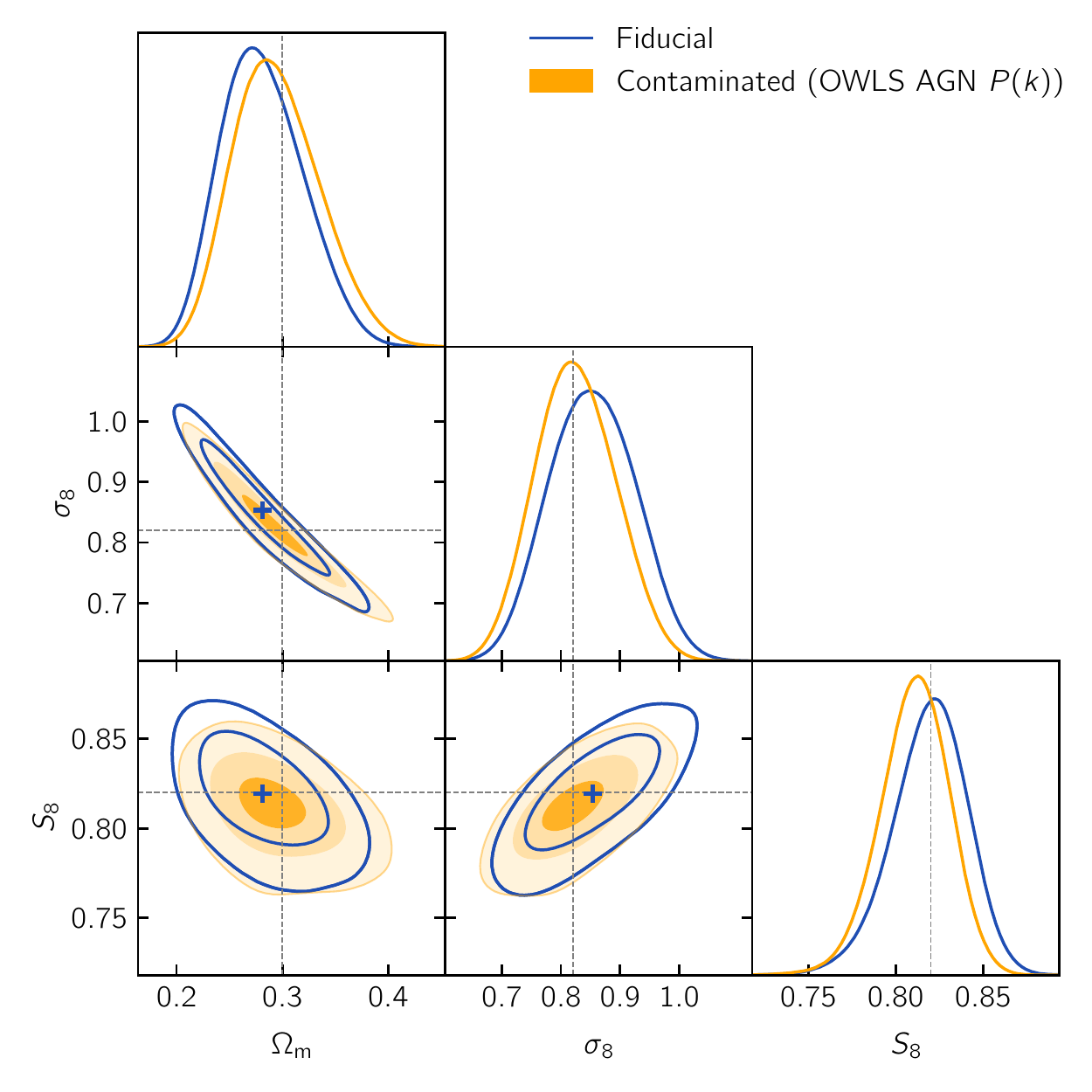}
    \caption{Validation of the $\Delta\chi^2<1$ scale cuts. We compare constraints from a noiseless data vector produced at the fiducial cosmology (dark blue) to those obtained from a contaminated data vector obtained by rescaling the matter power spectrum using \cref{eq:baryon_Pk_ratio} with the OWLS AGN simulation, both using the fiducial model. The nested, filled regions show the \SI{0.3}{$\sigma$}, $1\sigma$ and $2\sigma$ contours, corresponding to roughly 24\%, 68\% and 95\% confidence regions. The mean of the fiducial posterior, which is represented by the blue plus sign, lies within the \SI{0.3}{$\sigma$} contour of the contaminated posterior.}
    \label{fig:scale_cuts}
\end{figure}

To do so, we compare two synthetic, noiseless data vectors computed at the fiducial cosmology: one computed with the power spectrum from \halofit, and one where the power spectrum has been rescaled by the ratio of the power spectra measured in OWLS simulations \citep{2011MNRAS.415.3649V} with dark matter only and with AGN feedback, as in \cref{eq:baryon_Pk_ratio}. We then compute, {using the fiducial covariance matrix,} the $\chi^2$ distances between the two data vectors for each redshift bin pair and determine small-scale cuts by requiring that all $\chi^2$ distances be smaller than a threshold value $\Delta\chi^2/N_{\rm pair}$, where $N_{\rm pair}=10$ is the number of redshift bin pairs.
We then follow the iterative procedure laid out in \citet*{y3-cosmicshear2} and choose the threshold value $\Delta\chi^2$ such that
the bias due to baryons in the $(S_8,\Om)$ plane is less than \SI{0.3}{$\sigma$}. Specifically, we require that the maximum posterior point for the fiducial data vector lies within the two-dimensional \SI{0.3}{$\sigma$} confidence region of the marginal posterior for the contaminated data vector, as shown in \cref{fig:scale_cuts}, using the same scale cuts being tested for both runs.
We find $\Delta\chi^2=1$ allows to reach that goal\footnote{{Note that since power spectra for different redshift bin pairs are correlated, the requirement that each pair $ab$ verifies $\Delta\chi^2_{ab}<0.1$ yields a global ${\Delta\chi^2\approx0.34}$.}} and adopt the corresponding maximum multipoles as our fiducial scale cuts, as shown by the grayed area in \cref{fig:clobs,fig:clres}. This leaves 119 data points out of the 320 in total.

In comparison, the real-space analysis presented in \citet*{y3-cosmicshear1,y3-cosmicshear2} uses scale cuts that account for the full analysis of DES~Y3 lensing and clustering data (the so-called $3\times2$pt analysis), including shear ratios. In order to make our analysis comparable, when using shear ratios, we will use slightly more conservative cuts, with $\Delta\chi^2=0.5$, similar to the real-space analysis, which results in similar biases in the $(S_8,\Om)$ plane of about \SI{0.15}{$\sigma$}. This removes between one and two additional data points for each bin pair, leaving a total of 102 data points. Finally, we keep bandpowers $L$ for which the mean multipole, $\bar{L}$, is below $\lmax$.

We note that these multipoles $\lmax$ are in the range \numrange{200}{400} (except for bin $1,1$, which has larger error bars), corresponding to significantly larger angular scales than the cuts used in the HSC~Y1 \citep{2019PASJ...71...43H} and KiDS-450 \citep{2017MNRAS.471.4412K} analyses, who used redshift-independent multipole cuts at ${\lmax=1900}$ and ${\lmax=1300}$, respectively. Both analyses tested these choices and extensively demonstrated the robustness of their final cosmological constraints. These varying approaches on scale cut choices, discussed in \citet{2020arXiv201106469D}, motivate us to consider alternative scale cuts in the next section.

\subsubsection{Alternative scale cuts ($\kmax$)}
\label{sec:scalecuts_kmax}

We consider a second kind of multipole cuts derived from approximate, small-scale cuts of three-dimensional Fourier modes, which is motivated by theoretical considerations. Namely, assuming that the model for the matter power spectrum is valid up to a certain wavenumber $\kmax$, we aim at discarding multipoles $\ell$ receiving significant contributions from smaller scales (\ie for $k>\kmax$). To do so, we follow \citet{2020arXiv201106469D} and rewrite \cref{eq:limber} as an integral over $k$-modes, using the change of variables $k=\flatfrac{\qty(\ell+1/2)}{\chi(z)}$. We then define the scale $k_{>\alpha}(\ell)$ at which the integral for $C_\ell$ reaches a fraction $\alpha<1$ of its total value, such that
\begin{equation}
    \int_{- \infty}^{\ln k_{>\alpha}(\ell)} \dd{\ln k} {\dv{C_\ell}{\ln k}} = \alpha C_\ell.
\end{equation}
For a given choice of $\alpha$ and $\kmax$, we then obtain the small-scale multipoles cut by numerically solving for $\lmax$ such that ${k_{>\alpha}(\lmax)=\kmax}$. Here, we set $\alpha=0.95$, such that scales at wavenumbers $k$ larger than $k_{>\alpha}(\ell)$ contribute 5\% of the total signal. We will consider different values of $\kmax$ in the range \SIrange{1}{5}{\h\per\mega\parsec}.

{Note that, in general, the validity of the model depends on redshift, as non-linearities increase at lower redshift. However, we will use the same $\kmax$ value for all ten redshift bin pairs, which in practice is limited by the low redshift bin}. We show the cuts corresponding to ${\kmax=\text{\SIlist{1;3;5}{\h\per\mega\parsec}}}$ with dashed lines in \cref{fig:clobs,fig:clres}. These cuts leave \numlist{71;156;228} data points, respectively. The highest multipole used in this work is $\lmax\approx1600$ for redshift bin~4, for $\kmax=\text{\SI{5}{\h\per\mega\parsec}}$.

\subsection{Sampling, parameter inference and tensions}
\label{sec:sampling}

{Throughout this work, we assume a multivariate Gaussian likelihood, as detailed in \cref{sec:like_cov}, to carry out a Bayesian analysis of our data. The theoretical calculations are performed with the \cosmosis framework \citep{2015A&C....12...45Z}. We sample the posterior distributions using \polychord \citep{2015MNRAS.450L..61H}, a sophisticated implementation of nested sampling, with 500 live points and a tolerance of 0.01 on the estimated evidence. We report parameter constraints through one-dimensional marginal summary statistics computed and plotted with \textsc{GetDist} \citep{2019arXiv191013970L}, as
\begin{equation*}
    \textrm{Parameter = 1D mean}_{-\textrm{lower 34\% bound}}^{+\textrm{upper 34\% bound}}\textrm{ (MAP value)},
\end{equation*}
where the maximum a posterior (MAP) is reported in parenthesis.}

{We will compute a number of metrics to characterize and interpret the inferred posterior distributions.
For a number $N_{\rm param}$ of varied parameters, the number of parameters effectively constrained by the data is given by
\begin{equation}
    N_{\rm eff}=N_{\rm param}-\Tr(\mathcal{C}_\Pi^{-1} \mathcal{C}_p),
\end{equation}
where $\mathcal{C}_\Pi$ and $\mathcal{C}_p$ are the covariance matrices of the prior and posterior, approximated as Gaussian distributions, and $\Tr$ is the trace operator \citep{2019PhRvD..99d3506R}.
For a given posterior and its corresponding prior, we will also compute the Karhunen–Loève (KL) decomposition that measures the improvement of the posterior with respect to the prior \citep{2019PhRvD..99d3506R,2020PhRvD.101j3527R}. We can then project the observed improvement onto a set of modes, that we restrict to power laws in the cosmological parameters.
Finally, we will characterize the level of disagreement between posterior distributions using the posterior shift probability, as described in \citet{2021arXiv210503324R}. This metric is based on the parameter difference distribution obtained by differenciating samples from two independent posteriors, and computing the volume with the isocontour of a null difference.
To do so, we will use the \texttt{tensiometer}\footnote{\url{https://tensiometer.readthedocs.io}} package \citep[see previous references and][]{2021arXiv211205737D}, which fully handles the non-Gaussian nature of the derived posteriors.
}

\section{Validation}
\label{sec:validation}

In this section, we present a number of tests of our analysis framework. In \cref{sec:sims}, we introduce simulations that we use to verify that measured spectra are not significantly impacted by known systematic effects ($B$-modes and PSF leakage) in \cref{sec:bmodes_and_psf}, to validate the measurement pipeline and the covariance in \cref{sec:cov_validation}, and to test the accuracy of our theoretical model and its impact on cosmological parameter inferences in \cref{sec:modeling_validation}.

\subsection{Simulations}
\label{sec:sims}

\subsubsection{Gaussian simulations with DES~Y3 data}
\label{sec:sims_gaussian}

In the following sections, we use a large number of Gaussian simulations to validate the cosmic shear power spectra measurements, obtain a covariance matrix for $B$-modes spectra and cross-spectra with the PSF ellipticities. To make them as close as possible to DES~Y3 data, we use the actual positions and randomly rotated shapes of the galaxies in the DES~Y3 catalog. This ensures that the masks and the noise power spectra are identical to those of the real data measurements.

The generation of a single simulation proceeds as follows. Given predictions for the shear $E$-mode spectra at the fiducial model, $C_\ell^{ab}$, we generate a full-sky realization of the four correlated shear fields at a resolution of $\nside=1024$. To do so, we use the definition of the spectra, \cref{eq:defclEE}, as the covariance of the spherical harmonic coefficients of the fields to sample four-dimensional vectors, $(E_{\ell m}^1, E_{\ell m}^2, E_{\ell m}^3, E_{\ell m}^4)$, for $0\leq\ell<3\nside$, $-{\ell \leq m \leq +\ell}$, which are independent for different $(\ell,m)$. We then use the \texttt{alm2map} function of \healpy \citep{2019JOSS....4.1298Z} in polarization mode, with $T_{\ell m}^i=B_{\ell m}^i=0$, to generate the four correlated, true (but pixelated) shear maps. The next step consists in sampling these fields. As explained above, we use the DES~Y3 catalog of (mean- and response-corrected) ellipticities, to which we apply random rotations, and the positions of the galaxies as input. The random rotations are obtained by multiplying the complex ellipticities, $\vb{e}=e_1+ie_2$, by $e^{2i\theta}$, where $\theta$ is the random rotation angle. For a galaxy $i$ in redshift bin $a$, the ellipticity in the mock catalog is given by
\begin{equation}
    \vb{e}_i^\prime = \frac{\bm{\gamma}_i^a + e^{2i\theta}\vb{e}_i}{1+ e^{2i\theta} \bm{\gamma}_i^{a*} \vb{e}_i},
\end{equation}
where $\bm{\gamma}_i^a$ is the value of the (complex) shear field corresponding to the $a$-th redshift bin at the position of galaxy $i$.
This procedure is justified by the fact that the variance of the shear fields is about $10^3$ times smaller than the variance due to intrinsic shapes, $\sigma_e^2\sim0.3^2$, such that the variance of the new ellipticities remains extremely close to that of the true ellipticities.

We then perform power spectra measurements on these mock catalogs with the same pipeline that is used on data, except that these spectra need not be corrected for the pixel window function. The mean residuals with respect to the expected ($E$-mode) power spectra computed with \cref{eq:bpws_cl} using mixing matrices are shown in the lower left panel of \cref{fig:cov} for \num{10000} simulations, showing agreement within 5\% of the error bars. We also find that the (small but non-zero) $B$-mode power spectra measured in these simulations are consistent, at the same level, with expectations from $E$-mode leakage computed using \cref{eq:bpws_cl}.

{Note that the real space analysis of DES~Y3 lensing and clustering data \citep*{y3-3x2ptkp} relied on log-normal simulations using \textsc{Flask} \citep{2016MNRAS.459.3693X} to partially validate the covariance, as detailed in \citet*{y3-covariances}. However, those were mainly used to evaluate the effect of the survey geometry, which is already accounted for by \namaster \citep{2019MNRAS.484.4127A}, and need not be validated here. Therefore, we use simpler, Gaussian simulations to validate the measurement pipeline and obtain empirical covariance matrices (for $B$-mode and PSF tests). In order to validate the full covariance matrix, including the non-Gaussian contributions, we will rely on the \dgv suite of simulations (see \cref{sec:sims_pkd}), which rely on full $N$-body simulations and are tailored for lensing studies.}

\subsubsection{\dgv suite of simulations}
\label{sec:sims_pkd}

The DES~Y3 analysis of the convergence peaks and power spectrum presented in \citet{2022MNRAS.tmp..151Z} relied on the \dgv suite of weak lensing simulations. They were obtained from fifty $N$-body, dark matter-only simulations produced using the \pkd code \citep{2017ComAC...4....2P}. Each of these consists of $768^3$ particles in a \SI{900}{\per\h\mega\parsec} box, which is replicated $14^3$ times to reach a redshift of~3. Snapshots are assembled to produce density shells and the corresponding (true) convergence maps for the four DES~Y3 redshift bins. These simulations are then populated with DES~Y3 galaxies, in a way similar to what is done for Gaussian simulations (see \cref{sec:sims_gaussian}). This operation is repeated with a hundred noise realizations per simulation, thus producing \num{5000} power spectra measurements.

We will use these measurements to compute an empirical covariance matrix that includes non-Gaussian contributions, and that can be compared to our analytic covariance matrix, thus providing a useful cross-check.

\subsubsection{Buzzard v2.0 simulations}
\label{sec:sims_buzzard}

The \textsc{Buzzard} v2.0 simulations are a suite of simulated galaxy catalogs built on $N$-body simulations and designed to match important properties of DES~Y3 data. These simulations were used to validate the configuration space analysis of galaxy lensing and galaxy clustering within the DES~Y3 analysis and we refer the reader to \citet*{y3-simvalidation} for greater details.

In brief, the lightcones were obtained by evolving particles initialized at redshift $z=50$ with an optimized version of the \gadget $N$-body code \citep{2005MNRAS.364.1105S}. The lensing fields (convergence, lensing, magnification) were computed by ray-tracing the simulations with the \calclens code \citep{2013MNRAS.435..115B}, over 160 lens planes in the redshift range ${0\leq z \leq 2.35}$, and with a resolution of $\nside=8192$. The simulations were then populated with source galaxies so as to mimic the density, the ellipticity dispersion and photometric properties of the DES~Y3 sample.
The \sompz method was applied to these mock catalogs so as to divide them into four tomographic bins of approximately equal density, thus producing ensemble of redshift distributions that were validated against the known true redshift distributions \citep*[see][for details]{y3-sompz}.

We will use sixteen \textsc{Buzzard} simulations to perform an end-to-end validation of our measurement and inference pipelines in \cref{sec:modeling_validation_buzz}.
It is worth noting that these simulations do not incorporate the effects of massive neutrinos on the matter power spectrum, nor those imparted to intrinsic alignments. When analyzing these simulations, we will therefore fix the total mass of neutrinos to zero, and assume null fiducial values of the IA parameters (though they will be varied with the same flat priors).

\subsection{Validation of power spectrum measurements}
\label{sec:bmodes_and_psf}

In this section, we study the potential contamination of the signal with two measurements. First, we verify that the $B$-mode component of the power spectra is consistent with the null hypothesis of no $B$-mode, as any cosmological or astrophysical source of $B$-mode is expected to be very small. Second, we estimate the contamination of the signal by the PSF, which, if incorrectly modeled, would leak into the estimated cosmic shear $E$-mode spectra, and therefore bias cosmology.

\subsubsection{$B$-modes}
\label{sec:bmodes}

As mentioned in \cref{sec:pseudocl}, gravitational lensing does not produce $B$-modes, to first order in the shear field and under the Born approximation, \ie when the signal is integrated along the line of sight instead of following distorted photon trajectories. Second- and higher-order effects as well as source clustering and intrinsic alignments are expected to produce non-zero, but very small $B$-modes. However, the contamination of the ellipticities by various systematic effects, first and foremost by errors in the PSF model, are expected to produce much larger $B$-modes in practice. Indeed, the PSF does not possess the same symmetries as cosmological lensing, and its $E$- and $B$-mode spectra are almost identical. Therefore, any leakage due to a mis-estimation of the PSF could induce $B$-modes in galaxy ellipticities. As a consequence, measuring $B$-modes in the estimated shear maps and verifying that they are consistent with a non-detection (or pure shape-noise) constitutes a non-sufficient but nevertheless useful test of systematic effects \citep{2016MNRAS.457..304B,2017MNRAS.464.1676A,2019MNRAS.484L..59A,2019A&A...624A.134A}.

\begin{figure*}
    \centering
    \includegraphics[scale=0.65]{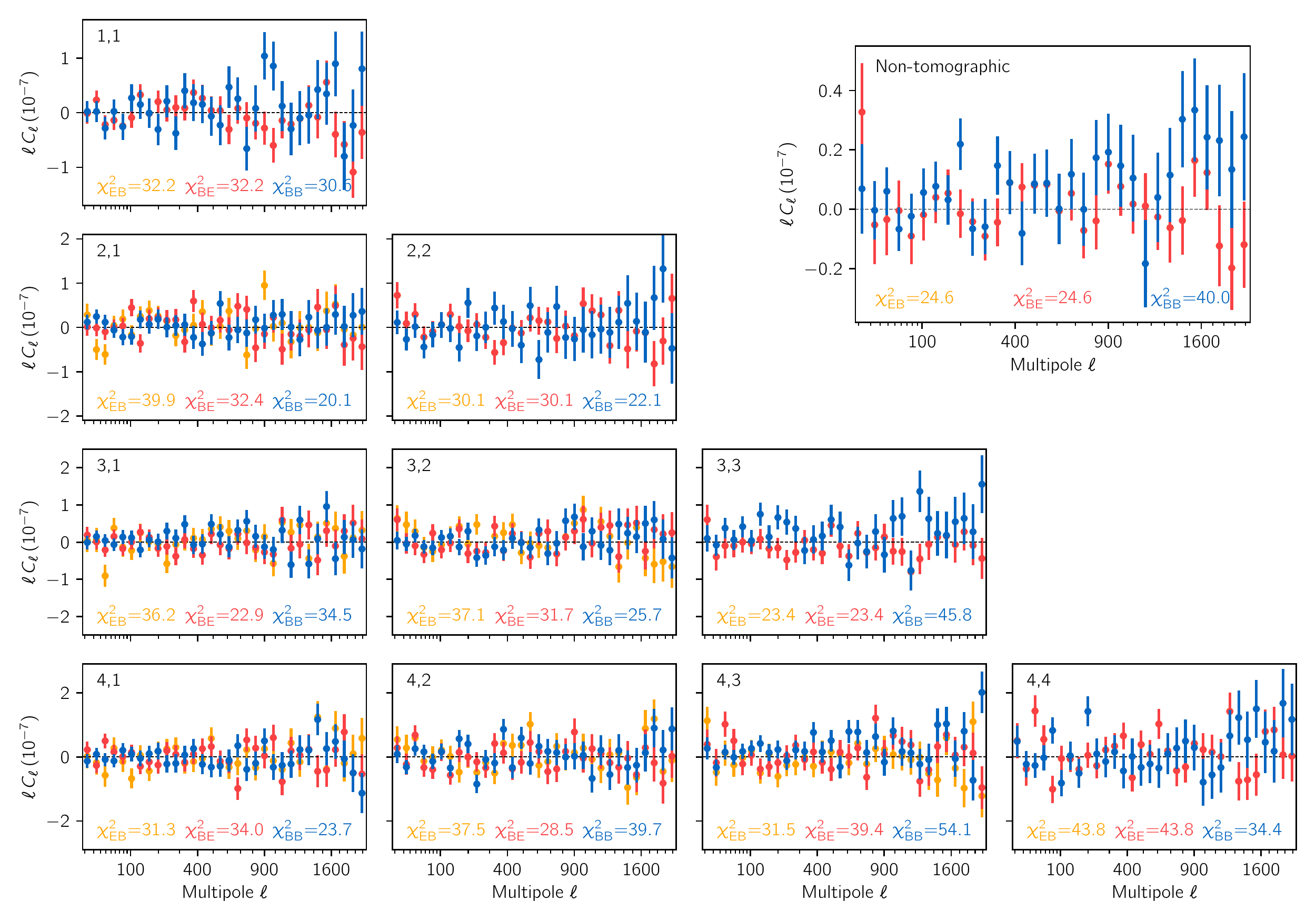}
    \caption{$EB$ and $BB$ cosmic shear power spectra measured with DES~Y3 data for each pair of tomographic bins in the lower triangle, and the entire sample in the upper right panel (note that the $EB$ and $BE$ power spectra are different only for cross-redshift bin spectra). Error bars are computed from \num{10000} Gaussian simulations using the DES~Y3 catalog ellipticities and positions, as explained in \cref{sec:sims_gaussian}. We find a $\chi^2$ of \num{344.0} for 320 degrees of freedom for {tomographic} $B$-mode power spectra, corresponding to a probability-to-exceed of 0.17. We find a $\chi^2$ of \num{535.4} for 512 degrees of freedom for $EB$ {tomographic} cross-power spectra (counting all 16 independent bin pairs), corresponding to a probability-to-exceed of 0.23. Individual $\chi^2$ are reported for each redshift bin pairs in the corresponding panels. In the non-tomographic case, we find, for the $B$-mode power spectrum, a $\chi^2$ of \num{40.0} for 32 degrees of freedom, corresponding to a probability-to-exceed of 0.16.}
    \label{fig:clobs_BB}
\end{figure*}

\Cref{fig:clobs_BB} shows measurements of the tomographic $B$-mode power spectra in blue {for DES~Y3 data}. We use \num{10000} Gaussian simulations presented in \cref{sec:sims_gaussian} to compute the covariance matrix (we have verified convergence) and obtain a total $\chi^2$, for the stacked data vector of $B$-mode spectra, of \num{344.0} for 320 degrees of freedom, corresponding to a probability-to-exceed of 0.17. This is consistent with the null hypothesis of no $B$-modes. In addition, we show $EB$ cross-spectra in \cref{fig:clobs_BB} for completeness, finding a $\chi^2$ of \num{535.4} for 512 degrees of freedom, and a probability-to-exceed of 0.23.
{We also show, for completeness, measurements of the non-tomographic $B$-mode power spectrum, already presented in \citet*{y3-shapecatalog}. In this case, we find a $\chi^2$ of \num{40.0} for 32 degrees of freedom and a probability-to-exceed of 0.16. Note that \citet*{y3-shapecatalog} also included a test where the galaxy sample was split in three bins, as a function of the PSF size at the positions of the galaxies, and found agreement with the hypothesis of no $B$-mode.}

\subsubsection{Point spread function}
\label{sec:psf}

\citet*{y3-piff} introduced the new software \textsc{Piff} to model the point spread function (PSF) of DES~Y3 data, using interpolation in sky coordinates with improved astrometric solutions.
Although the impact of the PSF on DES Y3 shapes and real-space shear two-point functions was already investigated in \citet*{y3-shapecatalog} and \citet*{y3-cosmicshear1}, we investigate PSF contamination in harmonic space as the leakage of PSF residuals might differ from those in real space. We do so by measuring $\rho$-statistics \citep{10.1111/j.1365-2966.2010.16277.x} in harmonic space and estimate the potential level of contamination of the data vector.

Our detailed results are presented in \cref{app:psf}. We conclude that we find no significant contamination and that the residual contamination has negligible impact on cosmological constraints.

\subsection{Validation of the covariance matrix}
\label{sec:cov_validation}

We compare the fiducial covariance matrix to the covariances estimated from Gaussian simulations described in \cref{sec:sims_gaussian} as well as the \dgv simulations described in \cref{sec:sims_pkd}.

The middle left panel of \cref{fig:cov} shows the ratios of the square-root of the diagonals of those covariance matrices. When compared to the covariance estimated from Gaussian simulations, we find excellent agreement, at the 5\% level across all scales and redshift bin pairs. Our fiducial, semi-analytical covariance predicts only slightly larger error bars, at the \numrange{2}{3}\% level. We also find very good agreement with the covariance matrix computed from \dgv simulations, with the fiducial covariance matrix showing smaller error bars, at the 15\% level, for the largest scales only. For both sets of simulations, we also compared diagonals of the off-diagonal blocks (\ie the terms $\cov(C_\ell^{ab},C_{\ell'}^{cd})$ with $ab\ne cd$ but $\ell=\ell'$) and found good agreement, up to the uncertainty due to the finite number of simulations. Finally, we verified that replacing the analytic covariance matrix by the \dgv covariance matrix has negligible impact on cosmological constraints inferred from the fiducial data vector {(shifts below $0.1\sigma$), as shown in \cref{app:ic_cont}}.

\subsection{Validation of the robustness of the models}
\label{sec:modeling_validation}

In this section, we demonstrate the robustness of our modeling using synthetic data in \cref{sec:modeling_validation_synth}, and using Buzzard simulations in \cref{sec:modeling_validation_buzz}.

\subsubsection{Validation with synthetic data}
\label{sec:modeling_validation_synth}

Our fiducial scale cuts, as explained in \cref{sec:scalecuts_fid}, are constructed in such a way as to minimize the impact on cosmology from uncertainties in the small-scale matter power spectrum {due to baryonic feedback}, as shown in \cref{fig:scale_cuts}.

{We further test the robustness of our fiducial model, based on \halofit, by testing other prescriptions for the non-linear matter power spectrum. To do so, we compare constraints, inferred with the same model, but for different synthetic data vectors computed}
\begin{inparaenum}
    \item with \halofit, 
    \item with \hmcode with dark matter only (\ie using $\Ahm=3.13$), and
    \item with the \textsc{Euclid Emulator} \citep{2019MNRAS.484.5509E}.
\end{inparaenum}
These data vectors are compared in \cref{fig:clres} and the constraints are shown in \cref{fig:test_Pk}, which shows that contours are shifted by less than \SI{0.3}{$\sigma$} in the $(S_8,\Om)$ plane.

We also aim at constraining the effect of baryonic feedback using alternative scale cuts based on a $\kmax$ cut-off in Fourier space, as explained in \cref{sec:scalecuts_kmax}.
In order to validate the robustness of this alternative model,
we follow a similar approach and consider predictions for the shear power spectra from four hydrodynamical simulations (Illustris, OWLS AGN, Horizon AGN and MassiveBlack~II), as shown in \cref{fig:clres}. We then build corresponding data vectors using \halofit and a rescaling of the matter power spectrum, as in \cref{eq:baryon_Pk_ratio}. Next, we analyze those data vectors using
\begin{inparaenum}
    \item the true model considered here (\ie \halofit and rescaling), and then
    \item \hmcode with one free parameter.
\end{inparaenum}
We finally test whether the $(S_8,\Om)$ best-fit parameters for the true model are within the \SI{0.3}{$\sigma$} contours of the posterior assuming \hmcode.

When varying only $\Ahm$, we do find that this test passes for $\kmax=\text{\SIlist{1;3;5}{\h\per\mega\parsec}}$ with biases of \SI{0.22}{$\sigma$} at most (and typically \SI{0.1}{$\sigma$}), even though the inferred $\Ahm$ parameter largely varies across simulations (we find posterior means of \numlist{2.2;2.7;3.4;3.6} for Illustris, OWLS AGN, Horizon AGN and MassiveBlack~II, respectively). This means that biases introduced by \hmcode, if any, are not worse than potential projection effects found when using the true model, all of which are found to be below the level of \SI{0.3}{$\sigma$}.

\subsubsection{Validation with Buzzard simulations}
\label{sec:modeling_validation_buzz}

In this section, we use Buzzard simulations (see \cref{sec:sims_buzzard}) to validate our measurement and analysis pipelines together. Precisely, we verify that
\begin{inparaenum}
    \item we are able to recover the true cosmology used when generating Buzzard simulations and
    \item the model yields a reasonable fit to the measured shear spectra.
\end{inparaenum}

We start by measuring cosmic shear power spectra and verify that the mean measurement (not shown) is consistent with the theoretical prediction from our fiducial model at the Buzzard cosmology, using the true Buzzard redshift distributions, and with a covariance recomputed with these inputs.

We then run our inference pipeline on the mean data vector, first with the covariance corresponding to a single realization, and then with a covariance rescaled by a factor of $1/16$, to reflect the uncertainty on the average of the measurements.
The first case is testing whether we can recover the true cosmology on average, while the second is a stringent test of the accuracy of the model, given that error bars are divided by $\sqrt{16}=4$ with respect to observations {with the DES~Y3 statistical power.}
For these tests, the priors on shear and redshift biases are centered at zero, with a standard deviation of 0.005.

\begin{figure}
    \centering
    \includegraphics[scale=0.65]{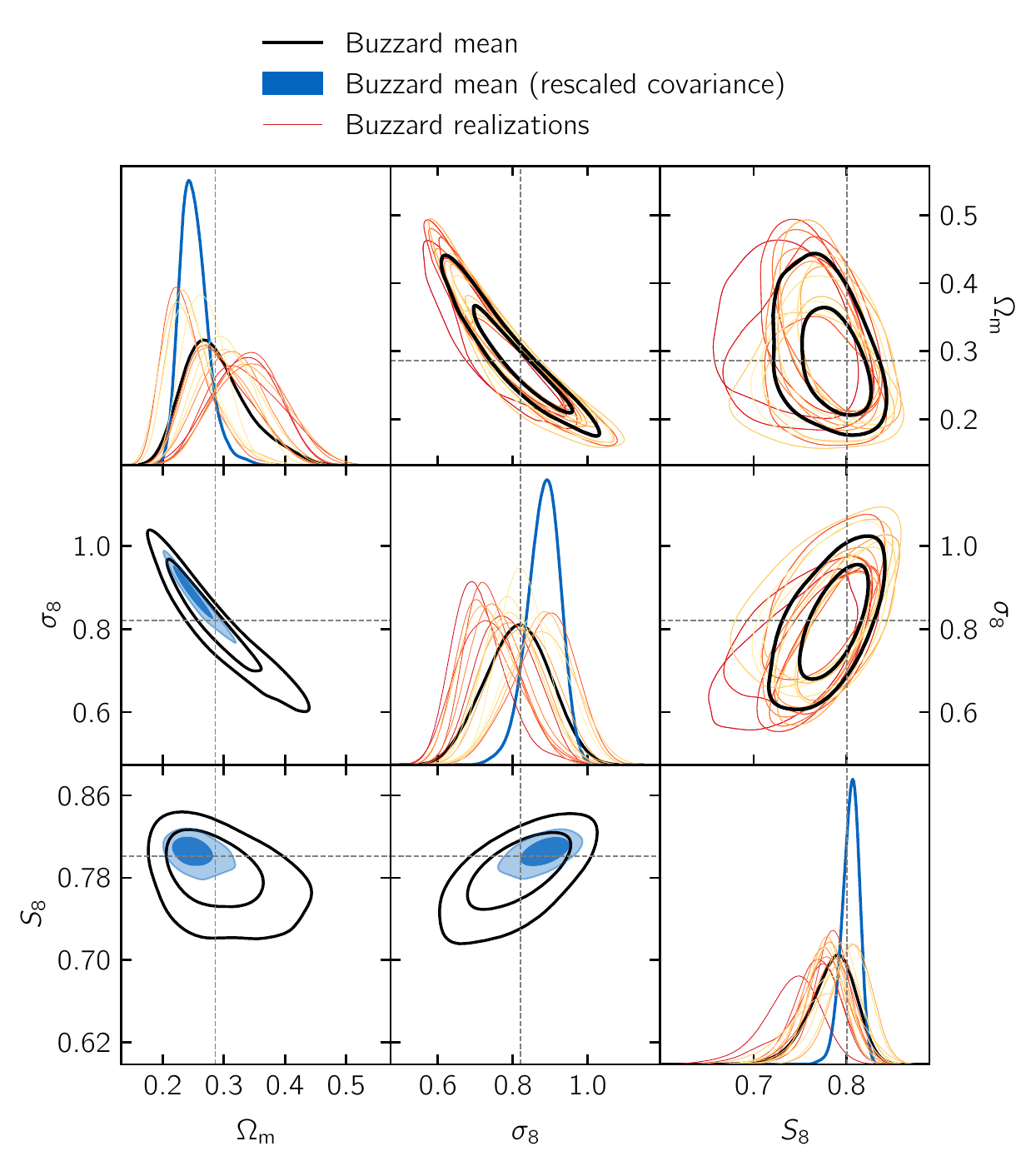}
    \caption{Validation of the analysis framework with Buzzard simulations. We show the one- and two-dimensional marginal posterior distributions corresponding to the mean Buzzard data vector with the data covariance (black) and the same covariance rescaled by a factor $1/16$ (blue). The posteriors obtained for each realization are shown in yellow to red.}
    \label{fig:bz_combined}
\end{figure}

The 68\% and 95\% confidence contours are shown in \cref{fig:bz_combined} for both covariances, using the fiducial $\chi^2<1$ scale cuts. We only show the contours for the best constrained parameters ($\Om$, $\sigma_8$ and $S_8$) but we verified that the true cosmology is recovered in the full parameter space. We find that it is perfectly recovered in the first case and within $1\sigma$ contours in the second case, consistent with fluctuations on the mean Buzzard data vector.
{We find {that the effective number of constrained parameters is} $N_{\rm eff}\approx\num{7.8}$ in the first case, whereas,}
in the second case, we find $N_{\rm eff}\approx\num{9.6}$ (recall we fix the neutrino mass to zero for tests on Buzzard, so $N_{\rm param}=18$ here). In the second test, we find that $\chi^2=139.4$ at the best-fit parameters (maximum a posteriori) for $N=119$ data points, and $N-N_{\rm eff}$ degrees of freedom, such that the best-fit $\chi^2$ corresponds to a probability-to-exceed of \num{2.7}\%.
For $\kmax$ cuts, we also recover the input cosmology within error bars and find $\chi^2/(N-N_{\rm eff})$ of $98.4/61.7$, $191.6/146.1$ and $254.5/217.8$ respectively for {$\kmax$ of \SIlist{1;3;5}{\h\per\mega\parsec} (although note we will not use this combination of model and scale cuts on data)}. Together, these tests suggest that {the accuracy of} our fiducial model {exceeds that required by} the statistical power of DES~Y3 data.

We then run our inference pipeline on each realization to visualize the scatter in the posteriors due to statistical fluctuations. This exercise allows us to verify that the model does not feature catastrophic degeneracies that have the potential to bias the marginal posterior distributions over cosmological parameters, in particular in the $(S_8,\Om)$ plane. The contours are shown in \cref{fig:bz_combined}, along with the contours obtained from the mean Buzzard data vector. We also compute the $\chi^2$ at best fit for each realization and find that the distribution is perfectly consistent with a $\chi^2$ distribution with $N-N_{\rm eff}$ degrees of freedom, where we find $N_{\rm eff}\approx\num{7.8(2)}$ in these cases.

\section{Blinding}
\label{sec:blinding}

We follow a blinding procedure, decided beforehand, that is meant to prevent confirmation and observer biases, as well as fine tuning of analysis choices based on cosmological information from the data itself.
After performing sanity checks of our measurement and modeling pipelines that only drew from the data basic properties such as its footprint and noise properties, we proceeded to unblind our results in three successive stages as {described below. It is worth noting, though, that as this work follows the real space analysis of \citet*{y3-cosmicshear1,y3-cosmicshear2}, the blinding procedure is meant to validate the components of the analysis that are different, such as the cosmic shear power spectrum measurements, the scale cuts, and the covariance matrix.}

\textbf{Stage 1.} The shape catalog was blinded by a random rescaling of the measured conformal shears of galaxies, as detailed in \citet*{y3-shapecatalog}. This step preserves the statistical properties of systematic tests while shifting the inferred cosmology. A number of null tests were presented in \citet*{y3-shapecatalog} to test for potential additive and multiplicative biases before deeming the catalog as science-ready and unblinding it. In the previous section \cref{sec:bmodes_and_psf}, we repeated two of these tests in harmonic space, namely the test of the presence of $B$-modes and the test of the contamination by the PSF.

Once all these tests had passed, we used the unblinded catalog to measure the shape noise power spectrum and compute the Gaussian contribution to the covariance matrix. We then repeated the systematic and validation tests, in particular those based on Gaussian simulations where shape noise is inferred from the data. 

\textbf{Stage 2.} Using the {updated} covariance matrix, we proceeded to validate analysis choices with synthetic data. We first determined fiducial scale cuts based on the requirement that baryonic feedback effects do not bias cosmology at a level greater than \SI{0.3}{$\sigma$}, as detailed in \cref{sec:scalecuts_fid}. We then verified that baryonic effects as predicted from a range of hydrodynamical simulations do not bias cosmology for alternative scale cuts, provided that \hmcode (with a free baryonic amplitude parameter) is used instead of \halofit, as detailed in \cref{sec:modeling_validation_synth}. Finally, we verified that effects that are not accounted for in the model do not bias cosmology, \eg PSF residual contamination in \cref{app:psf}, and higher-order lensing effects and uncertainties in the matter power spectrum using the $N$-body Buzzard simulations \cref{sec:modeling_validation_buzz}.

\textbf{Stage 3.} Before unblinding the data vector and cosmological constraints, we performed a last series of sanity checks. In particular, we verified that the model is a good fit to the data {by asserting} that the $\chi^2$ statistic at the best-fit parameters corresponds to a probability-to-exceed above 1\%. We found that the best-fit $\chi^2$ is 129.3 for 119 data points and $N_{\rm eff}\approx5.6$ constrained parameters, corresponding to a probability-to-exceed of 14.6\%. We also verified that the marginal posteriors of nuisance parameters were consistent with their priors. {Finally, we performed two sets of internal consistency tests, in parameter space and in data space. For the tests in parameter space, we compared, with blinded axes, constraints for $(S_8,\Om)$ from the fiducial data vector with constraints from subsets of the data vector, first removing one redshift bin at a time, and then removing large or small angular scales, as detailed in \cref{item:ictests_z,item:ictests_scales} of \cref{app:ic_cont}. The tests in data space, presented in \cref{app:ic_ppd}, are} based on the posterior predictive distribution (PPD), and follow the methodology presented in \citet*{y3-inttensions}. The PPD goodness-of-fit test yields a calibrated probability-to-exceed of 11.6\%. These tests are detailed in \cref{app:ic}, along with other post-unblinding internal consistency tests.

After this series of tests all passed, we plotted the data and compared it to the best-fit model, as shown in \cref{fig:clobs}, and finally unblinded the cosmological constraints, presented in the next section.

\section{Cosmological constraints}
\label{sec:res}

This section presents our main results. We use measurements of cosmic shear power spectra from DES~Y3 data to constrain the \lcdm model in \cref{sec:res_lcdm}. We then explore alternative analysis choices to constrain intrinsic alignments in \cref{sec:res_IA} and baryonic feedback in \cref{sec:res_baryons}. We compare our results to other weak lensing analyses of DES~Y3 data in \cref{sec:res_consistency_kp}, namely the comic shear two-point functions \citep*{y3-cosmicshear1,y3-cosmicshear2}, convergence peaks and power spectra \citep{2022MNRAS.tmp..151Z} and convergence second- and third-order moments \citep{2021arXiv211010141G}, and to weak lensing analyses from the KiDS and HSC collaborations in \cref{sec:res_other_wl}. Finally, as an illustrative exercise, we reconstruct the matter power spectrum from DES~Y3 cosmic shear power spectra using the method of \citet{2002PhRvD..66j3508T} in \cref{sec:res_Pk}. A number of internal consistency tests are also presented in \cref{app:ic} and the full posterior distribution is shown in \cref{app:full_posterior}.

Note that, for all the constraints that are presented in the following sections, we have recomputed the effective number of constrained parameters and verified that the $\chi^2$ statistic at best fit corresponds to a probability-to-exceed above 1\%.

\subsection{Constraints on \lcdm}
\label{sec:res_lcdm}

\begin{figure}
    \centering
    \includegraphics[scale=0.65]{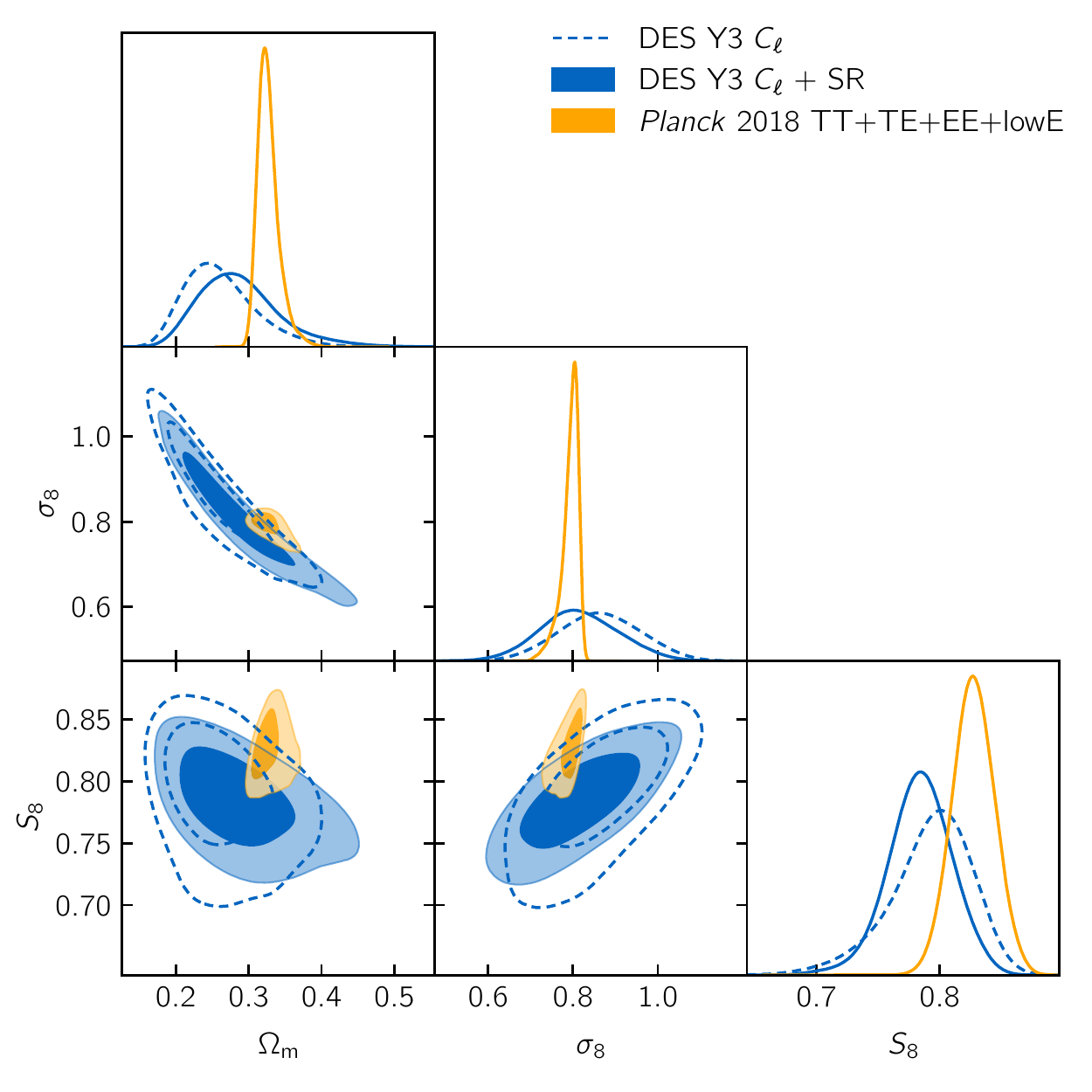}
    \caption{Cosmological constraints on the amplitude of structure $\sigma_8$, the total matter density $\Om$ and their combination ${S_8\equiv\sigma_8\sqrt{\Om/0.3}}$. The inner (outer) contours show 68\% (95\%) confidence regions. Constraints from DES~Y3 cosmic shear power spectra with the two sets of fiducial scale cuts are shown in blue, with (solid) and without (dashed) shear ratios \citep*{y3-shearratio}. Constraints obtained from \planck 2018 measurements of cosmic microwave background temperature and polarization anisotropies are shown in yellow \citep{2020A&A...641A...6P}.}
    \label{fig:cont_lcdm_cl_planck}
\end{figure}

\begin{figure*}
    \centering
    \includegraphics[scale=0.65]{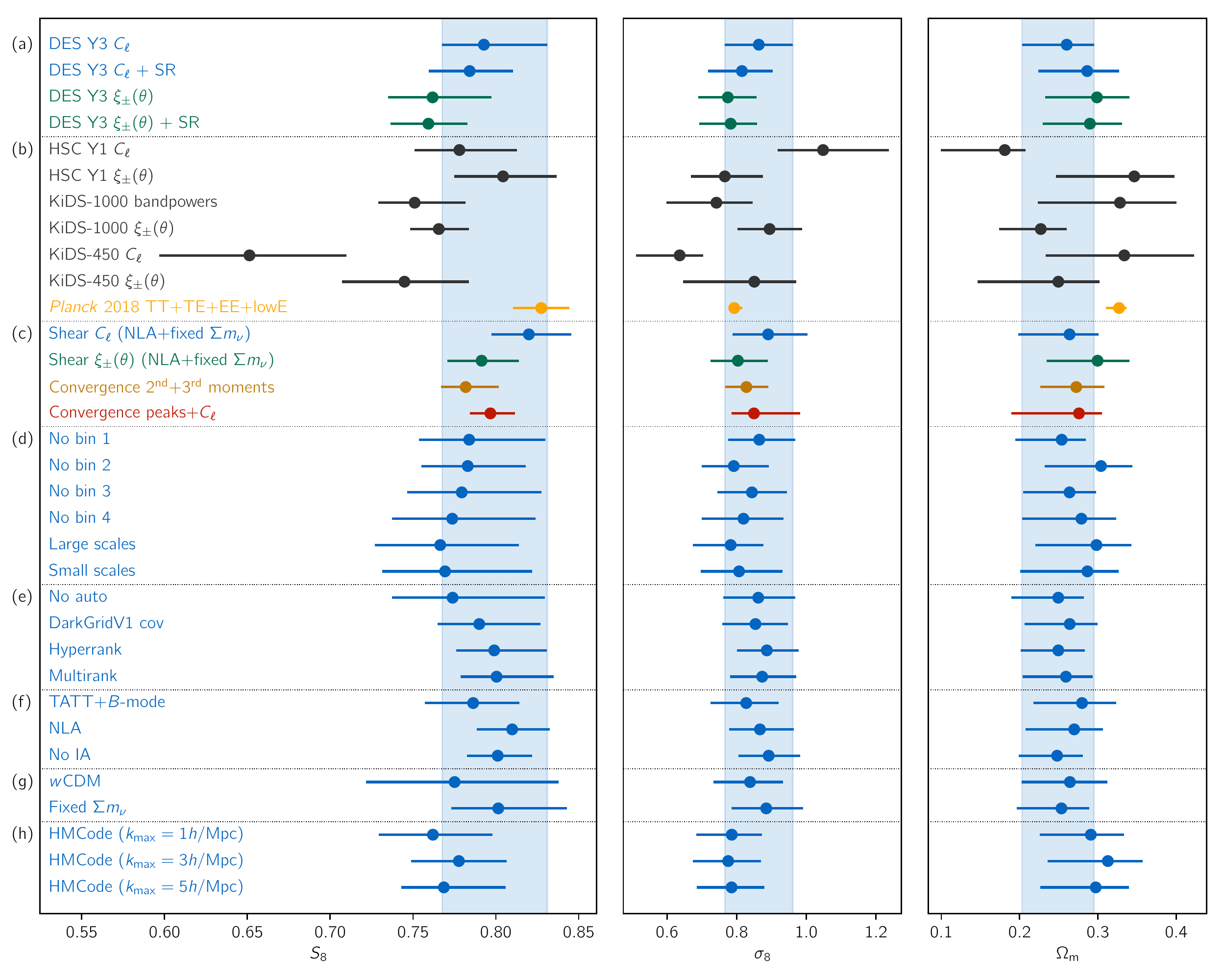}
    \caption{Comparison of one-dimensional marginal posterior distributions over the parameters $S_8\equiv\sigma_8(\Om/0.3)^{0.5}$, $\sigma_8$ and $\Om$, from DES~Y3 data as well as other experiments, and consistency tests for this work (in blue).
    \begin{inparaenum}[(a)]
        \protect\item Constraints obtained from the harmonic (this work) and real \citep*{y3-cosmicshear1,y3-cosmicshear2} space analyses of DES~Y3 data are shown in blue and green (see also \cref{fig:cont_lcdm_xipm}), both with and without shear ratio information \citep*[SR,][]{y3-shearratio}.
        \protect\item Constraints from other weak lensing surveys, namely HSC~Y1 \citep{2019PASJ...71...43H,2020PASJ...72...16H}, KiDS-1000 \citep{2021A&A...645A.104A} and KiDS-450 \citep{2017MNRAS.465.1454H,2017MNRAS.471.4412K} are shown in gray, and constraints from cosmic microwave background observations from \planck 2018 are shown in yellow \citep{2020A&A...641A...6P}.
        \protect\item Constraints from four weak lensing analyses of DES~Y3 data are compared, including the analysis of mass map moments \citep{2021arXiv211010141G} and peaks \citep{2022MNRAS.tmp..151Z}, and illustrating a high level of consistency (see also \cref{fig:cont_lcdm_all}).
        \protect\item Consistency tests where redshift bins are removed one at a time (first four) and where the data vector is split into its large- and small-scale data points (last two) (see also \cref{app:ic}).
        \protect\item Various other consistency tests: removing auto power spectra, swapping the covariance matrix, and marginalizing over redshift distribution uncertainties with \hyperrank and \multirank (see also \cref{app:ic}).
        \protect\item Modeling robustness test for intrinsic alignment (IA), including $B$-mode power spectra, or replacing TATT by NLA, or removing IA contributions altogether (see also \cref{sec:res_IA} \cref{fig:cont_ia}).
        \protect\item Other robustness test, freeing the dark energy equation-of-state $w$ or fixing the neutrino mass to \SI{0.06}{\eV}.
        \protect\item Baryonic feedback tests where the matter power spectrum is computed with \hmcode instead of \halofit, and fiducial scale cuts are replaced with $\kmax=\text{\SIlist{1;3;5}{\h\per\mega\parsec}}$ scale cuts (see also \cref{sec:res_baryons} and \cref{fig:cont_hm1}).
    \end{inparaenum}}
    \label{fig:1d_all}
\end{figure*}

We present here our constraints on \lcdm assuming the fiducial model presented in \cref{sec:modeling}, that is, using \halofit for the matter power spectrum and TATT for intrinsic alignments. Constraints are shown in blue in \cref{fig:cont_lcdm_cl_planck} and compared to constraints from \planck 2018 measurements of cosmic microwave background temperature and polarization anisotropies \citep[\planck 2018 TT+TE+EE+lowE,][]{2020A&A...641A...6P}, in yellow. The one-dimensional marginal constraints are also shown in \cref{fig:1d_all} along with constraints for all variations of the analysis, and the full posterior is shown in \cref{fig:full_posterior}. Using only shear power spectra (\ie no shear ratio information), we find
\begin{align*}
    \Omega_{\rm m} & = 0.260^{+0.035}_{-0.057} \; (0.242), && \text{[$C_\ell$ TATT]} \\
    \sigma_8       & = 0.863\pm 0.096 \; (0.902), && \text{[$C_\ell$ TATT]} \\
    S_8            & = 0.793^{+0.038}_{-0.025} \; (0.810), && \text{[$C_\ell$ TATT]}
\end{align*}
where we report the mean, the 68\% confidence intervals of the posterior, and the best-fit parameter values, \ie the mode of the posterior, in parenthesis. The corresponding theoretical shear power spectra are shown in \cref{fig:clobs}, showing good agreement with data, consistent with the $\chi^2$ at best-fit of 129.3. The best constrained combination of parameters ${\sigma_8 \qty(\Omega_{\rm m}/0.3)^{\alpha}}$, inferred from a principal component analysis, is given by
\begin{align*}
    \sigma_8 \qty(\Omega_{\rm m}/0.3)^{0.595} & = 0.781\pm 0.032 \; (0.794). && \text{[$C_\ell$ TATT]}
\end{align*}
We also compute the Karhunen–Loève (KL) decomposition to quantify the improvement of the posterior with respect to the prior using \texttt{tensiometer} (see \cref{sec:sampling}). We find that the KL mode that is best constrained by the data corresponds to ${\alpha={0.521}}$, which is remarkably close to the $S_8$ ($\alpha=0.5$) parameter theoretically inferred in \citet{1997ApJ...484..560J}. A visualization of the KL decomposition is also given in \cref{app:full_posterior}.

We then include shear ratio information \citep*{y3-shearratio} to further reduce the uncertainty on $S_8$, as shown by the filled contours in \cref{fig:cont_lcdm_cl_planck}. We find this addition improves constraints on $S_8$ by about 18\% and yields a more symmetric marginal posterior, with
\begin{align*}
    S_8 & = 0.784\pm 0.026 \; (0.798), && \text{[$C_\ell$+SR TATT]}\\
    \sigma_8 \qty(\Omega_{\rm m}/0.3)^{0.598} & = 0.783\pm 0.021 \; (0.788). && \text{[$C_\ell$+SR TATT]}
\end{align*}
This additional data noticeably removes part of the lower tail in $S_8$, which is due to a degeneracy with IA parameters, as will be seen in \cref{sec:res_IA}, and also improves constraints on redshift distributions uncertainties by 10-30\%. The volume of the two-dimensional marginal $(S_8,\Om)$ posterior, as approximated from the sample covariance, is reduced by about 20\% when including shear ratios.  

In comparison to constraints from \planck 2018, we find a lower amplitude of structure $S_8$. We estimate the tension with the parameter shift probability metric using the \texttt{tensiometer} package, which accounts for the non-Gaussianity of the posterior distributions \citep{2021arXiv210503324R}, and find tensions of about \SI{1.4}{$\sigma$} and \SI{1.5}{$\sigma$} with and without shear ratios, respectively.

Finally, we note that DES~Y3 shear data alone is not able to constrain the dark energy equation-of-state $w$. We find that the evidence ratio between \wcdm and \lcdm is $R_{w/\Lambda}=\num{0.68(18)}$, which is inconclusive, based on the Jeffreys scale. We thus find no evidence of a departure from \lcdm, consistent with \citet*{y3-cosmicshear1} and \citet*{y3-cosmicshear2}.

\subsection{Constraints on intrinsic alignments}
\label{sec:res_IA}

\begin{figure}
    \centering
    \includegraphics[scale=0.65]{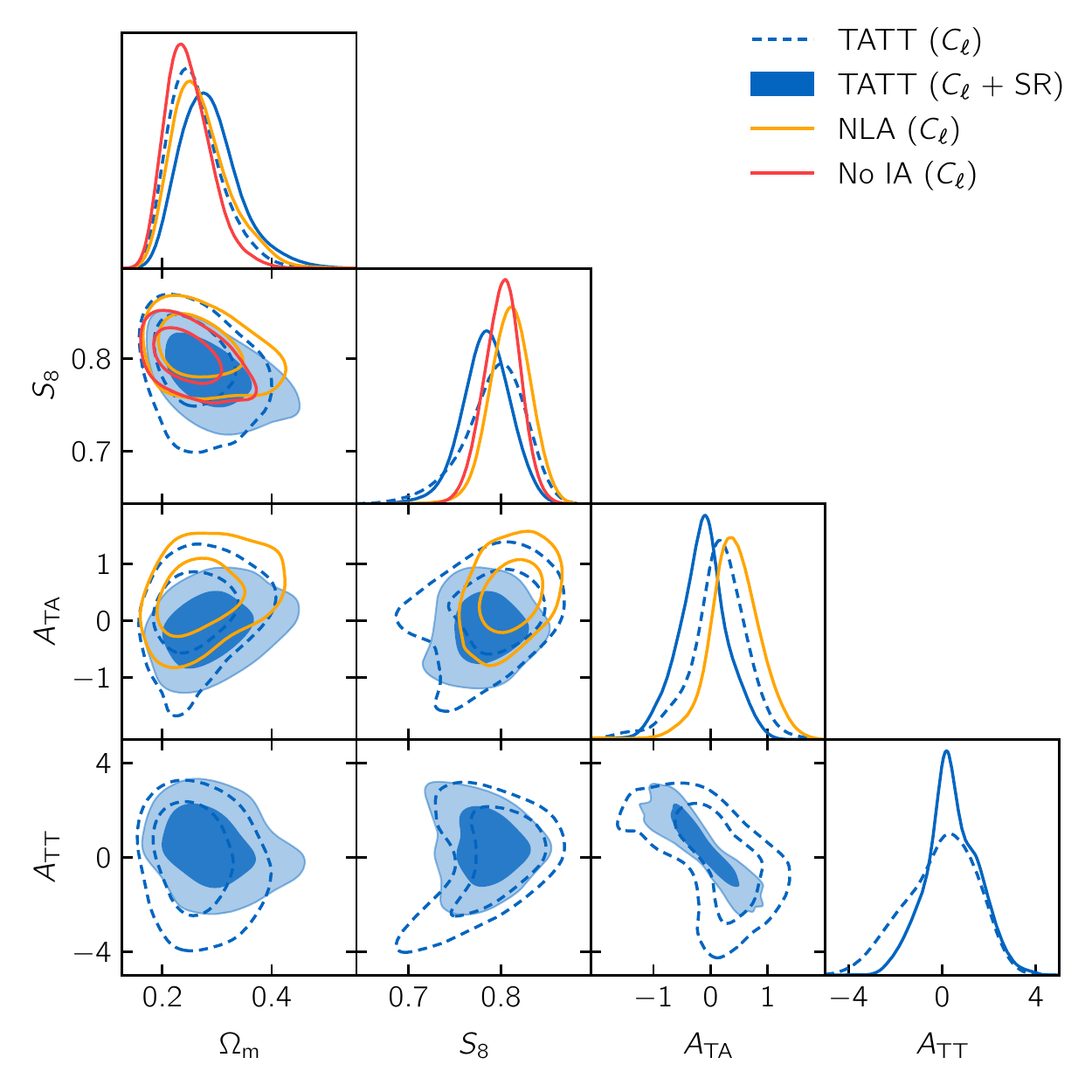}
    \caption{Constraints on cosmological and intrinsic alignment (IA) parameters from DES~Y3 cosmic shear power spectra. The three colors refer to the assumed IA model: TATT in blue, NLA in orange and no IA in red. The filled blue contours include information from shear ratios while the dashed ones do not. Shear ratios are not included for the NLA and no IA models.}
    \label{fig:cont_ia}
\end{figure}

In this section, we focus on constraints on intrinsic alignments (IA) and explore the robustness of cosmological constraints with respect to the IA model.

The fiducial model, TATT, accounts for the possibility of tidal torquing and has five free parameters in the DES~Y3 implementation (see \cref{tab:params}). \Cref{fig:cont_ia} shows constraints on the amplitude parameters for the tidal alignment and tidal torquing components. As stated in \citet{2019PhRvD.100j3506B}, the II component of the TATT model, which is found to dominate over the GI and IG components \citep*[see fig. 16 of][]{y3-cosmicshear2}, receives contributions that are proportional to $\Ata^2$, $\Att^2$ and $\Ata\Att$. There is therefore a partial sign degeneracy between those parameters, which can be observed in the corresponding panel of \cref{fig:cont_ia}. We then find that including shear ratios significantly reduces the marginal $(\Ata,\Att)$ posterior volume by a factor of about 3, which in turn improves cosmological constraints, as reported in the previous section. In this case, we obtain 
\begin{align*}
    \Ata & = -0.14\pm 0.43 \; (-0.398), && \text{[$C_\ell$+SR TATT]} \\
    \Att & = 0.4\pm 1.1 \; (1.714). && \text{[$C_\ell$+SR TATT]}
\end{align*}
These constraints alone do not exclude zero, potentially due to the aforementioned sign degeneracy. If we restrict the prior to $\Ata>0$, we find ${\Ata = 0.30^{+0.12}_{-0.30}}$ and ${\Att = -0.69^{+0.83}_{-0.43}}$, with essentially unchanged cosmological constraints.
We do not show constraints on the redshift tilt parameters $\ata$ and $\att$, which are unconstrained by the data (which might be due to amplitude parameters being consistent with zero).

We also report constraints on the NLA model in \cref{fig:cont_ia}, a subset of TATT where $\Att=\bta=0$, which is not excluded by the data. {We exclude shear ratio information here, so as to compare constraints obtained with shear power spectra alone (TATT constraints are shown by dashed lines in \cref{fig:cont_ia})}. Because of the complex degeneracy between $S_8$ and $\Att$, visible in \cref{fig:cont_ia}, fixing the tidal torquing component to zero results in cosmological constraints that are improved by about 27\% on $S_8$, and which are found to be consistent with the TATT case. Assuming the NLA model, we find
\begin{align*}
    S_8 & = 0.810\pm 0.023 \; (0.834), && \text{[$C_\ell$ NLA]} \\
    \Ata & = 0.40\pm 0.51 \; (0.701), && \text{[$C_\ell$ NLA]},
\end{align*}
\ie a slightly larger value of $S_8$, albeit within uncertainties of the fiducial model.
Finally, we note that removing IA contributions altogether further improves the constraint on $S_8$ by about 16\%, yielding
\begin{align*}
    S_8 & = 0.801^{+0.021}_{-0.018} \; (0.836), && \text{[$C_\ell$ no IA]},
\end{align*}
also consistent with the NLA and TATT cases.

In terms of model selection, we find that going from no IA to NLA, and then from NLA to TATT improves fits by ${\Delta\chi^2=-0.3}$ and ${\Delta\chi^2=-1.1}$ respectively, while introducing two and three more parameters. The evidence ratios are given by
${R_{{\rm NLA}/{\rm TATT}}=\num{3.59(93)}}$,
${R_{{\rm no IA}/{\rm TATT}}=\num{17.5(43)}}$ and
${R_{{\rm no IA}/{\rm NLA}}=\num{4.88(11)}}$,
marking a weak preference for NLA over TATT, but a substantial preference for no IA over TATT, according to the Jeffreys scale.

Cosmic shear analyses in harmonic space usually only exploit the $E$-mode part of the power spectrum. However, as detailed in \cref{sec:ia_th}, tidal torquing generates a small $B$-mode signal, which may at least be constrained by our $B$-mode data. We validated our analysis pipeline by checking that
\begin{inparaenum}[(i)]
    \item the $E$-to-$B$-mode leakage measured in our Gaussian simulations (see \cref{sec:sims_gaussian}) is consistent with expectations from mixing matrices, 
    \item we do recover correct IA parameters, with tighter constraints, for synthetic data vectors for different values of the IA parameters (including non-zero $\Att$).
\end{inparaenum}
We obtain constraints that are consistent for cosmological parameters inferred without $B$-mode data. However, they seem to strongly prefer non-zero $\Att$, and are not consistent across redshift bins. This preference is indeed entirely supported by bin pairs $3,3$ an $3,4$, that have the highest $\chi^2$ with respect to no $B$-mode, as shown in \cref{fig:clobs_BB}. Including $B$-mode data and freeing TATT parameters, the $\chi^2$ for those bins are reduced by \num{13.5} and \num{17.4} respectively, while all other bin pairs are unaffected ($\chi^2$ changed by less than 1). Indeed, we find that removing bin~3 entirely makes the preference for non-zero $\Att$ disappear, with very small impact on the cosmology. We obtain very similar results when including shear ratios. We conclude from this experiment that DES~Y3 data is not able to constrain the contribution of tidal torquing to the TATT model efficiently, leading to the model picking up potential flukes in the $B$-mode data, which has been verified to be globally consistent with no $B$-modes.
{Future data will place stronger constraints on $B$-modes and its potential cosmological sources.}

\subsection{Constraints on baryons}
\label{sec:res_baryons}

\begin{figure}
    \centering
    \includegraphics[scale=0.65]{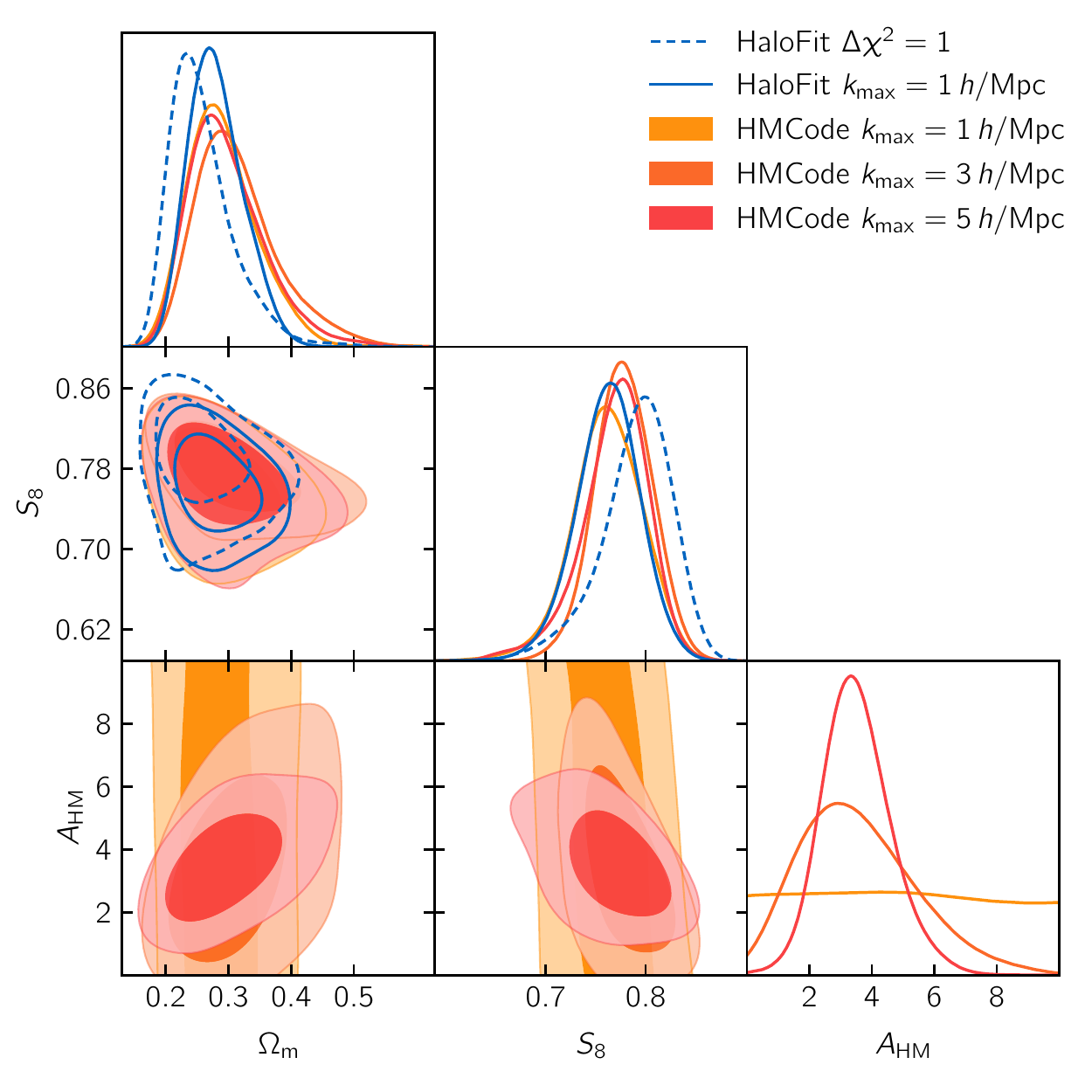}
    \caption{Constraints on cosmological and baryonic feedback parameters from DES~Y3 cosmic shear power spectra. In blue, we show constraints for the fiducial model, \ie using \halofit. In orange to red, we show constraints using \hmcode with one free parameter, while varying the $\kmax$ cut-off from 1~to~ \SI{5}{\h\per\mega\parsec} (see \cref{fig:clobs}). We also show, with dashed lines, the constraints for the fiducial \halofit model and the $\kmax=\SI{1}{\h\per\mega\parsec}$ cut, which is even more conservative than our fiducial $\Delta\chi^2=1$ cut. Note that all constraints shown here use TATT to model intrinsic alignments and none include shear ratio information.}
    \label{fig:cont_hm1}
\end{figure}

We now turn our attention towards baryonic feedback. Our fiducial analysis discards scales where baryonic feedback is expected to impact the shear power spectrum. However, we have shown in \cref{sec:modeling_validation_synth} that \hmcode provides a model that is both accurate and flexible enough for our analysis, for scale cuts with $\kmax$ in the range \SIrange{1}{5}{\h\per\mega\parsec}.

\Cref{fig:cont_hm1} shows constraints obtained assuming \hmcode with one free parameter, for varying scale cuts, as well as a comparison to the fiducial \halofit model. We find cosmological constraints to be robust to the choice of $\kmax$, with deviations below \SI{0.5}{$\sigma$}. In particular, in \cref{fig:cont_hm1} we show contours for both models for $\kmax=\SI{1}{\h\per\mega\parsec}$, which is more conservative than our fiducial ${\Delta\chi^2=1}$ scale cut, and find very good agreement.
We then find that extra data points included when raising $\kmax$ from \SIrange{1}{5}{\h\per\mega\parsec} (71 to 228) do constrain the \hmcode baryonic feedback parameter $\Ahm$, but have a relatively little impact on cosmological constraints, both in position and width. In other words, given our current error bars, cosmological information at small scales is partially lost by marginalizing over uncertainties in the baryonic feedback model.
For the $\kmax=\SI{5}{\h\per\mega\parsec}$ cut, we find $\chi^2=235.2$ ($p=0.25$) at best-fit, and constraints given by
\begin{align*}
    \Omega_{\rm m} & = 0.297^{+0.043}_{-0.071} \; (0.246), && \text{[$C_\ell$ \hmcode TATT]} \\
    S_8 & = 0.769^{+0.037}_{-0.026} \; (0.762), && \text{[$C_\ell$ \hmcode TATT]} \\
    \Ahm & = 3.52^{+0.94}_{-1.2} \; (1.620). && \text{[$C_\ell$ \hmcode TATT]}
\end{align*}
This is in good agreement with cosmological constraints reported for the \halofit model in \cref{sec:res_lcdm}, although this model does favor slightly lower $S_8$ and $\sigma_8$ values, and a higher $\Om$ value, which happens to be closer to the \planck value, as seen in \cref{fig:1d_all}. As a consequence, the tension with \planck rises to \SI{1.7}{$\sigma$} in this case. The corresponding best-fit model is represented by dashed lines in \cref{fig:clobs}, where we observe that, on large scales, \ie for multipoles below the fiducial scale cuts, both models agree very well. However, on smaller scales, \hmcode yields shear power spectra 10-20\% lower, which, visually, seems to provide a better fit to data (again, those scales are excluded in the fiducial model).

When using \hmcode with two free parameters, we find that the constraining power is entirely transferred to the second parameter, $\etahm$, with very little impact on cosmological constraints. For ${\kmax=\SI{5}{\h\per\mega\parsec}}$, we find ${\etahm=0.86^{+0.29}_{-0.35}}$ while $\Ahm$ is unconstrained.

The previous constraints are based on our fiducial IA model, TATT. However, we showed in the previous section that the NLA model seems favored by the data (using evidence ratios). If we use this model instead, {as done in the KiDS-1000 analysis \citep{2021A&A...645A.104A}}, we find ${S_8 = 0.790\pm 0.024}$ and ${\Ahm = 3.67^{+0.71}_{-0.92}}$, although we note immediately that we have not validated our scale cuts against this specific model and that these results should be interpreted with caution.

Our results do not allow exclusion of the dark matter only value of ${\Ahm=3.13}$ in either direction. In comparison to the hydrodynamical simulations we used in \cref{sec:baryons} to validate the model, constraints from data are closer to Massive~Black~II, although the uncertainty from shear power spectra alone is too large to discriminate between baryonic feedback prescriptions.
\Cref{fig:cont_hm1} suggests that a better understanding of the effect of baryons on the distribution of matter will be an important task in order to be able to capture cosmological information at small scales.
For the foreseeable future, this will likely require cross-correlating shear data with other probes that are sensitive to baryons, \eg~{Compton-$y$} maps of the thermal Sunyaev-Zeldovich (SZ) effect with CMB maps (see, \eg, \citealt{2021arXiv210801601P,2021arXiv210801600G} with DES~Y3 data and \citealt{2021arXiv210904458T} with KiDS-1000 data) or the kinetic SZ effect \citep{2021PhRvD.103f3513S,2021PhRvD.103f3514A}.
Another avenue is to exploit information from even smaller scales, \eg using a principal component analysis to span a variety of scenarios from hydrodynamical simulations (see \citealt{2019MNRAS.488.1652H} for the methodology and \citealt{2021MNRAS.502.6010H} for an application to DES~Y1 data) or a \textit{baryonification} model (see \citealt{2015JCAP...12..049S,2019JCAP...03..020S}, and Chen et al. in preparation for an application to DES~Y3 data).

\subsection{Consistency with other DES~Y3 weak lensing analyses}
\label{sec:res_consistency_kp}

\begin{figure}
    \centering
    \includegraphics[scale=0.65]{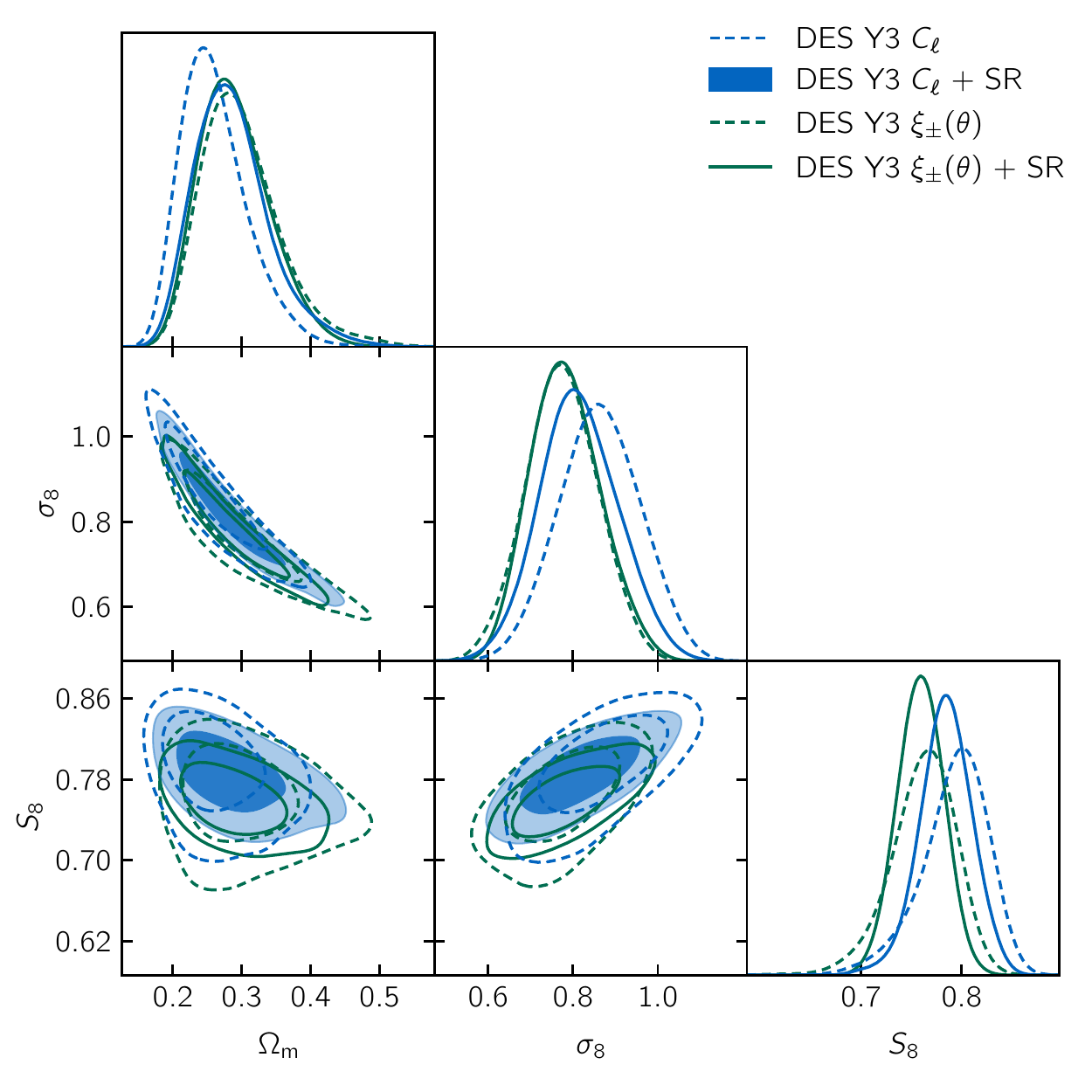}
    \caption{Comparison of cosmological constraints obtained from the analysis of cosmic shear two-point functions of DES~Y3 data in real \citep*[in green,][]{y3-cosmicshear1,y3-cosmicshear2} and harmonic space (in blue, this work).
    Solid contours indicate constraints that include shear ratio information \citet*{y3-shearratio}. We find ${\Delta S_8=0.025}$, with shear ratios, consistent with the expected statistical scatter ${\sigma (\Delta S_8) \sim 0.02}$ predicted in \protect\citet{2020arXiv201106469D}.}
    \label{fig:cont_lcdm_xipm}
\end{figure}

\begin{figure}
    \centering
    \includegraphics[scale=0.65]{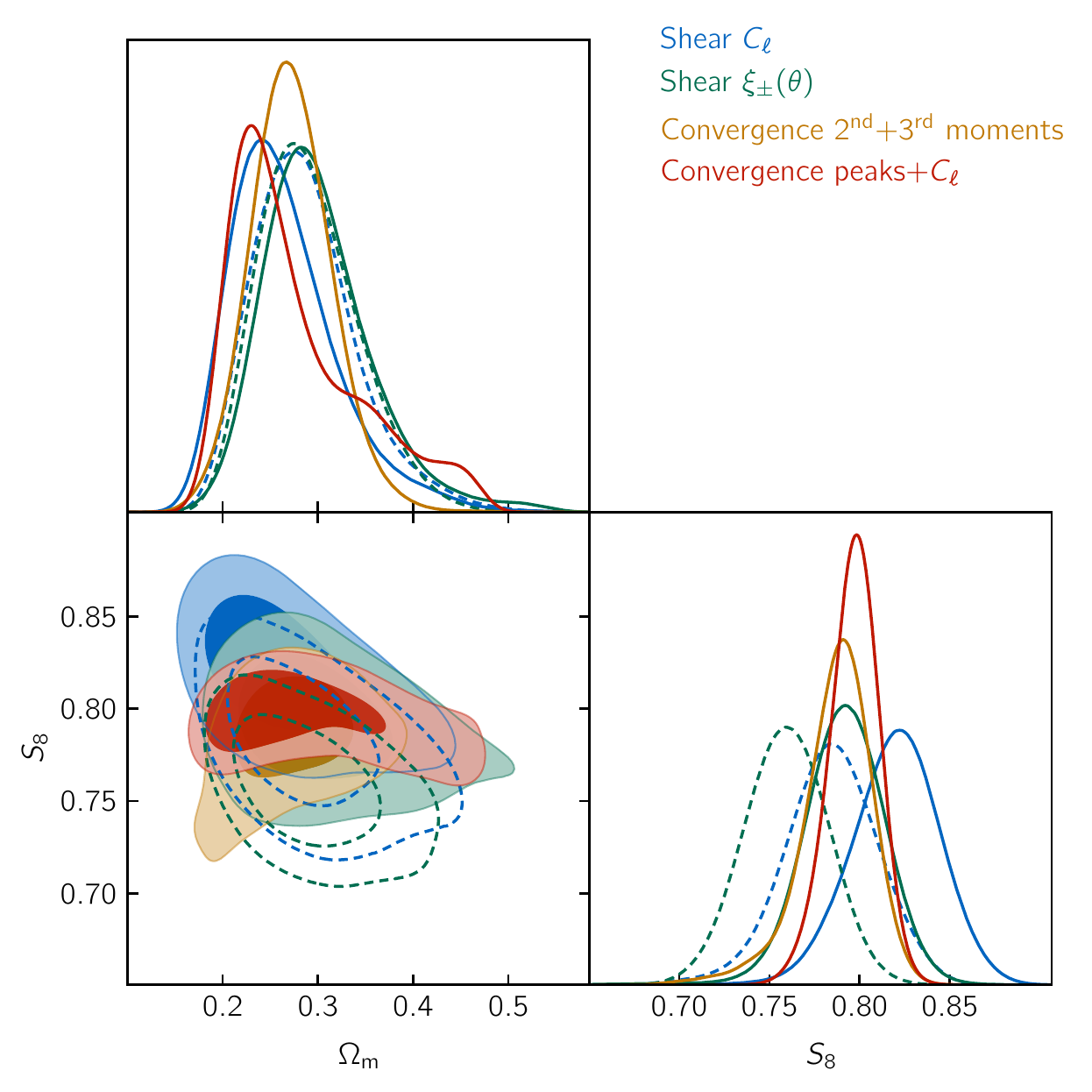}
    \caption{Comparison of cosmological constraints obtained from the analysis of DES~Y3 lensing data using four different statistics: shear power spectra (this work, in blue), shear two-point functions \citep*[][in green]{y3-cosmicshear1,y3-cosmicshear2}, convergence second and third order moments \citep[][in orange]{2021arXiv211010141G}, and convergence peaks and power spectra \citep[][in red]{2022MNRAS.tmp..151Z}. For the first two, we have matched the modeling to that adopted for the analysis of non-Gaussian convergence statistics, namely restricting the intrinsic alignment model to NLA and fixing the total mass of neutrinos (see main text for a discussion of possible caveats). These constraints are shown by solid contours, whereas constraints obtained with the fiducial model are shown by the dashed contours, for reference. None of the constraints shown here include shear ratio information. {Although the comparison requires some care, this figure highlights the overall consistency of DES~Y3 lensing data and existing analyses}.}
    \label{fig:cont_lcdm_all}
\end{figure}

In this section, we compare our results obtained from cosmic shear power spectra to other studies using DES~Y3 lensing data, as detailed below. We first focus on the comparison with the real-space analysis of shear two-point functions presented in \citet*{y3-cosmicshear1,y3-cosmicshear2}. The study presented here is its harmonic space counterpart, in the sense that we follow a very similar methodology and use the same fiducial model. We then extend the comparison to studies that incorporate non-Gaussian information from the DES~Y3 convergence (mass) map \citep*{y3-massmapping}, namely the analysis of peaks and power spectra from \citet{2022MNRAS.tmp..151Z}, and the analysis of second and third order moments from \citet{2021arXiv211010141G}. \Cref{fig:cont_lcdm_xipm,fig:cont_lcdm_all} show cosmological constraints obtained from those studies, which are found to be in very good agreement, illustrating the internal consistency of DES~Y3 shear analyses. See also \cref{fig:1d_all} for a comparisom of all one-dimensional marginal constraints.

\textbf{Real space two-point functions $\xipm$.}
\Cref{fig:cont_lcdm_xipm} shows cosmological constraints obtained from two-point functions in real space \citep*{y3-cosmicshear1,y3-cosmicshear2} and in harmonic space (this work), both with and without including shear ratio information. We find that both studies yield very consistent cosmological constraints, with a preference for slightly higher $S_8$ from shear power spectra. However, the difference between the means of the posteriors is ${{\Delta S_8}=0.031}$ when excluding shear ratios, which is fairly consistent with the expected statistical scatter $\sigma(\Delta{S_8})\sim\num{0.02}$ predicted\footnote{Note that this prediction depends strongly on the two sets of scale cuts and the survey configuration.} in \citet{2020arXiv201106469D}. The degeneracy directions are found to be slightly different, with $\alpha_{C_\ell}=0.595$ and $\alpha_{\xi_\pm}=0.552$ for harmonic and real space analyses, respectively. When including shear ratios, the difference narrows down to ${{\Delta S_8}=0.025}$ and the best constrained direction is almost identical, with $\alpha_{C_\ell}=0.598$ and $\alpha_{\xi_\pm}=0.586$. As a consequence of the higher value of $S_8$ found here, the tension with \planck is reduced from \SI{2.3}{$\sigma$} in \citep*{y3-cosmicshear1,y3-cosmicshear2} to \SI{1.5}{$\sigma$} in this work.

For IA parameters, we find an overall excellent agreement (not shown). Although the real-space analysis shows a weak preference for negative $\Ata$ and positive $\Att$, we observe the same degeneracy between those parameters, with almost perfect overlap. The two parameters that describe redshift evolution are unconstrained in both cases, but the posteriors are also nearly identical. {We also find that fixing the IA model to NLA results in a slightly higher value for $S_8$.}

\textbf{Non-Gaussian statistics from mass maps.}
\Cref{fig:cont_lcdm_all} presents cosmological constraints from all four lensing analyses. Due to difficulties in modeling non-Gaussian statistics, both analyses of moments and peaks \citep{2021arXiv211010141G,2022MNRAS.tmp..151Z} include IA contributions using a model based on NLA, and both fix the total mass of neutrinos to the minimum value of \SI{0.06}{\eV}. In order to make the comparison more meaningful, we therefore re-analyze shear two-point functions and power spectra with these two changes, which tends to favor slightly higher values of $S_8$ (either change individually also goes in this direction).
We warn the reader that
\begin{inparaenum}[(i)]
    \protect\item despite matching important modeling choices, there remain differences in the analysis in terms of priors, modeling pipeline technology (\eg \citealt{2022MNRAS.tmp..151Z} uses an emulator) and methodology, and
    \protect\item the scale cuts used for two-point functions were not validated for this specific model, and should be interpreted with caution.
\end{inparaenum}
Nevertheless, this figure illustrates the high level of consistency of these analyses -- all of which followed a similar blinding procedure -- and of DES~Y3 lensing data.

\subsection{Comparison with other lensing surveys}
\label{sec:res_other_wl}

\begin{figure}
    \centering
    \includegraphics[scale=0.65]{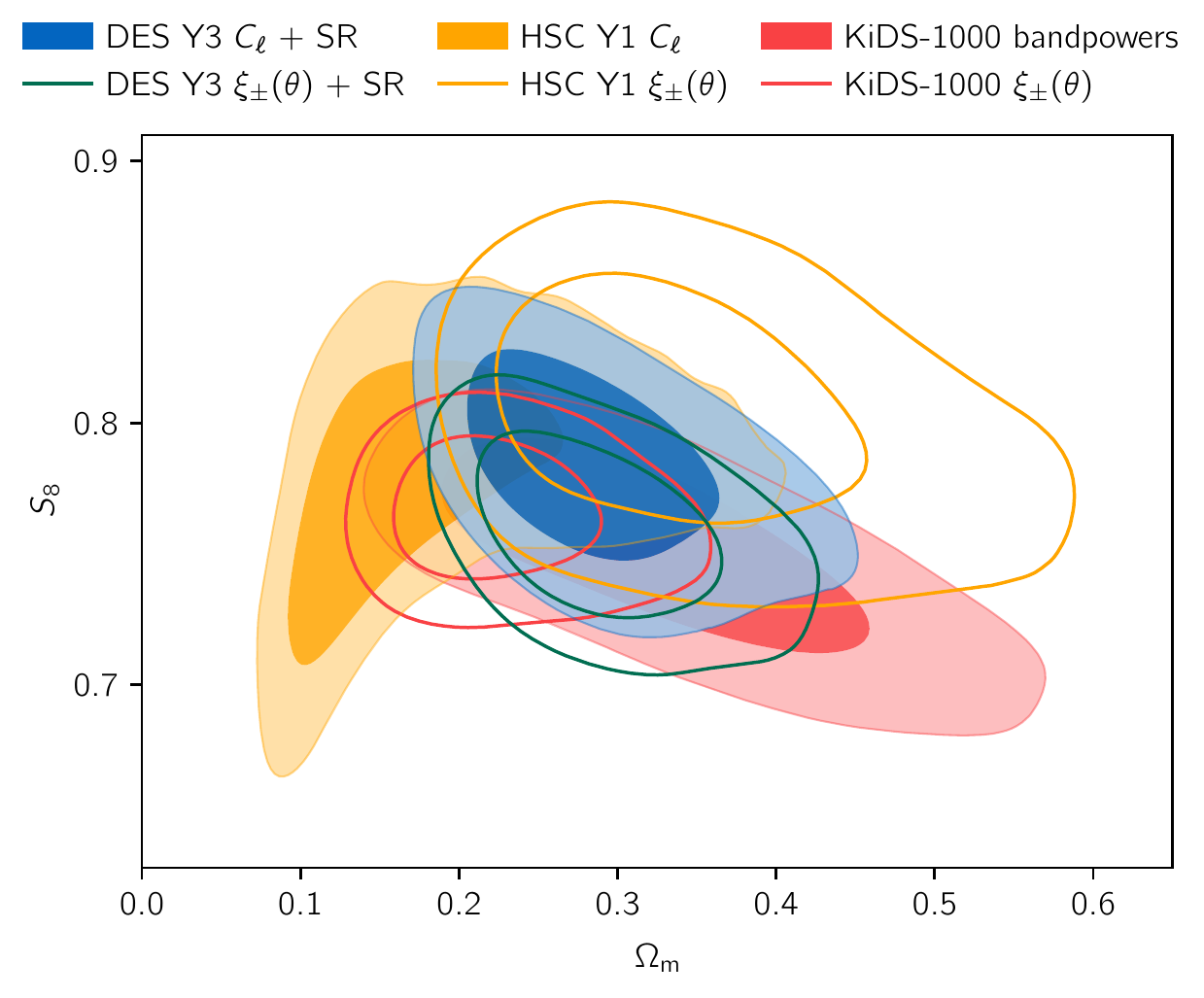}
    \caption{Comparison of cosmological constraints from the analysis of cosmic shear in harmonic (filled contours) and real space (contour lines) for DES~Y3 (this work in blue, \citealt*{y3-cosmicshear1,y3-cosmicshear2} in green), HSC~Y1 \citep[][in yellow]{2019PASJ...71...43H,2020PASJ...72...16H} and KiDS-1000 \citep[][in red]{2021A&A...645A.104A}. {We note that these results rely on different analysis and modeling choices.}}
    \label{fig:cont_lcdm_all_wl}
\end{figure}

In the past two years, both the HSC and KiDS collaborations have presented cosmic shear analyses of their data in harmonic and real space. \Cref{fig:cont_lcdm_all_wl} compares constraints obtained from DES~Y3 data to those obtained from KiDS-1000
\citep{2021A&A...645A.104A} and HSC~Y1 \citep{2019PASJ...71...43H,2020PASJ...72...16H}. Uni-dimensional marginal distributions are also shown in \cref{fig:1d_all}. {As shown in \citet{2020arXiv201106469D} on simulations, statistical fluctuations are not expected to bias one estimator over the other and shift constraints in any specific direction, while unmodeled systematic effects might. We do not find any clear trend here.}

Both KiDS-1000 and HSC analyses use NLA to model intrinsic alignments with fixed neutrino masses. However we decide to present constraints that were
{obtained from the fiducial models assumed by} each collaboration for simplicity. We also note that the KiDS-1000 analysis uses a ``bandpowers'' estimator of shear power spectra that stems from an original measurement of two-point functions in real space with a thin spacing. A recent analysis \citep{2021arXiv211006947L} applying a pseudo-$\cl$ estimator found very similar constraints {on $S_8 = 0.754^{+0.027}_{-0.029}$ between the bandpowers and pseudo-$\cl$ estimators, despite appreciable differences in the intrinsic alignment parameter, likely due to how the two estimator cut large-scale information}.
Ignoring potential correlations due to overlapping survey areas, we find our results to be in agreement at the \SIlist{0.7;0.4}{$\sigma$} levels with KiDS-1000 bandpowers and HSC~Y1 $\cl$ analyses. Finally, we find good agreement on the IA parameter $\Ata$ (not shown), although constraints remain broad for all three surveys.

\subsection{Reconstruction of the matter power spectrum}
\label{sec:res_Pk}

In this section, we apply the method of \citet{2002PhRvD..66j3508T} to approximately reconstruct the linear matter power spectrum at present time, $P(k)$, from DES~Y3 shear power spectra. We immediately note that this exercise is strongly model-dependent, in that it requires to assume a full cosmological model to relate shear power spectra to the matter power spectrum.
{Moreover, it presents subtleties in relating physical scales between the linear and non-linear power spectra, as discussed in \citet{2002PhRvD..66j3508T}, and we will employ a simplified approach presented in the next paragraph.}
Nevertheless, assuming the \planck 2018 cosmology \citep{2020A&A...641A...6P}, we may compare the power spectrum reconstructed from DES~Y3 data to the expectation from \planck, which is relevant in the context of the $\sigma_8$ tension found in previous weak lensing surveys \citep{,y3-cosmicshear1,y3-cosmicshear2,2019PASJ...71...43H,2020PASJ...72...16H,2021A&A...645A.104A}, and that we also observe in \cref{fig:cont_lcdm_cl_planck}.

To do so, we recast \cref{eq:limber} as an integral over three-dimensional Fourier $k$-modes, using the change of variable ${k=(\ell+1/2)/\chi(z)}$. We then define a window matrix, $\mathbfss{W}$, such that the expected value of our data vector, $\expval{\hat{\vb{C}}_L}$, may be expressed as a function of the linear matter power spectrum at $z=0$, $P(k)$, computed in log-spaced $k$-bins of width $\Delta_{\ln k}$, $\vb{P}$, such that
\begin{equation}
    \expval{\hat{\vb{C}}_L} \approx \mathbfss{W} \vb{P}.
    \label{eq:tz_cl}
\end{equation}
This window matrix is given, for the element corresponding to $k$ and $C_L^{ab}$, {and ignoring intrinsic alignments}, by
\begin{equation}
    \mathbfss{W}_{k,L,a,b} \approx  k \Delta_{\ln k} (L+1/2) q_a(\chi) q_b(\chi) \frac{P_{\rm NL}\qty(k, z(\chi))}{P_{\rm fid}(k)}
    \label{eq:tz_w}
\end{equation}
with $\chi=(L+1/2)/k$. Given the data covariance $\mathbfss{C}$, the reconstructed power spectrum has estimated value and covariance given by
\begin{align}
    \hat{\vb{P}} & = \mathbfss{S} \mathbfss{W}^\intercal \mathbfss{C}^{-1} \hat{\vb{C}}_L , \label{eq:pk_est} \\
    \mathbfss{S} & = \qty[ \mathbfss{W}^\intercal \mathbfss{C}^{-1} \mathbfss{W} + \sigma^{-2} \mathbfss{I} ]^{-1}, \label{eq:pk_cov}
\end{align}
where we have included a regularization term, $\sigma$, which enables inverting \cref{eq:tz_cl} at the price of accepting that certain $k$-modes may not be recovered from the data (the results have very low dependence on $\sigma$, if chosen large enough, in the range where the data is constraining).
To ensure numerical stability, we use 20~bins in the range ${k\sim\SIrange{1e-3}{1e2}{\h\per\mega\parsec}}$, and subsequently rebin the estimated power spectrum within 10 bins for better visualization as well as to suppress the anticorrelation of adjacent bins.
The simplification here comes from \cref{eq:tz_w}, where the dependence on the linear matter power spectrum is made explicit by simply multiplying the numerator and denominator by $P_{\rm fid}(k)$, the power spectrum at redshift zero for the fiducial \planck 2018 cosmology. Our exercise therefore amounts to a reconstruction of the integrand over $\ln{k}$ with respect to what is expected from \planck, rather than a reconstruction of the linear matter power spectrum itself.

\begin{figure}
    \centering
    \includegraphics[scale=0.65]{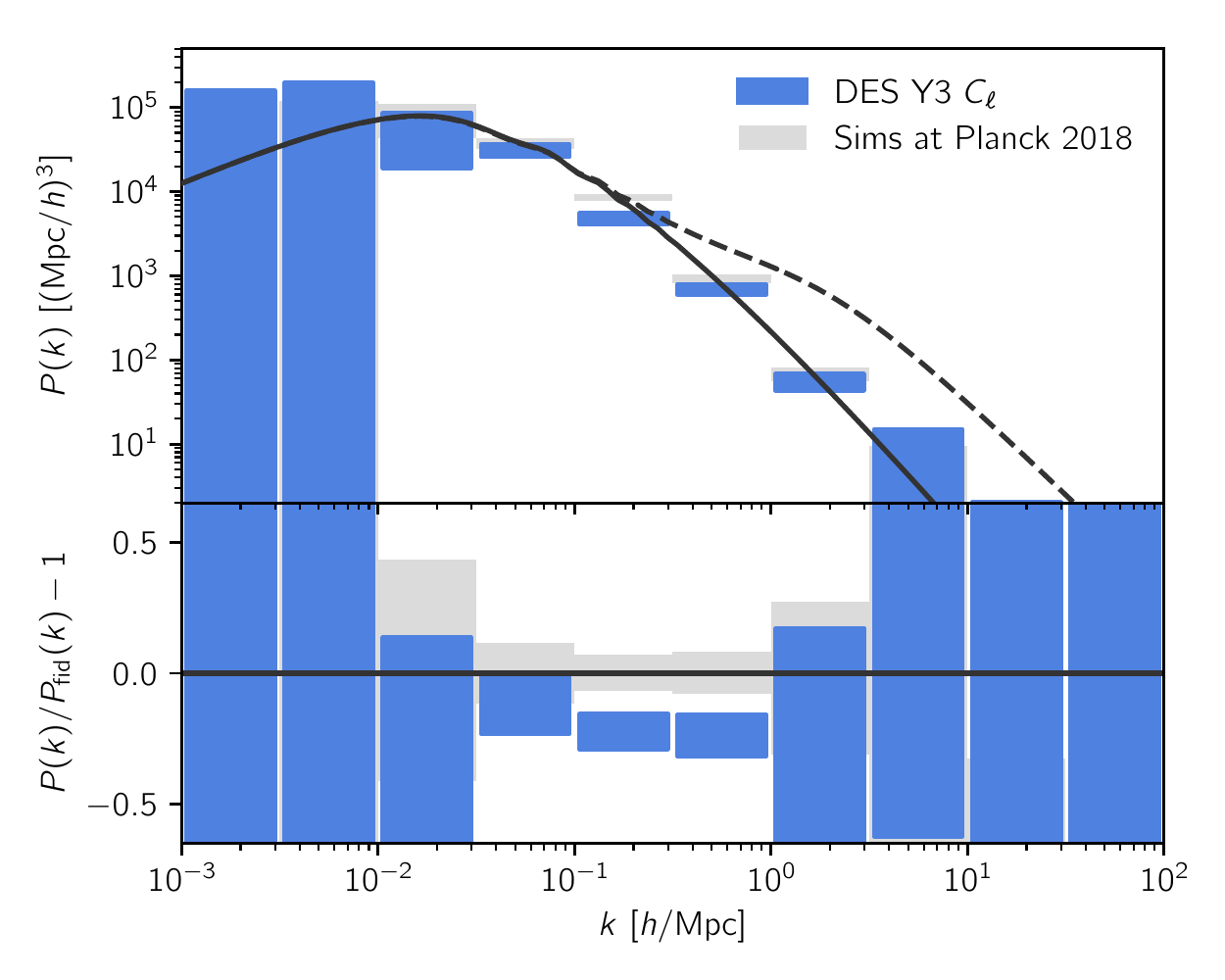}
    \caption{Matter power spectrum at redshift $z=0$ reconstructed from DES~Y3 shear power spectra, using a simplified version of the method of \citet{2002PhRvD..66j3508T}. The fiducial linear matter power spectrum, computed at \planck 2018 cosmology \citep{2020A&A...641A...6P}, is shown by the solid, black line (the corresponding non-linear power spectrum is shown by the dashed, black line).
    The blue boxes, centered on ${(k, \hat{\vb{P}})}$ (see \cref{eq:pk_est}) and of height given by the square-root of the diagonal of the covariance matrix $\mathbfss{S}$ (see \cref{eq:pk_cov}), show the reconstructed power spectrum within log-spaced $k$~bins.
    In the background, we show in gray the result of the reconstruction for \num{1000} simulated data vectors drawn from the likelihood at \planck cosmology; however, in this case, the height of the boxes represents the standard deviation of the results, offering a simple check for the covariance matrix. The reconstructed power spectrum is about 20\% (or roughly  \SI{2}{$\sigma$}) lower than the fiducial one around $k\sim\SI{0.3}{\h\per\mega\parsec}$.
    }
    \label{fig:pk_tz}
\end{figure}

The result is shown in \cref{fig:pk_tz}. The lower panel shows the reconstructed, binned ratio of the power spectrum with respect to the prediction from \planck 2018 (in blue), compared to the results obtained from simulated DES~Y3 data vectors generated by sampling the likelihood at the \planck 2018 cosmology (in gray). In the upper panel, we multiply these ratios by the fiducial linear power spectrum, shown in black. We find that the reconstructed spectrum is roughly 20\% lower than the prediction in the range ${k\sim\SIrange{0.03}{1}{\h\per\mega\parsec}}$ that is constrained by DES~Y3 data. In particular, the reconstruction is about \SI{2}{$\sigma$} low around $k\sim\SI{0.3}{\h\per\mega\parsec}$, which remains close to the linear regime.

\section{Conclusions}
\label{sec:conclusion}

In this work, we have used data from the first three years of observations by the Dark Energy Survey (DES~Y3), including a catalog of over a hundred million galaxy shape measurements \citep*{y3-shapecatalog} split into four redshift bins \citep*{y3-sompz}, to measure tomographic cosmic shear power spectra. {Our measurements over the DES~Y3 footprint of \SI{4143}{\deg\squared} are based on the pseudo-$\cl$ method, with a consistent spherical sky approach using the \namaster software \citep{2019MNRAS.484.4127A}.} We generally followed the DES~Y3 methodology laid out in \citet*{y3-cosmicshear1,y3-cosmicshear2} and the modeling choices presented in \citet*{y3-generalmethods} to infer cosmological constraints, and found ${S_8 \equiv \sigma_8 \sqrt{\Om/0.3} = 0.793^{+0.038}_{-0.025} \; (0.810)}$ {using cosmic shear alone}. We also included geometric information from small-scale galaxy-galaxy lensing ratios \citep*{y3-shearratio} to tighten the constraint to ${S_8 = 0.784\pm 0.026 \; (0.798)}$.

Following \citet*{y3-cosmicshear1,y3-cosmicshear2}, we modeled intrinsic alignments with TATT \citep{2019PhRvD.100j3506B} that coherently includes tidal alignment (TA) and tidal torquing (TT) mechanisms. We found, {as in \citet*{y3-cosmicshear2},} that the data does not strongly favor this model over the simpler non-linear alignment (NLA) model, as the data does not seem to constrain the TT contribution efficiently (even when including $B$-modes in the analysis, which may be sourced by TT). In all cases, we find consistent cosmological constraints, although using NLA tightens constraints on $S_8$ by about 25\%.

We include smaller scales that had been discarded in the fiducial analysis, switching from \halofit to \hmcode to model the non-linear matter power spectrum, thus including the effect of baryonic feedback, known to be a major source of uncertainty for cosmic shear at small scales \citep{2018MNRAS.480.3962C,2019MNRAS.488.1652H}. We derived a set of scale cuts that approximately map to a cut-off $\kmax$ in Fourier modes. When raising $\kmax$ from \SI{1}{\h\per\mega\parsec} to \SI{5}{\h\per\mega\parsec}, we found consistent cosmological constraints, while the extra statistical power appears to mainly constrain the baryonic feedback parameter, ${\Ahm = 3.52^{+0.94}_{-1.2} \; (1.620)}$. This result does not rule out the dark matter-only case ($\Ahm=3.13$) nor the predictions from the hydrodynamical simulations we considered in this work.
Given current error bars and theoretical uncertainties, it therefore remains difficult to extract small-scale cosmological information that is present in our cosmic shear data, thus highlighting the need to better understand the effect of baryonic processes on the clustering of matter, especially for future surveys \citep[see, \eg][]{2021A&A...649A.100M}.

This analysis complements other weak lensing analyses of DES~Y3 data, namely the analysis of cosmic shear two-point correlation functions presented in \citet*{y3-cosmicshear1,y3-cosmicshear2}, convergence second- and third-order moments \citep{2021arXiv211010141G}, and convergence peaks and power spectra \citep{2022MNRAS.tmp..151Z}, the latter two being based on maps from \citet*{y3-massmapping}.
With respect to the real-space two-point functions, we find very similar constraints, with a value of $S_8$ slightly higher by ${{\Delta S_8}=0.025}$ when including shear ratios, perfectly consistent with statistical fluctuations of order $\sigma(\Delta{S_8})\sim\num{0.02}$ predicted in \citet{2020arXiv201106469D}. The comparison of constraints from Gaussian and non-Gaussian statistics delivers an overall coherent picture, highlighting the cosmological information beyond two-point measurements and pointing towards the modeling improvements required for future analyses. This analysis thus provides an important consistency check of DES~Y3 lensing data. It also demonstrates the feasibility of conducting a harmonic space analysis over a wide survey footprint, which could be combined with other estimators, such as the real-space correlation functions, into a joint analysis in the future. To do so, one would need to compute an accurate estimate of the cross-covariance of the different statistics considered, or to perform a simulation-based, likelihood-free analysis \citep[see, \eg][]{2021MNRAS.501..954J}.

At last, we compared our results to those obtained by other weak lensing studies from the Hyper Suprime-Cam and Kilo-Degree Survey collaborations and found consistent constraints on cosmology. We also compared our results to constraints from observations of the cosmic microwave background. We found that the tension with \planck 2018 in $S_8$, {computed with the parameter shift probability \citep{2020PhRvD.101j3527R,2021arXiv210503324R}, is \SI{1.5}{$\sigma$} in this work, whereas it} is \SI{2.3}{$\sigma$} in \citet*{y3-cosmicshear1,y3-cosmicshear2}. This {shift} is reflected in the inferred linear matter power spectrum, in excess by about 20\% in the range ${k\sim\SIrange{3e-2}{1}{\h\per\mega\parsec}}$ for \planck with respect to DES~Y3.
Future observations, such as the complete data from the six-year program of the DES {and data from the next generation of surveys including LSST, Euclid and Roman, as well as } methodological improvements will be necessary to determine whether this apparent tension is the sign of an incorrect treatment of systematic effects, or of new physics.

\section*{Data availability}

A general description of DES data releases is available on the survey website at \url{https://www.darkenergysurvey.org/the-des-project/data-access/}. DES Y3 cosmological data has been partially released on the DES Data Management website hosted by the National Center for Supercomputing Applications at \url{https://des.ncsa.illinois.edu/releases/y3a2}. This includes Gold products, PSF modelling, Balrog catalogs, Deep Fields data and the Y3 galaxy catalogs, including the redshift distributions used in this analysis. The \cosmosis software \citep{2015A&C....12...45Z} is available at \url{https://bitbucket.org/joezuntz/cosmosis/wiki/Home}. The measurement code, used in this analysis to interface DES catalogs and \namaster, can be obtained upon request to the corresponding author.

\section*{Acknowledgments}

The authors would like to thank Masahiro Takada for useful discussions that motivated this work, and David Alonso and Andrina Nicola for discussions about the pseudo-$\cl$ method.

This research has made use of
NASA's Astrophysics Data System,
\textsc{adstex} (\url{https://github.com/yymao/adstex}),
\textsc{NumPy} \citep{2020Natur.585..357H},
\textsc{SciPy} \citep{2020NatMe..17..261V},
\textsc{Matplotlib} \citep{2007CSE.....9...90H},
\textsc{Numba} \citep{2021zndo...5524874L},
\textsc{AstroPy} \citep{2018AJ....156..123A,2013A&A...558A..33A},
\textsc{HealPy} \citep{2019JOSS....4.1298Z},
\namaster \citep{2019MNRAS.484.4127A},
\cosmosis software \citep{2015A&C....12...45Z},
\cosmolike \citep{2014MNRAS.440.1379E,2017MNRAS.470.2100K},
\textsc{GetDist} \citep{2019arXiv191013970L} and
\polychord \citep{2015MNRAS.450L..61H}.

Funding for the DES Projects has been provided by the U.S. Department of Energy, the U.S. National Science Foundation, the Ministry of Science and Education of Spain, 
the Science and Technology Facilities Council of the United Kingdom, the Higher Education Funding Council for England, the National Center for Supercomputing 
Applications at the University of Illinois at Urbana-Champaign, the Kavli Institute of Cosmological Physics at the University of Chicago, 
the Center for Cosmology and Astro-Particle Physics at the Ohio State University,
the Mitchell Institute for Fundamental Physics and Astronomy at Texas A\&M University, Financiadora de Estudos e Projetos, 
Funda{\c c}{\~a}o Carlos Chagas Filho de Amparo {\`a} Pesquisa do Estado do Rio de Janeiro, Conselho Nacional de Desenvolvimento Cient{\'i}fico e Tecnol{\'o}gico and 
the Minist{\'e}rio da Ci{\^e}ncia, Tecnologia e Inova{\c c}{\~a}o, the Deutsche Forschungsgemeinschaft and the Collaborating Institutions in the Dark Energy Survey. 

The Collaborating Institutions are Argonne National Laboratory, the University of California at Santa Cruz, the University of Cambridge, Centro de Investigaciones Energ{\'e}ticas, 
Medioambientales y Tecnol{\'o}gicas-Madrid, the University of Chicago, University College London, the DES-Brazil Consortium, the University of Edinburgh, 
the Eidgen{\"o}ssische Technische Hochschule (ETH) Z{\"u}rich, 
Fermi National Accelerator Laboratory, the University of Illinois at Urbana-Champaign, the Institut de Ci{\`e}ncies de l'Espai (IEEC/CSIC), 
the Institut de F{\'i}sica d'Altes Energies, Lawrence Berkeley National Laboratory, the Ludwig-Maximilians Universit{\"a}t M{\"u}nchen and the associated Excellence Cluster Universe, 
the University of Michigan, NSF's NOIRLab, the University of Nottingham, The Ohio State University, the University of Pennsylvania, the University of Portsmouth, 
SLAC National Accelerator Laboratory, Stanford University, the University of Sussex, Texas A\&M University, and the OzDES Membership Consortium.

Based in part on observations at Cerro Tololo Inter-American Observatory at NSF's NOIRLab (NOIRLab Prop. ID 2012B-0001; PI: J. Frieman), which is managed by the Association of Universities for Research in Astronomy (AURA) under a cooperative agreement with the National Science Foundation.

The DES data management system is supported by the National Science Foundation under Grant Numbers AST-1138766 and AST-1536171.
The DES participants from Spanish institutions are partially supported by MICINN under grants ESP2017-89838, PGC2018-094773, PGC2018-102021, SEV-2016-0588, SEV-2016-0597, and MDM-2015-0509, some of which include ERDF funds from the European Union. IFAE is partially funded by the CERCA program of the Generalitat de Catalunya.
Research leading to these results has received funding from the European Research
Council under the European Union's Seventh Framework Program (FP7/2007-2013) including ERC grant agreements 240672, 291329, and 306478.
We  acknowledge support from the Brazilian Instituto Nacional de Ci\^encia
e Tecnologia (INCT) do e-Universo (CNPq grant 465376/2014-2).

This manuscript has been authored by Fermi Research Alliance, LLC under Contract No. DE-AC02-07CH11359 with the U.S. Department of Energy, Office of Science, Office of High Energy Physics.





\bibliographystyle{mnras}
\bibliography{biblio,des_y3kp} 




\appendix

\section{Point spread function}
\label{app:psf}

This section presents the results of our tests for potential contamination of shear power spectra from the point spread function (PSF) and complements those presented in \citet*{y3-piff,y3-shapecatalog}.

We specifically focus on the additive biases due to PSF misestimation using $\rho$-statistics \citep{10.1111/j.1365-2966.2010.16277.x} following the same diagnostics as \citet*{y3-shapecatalog}. We expect other contributions like the brighter-fatter effect, dependencies of the PSF model residuals on star and galaxy colors, and tangential shear around stars to be negligible, as discussed in Section 5 of \citet*{y3-shapecatalog}.

The estimated shear $\bm{\gamma^{\text{est}}}$ is decomposed as 
\begin{equation}
    \bm{\gamma^{\text{est}}} = \bm{\gamma} + \delta\bm{e}_{\text{PSF}}+\delta \bm{e}_{\text{noise}}
    \label{eq:psf_bias}
\end{equation}
where $\bm{\gamma}$ represents the true shear, $\delta \bm{e}_{\text{noise}}$ denotes noise, and $\delta \bm{e}_{\text{PSF}}$ characterizes additive biases from PSF modeling errors. DES~Y3 uses a sample of reserved stars that were not used to obtain the PSF model, and for which we can compare the modeled PSF ellipticity $\bm{e}_{\text{model}}$ to the measured ellipticity $\bm{e}_{*}$ (and similarly for PSF sizes, with $T_{\text{model}}$ and $T_{*}$). The PSF bias term can be further modeled as
\begin{equation}
    \delta\bm{e}_{\text{PSF}} = \alpha \bm{p} + \beta \bm{q} + \eta \bm{w},
    \label{eq:psf_model}
\end{equation}
where $\bm{p} \equiv \bm{e}_{\text{model}}$, $\bm{q} \equiv \bm{e}_{\text{*}} - \bm{e}_{\text{model}}$, and $\bm{w} \equiv \bm{e}_{\text{*}} (T_{*}-T_{\text{model}})/T_{*}$.
Under the assumption that the true shear signal $\bm{\gamma}$ does not correlate with modeling errors, the cross power spectra of galaxy shear and the PSF parameters $\bm{p}$, $\bm{q}$, and $\bm{w}$ read
\begin{align}
    \vb{C}_\ell\qty(\bm{\gamma^{\text{est}}},\bm{p}) &= \alpha \vb{C}_\ell\qty(\bm{p},\bm{p}) + \beta \vb{C}_\ell\qty(\bm{q},\bm{p}) + \eta \vb{C}_\ell\qty(\bm{w},\bm{p}) \label{eq:psf_tau0},\\
    \vb{C}_\ell\qty(\bm{\gamma^{\text{est}}},\bm{q}) &= \alpha \vb{C}_\ell\qty(\bm{p},\bm{q}) + \beta \vb{C}_\ell\qty(\bm{q},\bm{q}) + \eta \vb{C}_\ell\qty(\bm{w},\bm{q}) \label{eq:psf_tau2},\\
    \vb{C}_\ell\qty(\bm{\gamma^{\text{est}}},\bm{w}) &= \alpha \vb{C}_\ell\qty(\bm{p},\bm{w}) + \beta \vb{C}_\ell\qty(\bm{q},\bm{w}) + \eta \vb{C}_\ell\qty(\bm{w},\bm{w}) \label{eq:psf_tau5}.
\end{align}

\begin{figure*}
    \centering
    \includegraphics[scale=0.65]{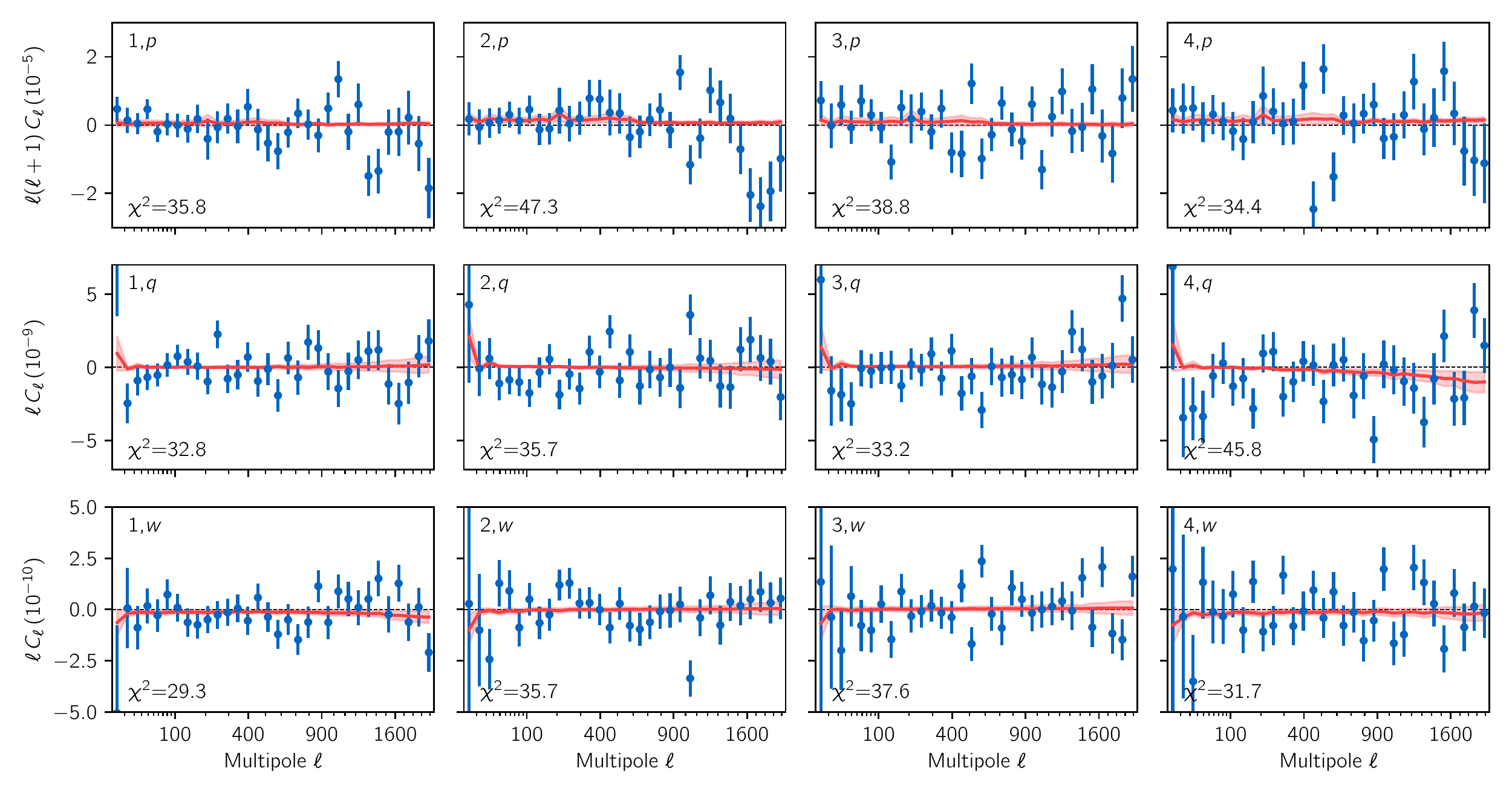}
    \caption{Cross-power spectra between galaxy shapes in the four redshift bins (from left to right) with PSF parameters $\bm{p}$, $\bm{q}$ and $\bm{w}$ (from top to bottom). The measurements are shown in blue, with error bars computed from \num{18000} Gaussian simulations using the DES~Y3 catalog ellipticities and positions, as explained in \cref{sec:sims_gaussian}. The model from \cref{eq:psf_tau0,eq:psf_tau2,eq:psf_tau5} at best-fit is shown by the red line, while the band shows the uncertainty. We find $\chi^2$ statistics with respect to the best fit between \num{29.3} and \num{45.8} (\num{29.3} to \num{47.3} for the null hypothesis) for 32 degrees of freedom, shown in the lower left corner for each panel, corresponding to a minimum probability-to-exceed of 0.04.}
    \label{fig:tau-model}
\end{figure*}

We first measured the cross power spectra of the shear and the PSF parameters $\bm{p}$, $\bm{q}$, and $\bm{w}$. We then repeated these measurements using 18,000 Gaussian simulations, as described in \cref{sec:sims_gaussian}, to obtain their covariance matrix.  
To calculate the cross power spectra between the PSF parameters (right-hand side of \cref{eq:psf_tau0,eq:psf_tau2,eq:psf_tau5}), we split the catalog into two halves that we cross-correlate, which effectively cancels out the shot noise. We then find the best-fit scalar parameters $\alpha$, $\beta$, $\eta$ over all scales and three cross-spectra types for each tomographic redshift bin using Markov chain Monte-Carlo (MCMC) samples generated with the public software package \textsc{emcee} \citep{2013PASP..125..306F}. {This approach is adapted from the measurements performed in the real space analysis \citep*{y3-cosmicshear1} using the same tomographic split, and the non-tomographic measurement from \citet*{y3-piff}}.

\begin{table}
    \centering
    \begin{tabular}{ccccc} 
        & Bin 1 &  Bin 2 &  Bin 3 & Bin 4 \\ \hline \rule{0pt}{3ex}
        $\alpha$ &  $0.003_{-0.007}^{+0.007}$ & $0.014_{-0.008}^{+0.008}$ & $0.008_{-0.010}^{+0.010}$ & $0.012_{-0.011}^{+0.011}$ \\ \rule{0pt}{3ex} 
        $\beta$ &  $0.02_{-0.36}^{+0.36}$ &  $-0.07_{-0.38}^{+0.38}$ &  $0.16_{-0.38}^{+0.39}$ & $-0.74_{-0.47}^{+0.46}$ \\ \rule{0pt}{3ex} 
        $\eta$ & $-5.4_{-4.4}^{+4.3}$ & $0.4_{-4.8}^{+4.8}$ &  $1.6_{-5.0}^{+5.1}$ & $-5.4_{-5.8}^{+5.9}$ \\ \rule{0pt}{3ex} 
        $\chi^2$ &  $99.5$ &  $116.3$ &  $113.4$ & $117.3$ \\ \hline
    \end{tabular}
    \caption{Values of the parameters $\alpha$, $\beta$ and $\eta$ for each redshift bin, estimated from fits to the cross-power spectra of galaxy and PSF shapes, according to equations \ref{eq:psf_tau0}, \ref{eq:psf_tau2}, \ref{eq:psf_tau5} as well as the goodness-of-fit, $\chi^{2}$, for ${96-3}$ degrees of freedom.}
    \label{tab:abe}
\end{table}

We present the best-fit $\alpha$, $\beta$, $\eta$ values in Table \ref{tab:abe}. While $\alpha$\ is consistent with the expected value of 0 and with real space results {from \citet*{y3-cosmicshear1}}, $\beta$ and $\eta$ values are different. We associate the difference to the fact that the real space analysis uses much smaller scales, down to the sub-arcminute range, while our harmonic space analysis only captures features larger than a few arcminutes.
The total goodness-of-fit on the stacked data vector of the shear and PSF cross spectra $\chi^2$ for $93$ degrees of freedom varies between 99.5 and 117.3 across redshift bins. As in the real space analysis, the $\chi^2$ values are rather large for all but the lowest redshift bin, with the probability-to-exceed being 0.045. Subsequently in \cref{fig:tau-model}, we show the best-fit model to the cross power spectra for each redshift bin and report the $\chi^2$ values for each shear and PSF parameter cross spectrum separately.

\begin{figure}
    \centering
    \includegraphics[scale=0.65]{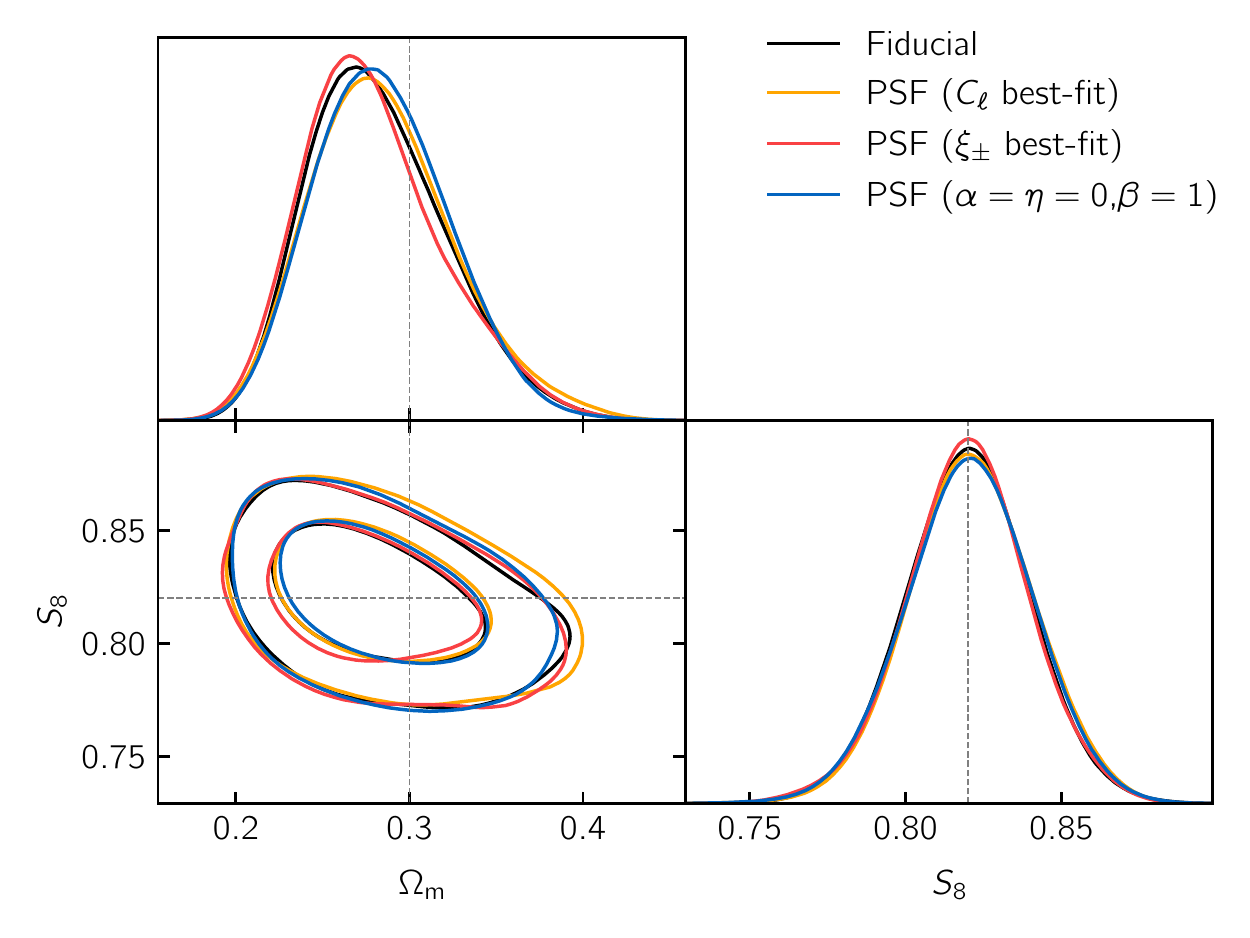}
    \caption{Impact of PSF contamination of the measured shear spectra on cosmological constraints. Fixing the values of the PSF model parameters ($\alpha$, $\beta$ and $\eta$) at the best-fit values inferred from power spectra (blue contours) or two-point functions (red), and at the expected values (orange), we contaminate a noiseless data vector using the model in \cref{eq:psf_model} and compare cosmological constraints to those obtained from the noiseless data vector (black).}
    \label{fig:psf_cont}
\end{figure}

Finally, we propagate the PSF bias in \cref{eq:psf_bias} to compute the expected contamination of the shear power spectra using the model of \cref{eq:psf_model}, in order to test its impact on cosmology. We do so using the best-fit values for the $\alpha$, $\beta$ and $\eta$ parameters from our analysis in harmonic space, the best-fit from the real space analysis in \citet*{y3-cosmicshear1} and the expected values $\alpha=\eta=0$ and $\beta=1$, consistent with non-tomographic results from \citet*{y3-piff}. \Cref{fig:psf_cont} shows that the impact on cosmological constraints is negligible.

\section{Validation on synthetic data}

\begin{figure}
    \centering
    \includegraphics[scale=0.65]{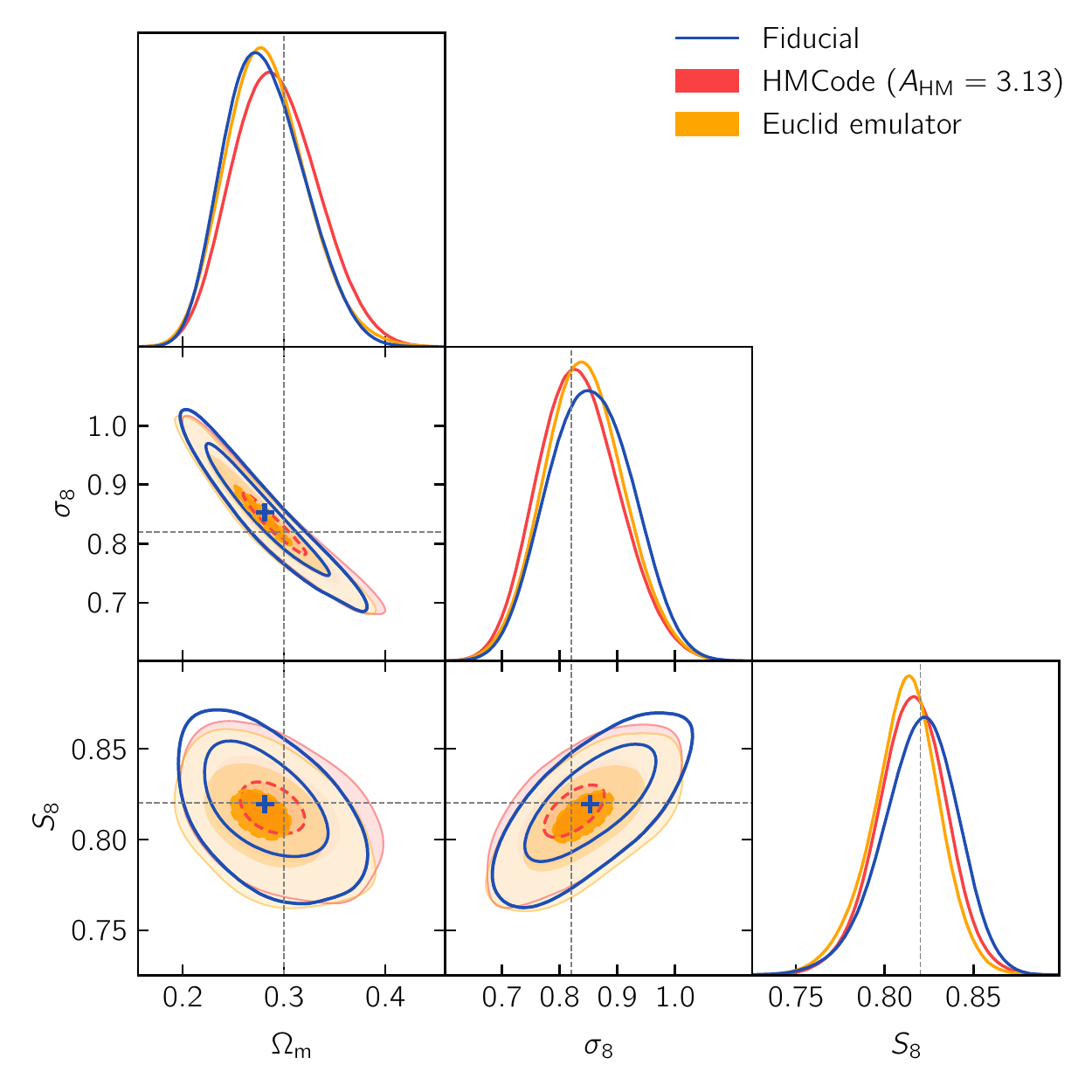}
    \caption{Test of the impact of the non-linear matter power spectrum on cosmological constraints. We analyze three synthetic data vectors with the fiducial model using \halofit and fiducial scale cuts. Constraints obtained from the fiducial data vector are shown in blue, with the mean of the posterior shown by the blue cross. These constraints are compared to those obtained from data vectors computed with \hmcode (red, $\Ahm=3.13$) and the \textsc{Euclid Emulator} (orange). The innermost \SI{0.3}{$\sigma$} contours (underlined in dashed lines) encompass the mean of the fiducial posterior.}
    \label{fig:test_Pk}
\end{figure}

\begin{figure}
    \centering
    \includegraphics[scale=0.65]{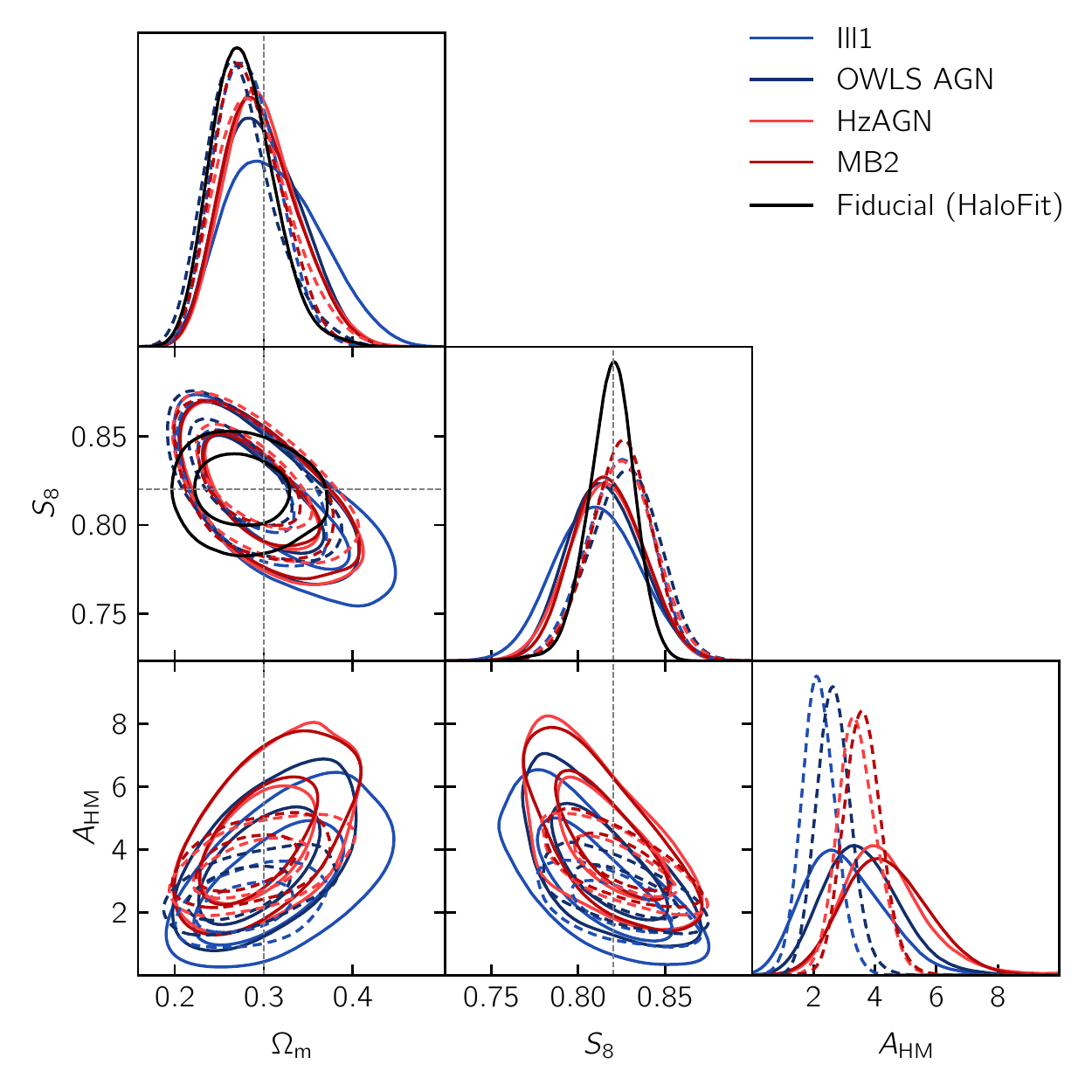}
    \caption{Validation of the baryonic feedback modeling with \hmcode. The four colored posteriors are obtained from shear power spectra that include the effect of baryons as predicted by four hydrodynamical simulations (see \cref{fig:clres}). Solid (dashed) lines were obtained using the scale cuts at $\kmax=\SI{3}{\h\per\mega\parsec}$ ($\kmax=\SI{5}{\h\per\mega\parsec}$). Despite preferring very different values of $\Ahm$ (the dark matter-only case corresponds to $\Ahm=3.13$), the cosmology is recovered in all cases. For comparison, the black contours show the posterior obtained from the fiducial data vector analyzed with \halofit with the scale cuts at $\kmax=\SI{3}{\h\per\mega\parsec}$.}
    \label{fig:hm_test}
\end{figure}

This section illustrates the validation of the modeling pipeline on synthetic data, as described in \cref{sec:modeling_validation_synth}. \Cref{fig:test_Pk} shows the impact of the choice for the non-linear matter power spectrum, whereas \cref{fig:hm_test} validates the use of \hmcode to probe the small-scale portion of our measurements, based on its robustness to various baryonic feedback prescriptions from four different hydrodynamical simulations.

\section{Internal consistency}
\label{app:ic}

This section presents a number of tests in parameter (\cref{app:ic_cont}) and data space (\cref{app:ic_ppd}) for the fiducial run, \ie using our fiducial \lcdm model and scale cuts, and excluding shear ratio information.

\subsection{Robustness of cosmological constraints}
\label{app:ic_cont}

\begin{figure*}
    \centering
    \includegraphics[scale=0.65]{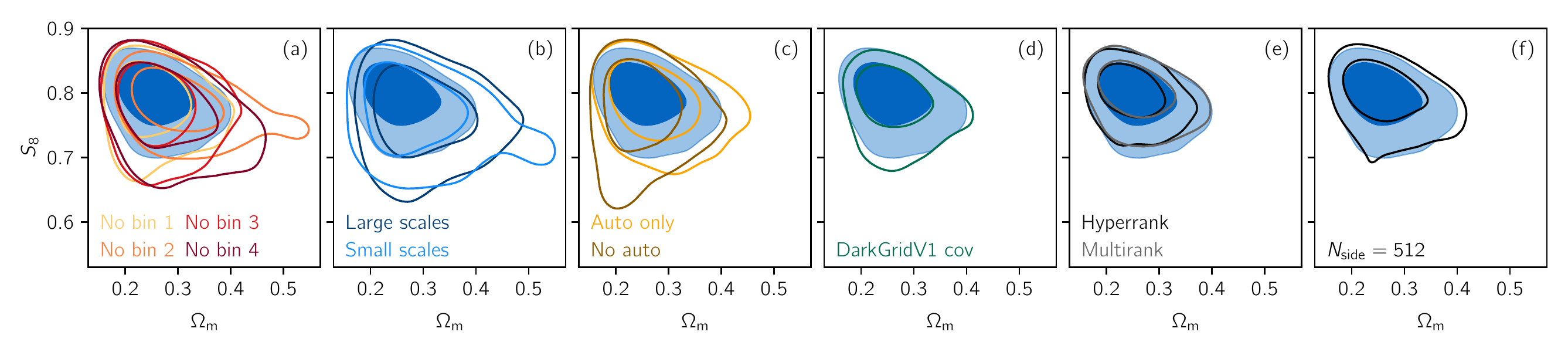}
    \caption{Robustness tests of cosmological constraints, comparing variations in analysis choices to the fiducial constraints in blue. We first repeat the analysis removing part of the data vector, according to
    \begin{inparaenum}[(a)]
        \protect\item redshift bins,
        \protect\item scales, and
        \protect\item auto-power spectra.
    \end{inparaenum}
    We then modify certain parts of the analysis, namely
    \begin{inparaenum}[(a)]\setcounter{enumi}{3}
        \protect\item the covariance matrix,
        \protect\item the methodology to marginalize over uncertainties in the redshift distributions, and
        \protect\item the measurement resolution.
    \end{inparaenum}
    See \cref{app:ic_cont} for details.}
    \label{fig:cont_consistency}
\end{figure*}

We first perform a series of tests, listed below, to assert the robustness of cosmological constraints presented in \cref{sec:res_lcdm}. \Cref{fig:1d_all} presents uni-dimensional marginal distributions for these tests in sections (d) and (e). We also show the two-dimensional marginal distributions in the $(S_8,\Om)$ plane in \cref{fig:cont_consistency}, in the following order: 
\begin{enumerate}[(a)]
    \item \textit{Redshift test.} \label{item:ictests_z} Many parts of the cosmological model (including intrinsic alignments) are redshift-dependent by construction, whereas systematic effects may differentially impact the four redshift bins. To test the robustness of the cosmological constraints to such effects, we therefore perform the analysis of cosmic shear power spectra removing one bin at a time (\eg, when removing bin~2, we remove the bin pairs {2,1}, {2,2}, {3,2} and {4,2} from the data vector), and show contours in \cref{fig:cont_consistency}, panel (a). While contours widen, as expected, {and some degeneracies with $\Ata$ appear to create some tails in the posteriors,} we find an overall excellent agreement, with no visible trend.
    \item \textit{Large \vs small scales.} \label{item:ictests_scales} As discussed throughout the paper, the non-linear scales play a crucial role in this analysis, as they contain a significant amount of cosmological information, but are also the most difficult to model. Using our fiducial set of scale cuts, we split the data vector between large and small scales as follows: for each redshift bin pair, we find the multipole $\ell_{\rm thr}$, within the scale cuts ${\lmin\leq\ell\leq\lmax}$, that results in approximately equal signal-to-noise ratio $S/N$ on both sides, \ie ${S/N_{\lmin\leq\ell\leq\ell_{\rm thr}} \approx S/N_{\ell_{\rm thr}\leq\ell\leq\lmax}}$. This procedure leaves us with 58 and 61 data points for large and small scales, respectively. We find that constraints using either only large scales or only small scales are very similar in width and in very good agreement with each other. {The broadening of the posteriors seems related to partial degeneracies with intrinsic alignment parameters, in particular $\Att$. Nevertheless, they are in very good agreement with} the constraints from the full analysis. 
    \item \textit{Auto-power spectra.} The pseudo-$\cl$ estimator we use here requires the subtraction of the noise power spectrum, which is estimated analytically from the shape catalog here, following \citet{2020arXiv201009717N}. In order to evaluate the potential impact of a misestimation, we analyze our data without auto-power spectra, \ie removing bin pairs {1,1}, {2,2}, {3,3} and {4,4} from the data vector (No auto), and then using only those pairs (Auto only). We find constraints that are wider but consistent with the full analysis, with no clear indication for an issue with noise spectrum subtraction. 
    \item \textit{Covariance.} As described in \cref{sec:like_cov}, our covariance matrix is a hybrid matrix that uses \namaster to evaluate the Gaussian contribution with the effects of the mask and binning properly accounted for, and \cosmolike to evaluate the non-Gaussian contribution, at the fiducial \planck 2018 cosmology. We have also used \dgv simulations \citep{2022MNRAS.tmp..151Z} to obtain an empirical estimate of the covariance matrix, for comparison and validation of our analytic (and therefore noiseless) estimate. We test the impact of this choice by using the empirical covariance in our cosmological analysis, and find that our constraints are almost insensitive to this choice, showing the excellent agreement of the two covariance matrices.
    \item \textit{\hyperrank.} Throughout this work, we have employed the fiducial approach over marginalizing over redshift distribution biases, $\Delta z_a$'s, in order to account for uncertainty in the redshift distributions. However, the DES~Y3 redshift pipeline produced samples of the redshift distributions that can be properly marginalized over using either the \multirank or \hyperrank methods, {by sampling, respectively, realizations themselves, or a set of hyperparameters used to rank and select realizations \citep*[for details, see][]{y3-hyperrank}}. We do so here and find cosmological constraints in excellent agreement with the fiducial analysis, with roughly 15\% smaller uncertainty on $S_8$ for both techniques.
    \item \textit{Resolution.} As detailed in \cref{sec:pseudocl}, the pseudo-$\cl$ estimator is based on pixelized \healpix maps of the shear catalog. However, as discussed in \citet{2020arXiv201009717N}, the effects of the pixelization of the shear field depend both on the density of galaxies and the chosen resolution. We used a resolution parameter of $\nside=1024$, which allows us to probe multipoles up to $\ell\sim2000$, while yielding a relatively complete mask, without too many empty pixels in the survey area, and with a mean number of galaxies per pixel of around \numrange{17.2}{17.5} for all four bins. This means that we are in the regime where the shear maps are that of the averaged shear field (as opposed to the sampled shear field) and that we may use standard \healpix window functions to correct for the smoothing that has taken place. In order to verify the impact on cosmological constraints, we repeat the measurements, including noise power spectrum and Gaussian covariance estimation, at $\nside=512$. We do observe expected differences in the shear power spectra -- almost negligible at large scales and growing up to about the size of the error bars at $\ell\sim1024$, with no clear trend -- but find negligible impact on cosmology.
\end{enumerate}

\subsection{Internal consistency of data with posterior predictive distributions}
\label{app:ic_ppd}

\begin{table}
    \centering
    \begin{tabular}{l c}
        Test & Calibrated $p$-value \\
        \hline
        Goodness-of-fit & 0.116 \\
        Bin 1 \vs no bin 1 & 0.998 \\
        Bin 2 \vs no bin 2 & 0.020 \\
        Bin 3 \vs no bin 3 & 0.080 \\
        Bin 4 \vs no bin 4 & 0.876 \\
        Small \vs large scales & 0.395 \\
        Large \vs small scales & 0.212 \\
        \hline
    \end{tabular}
    \caption{Internal consistency tests using the posterior predictive distribution method from \citet*{y3-inttensions}. See \cref{app:ic_ppd} for details.
}
    \label{tab:ppd}
\end{table}

\begin{figure*}
    \centering
    \includegraphics[scale=0.65]{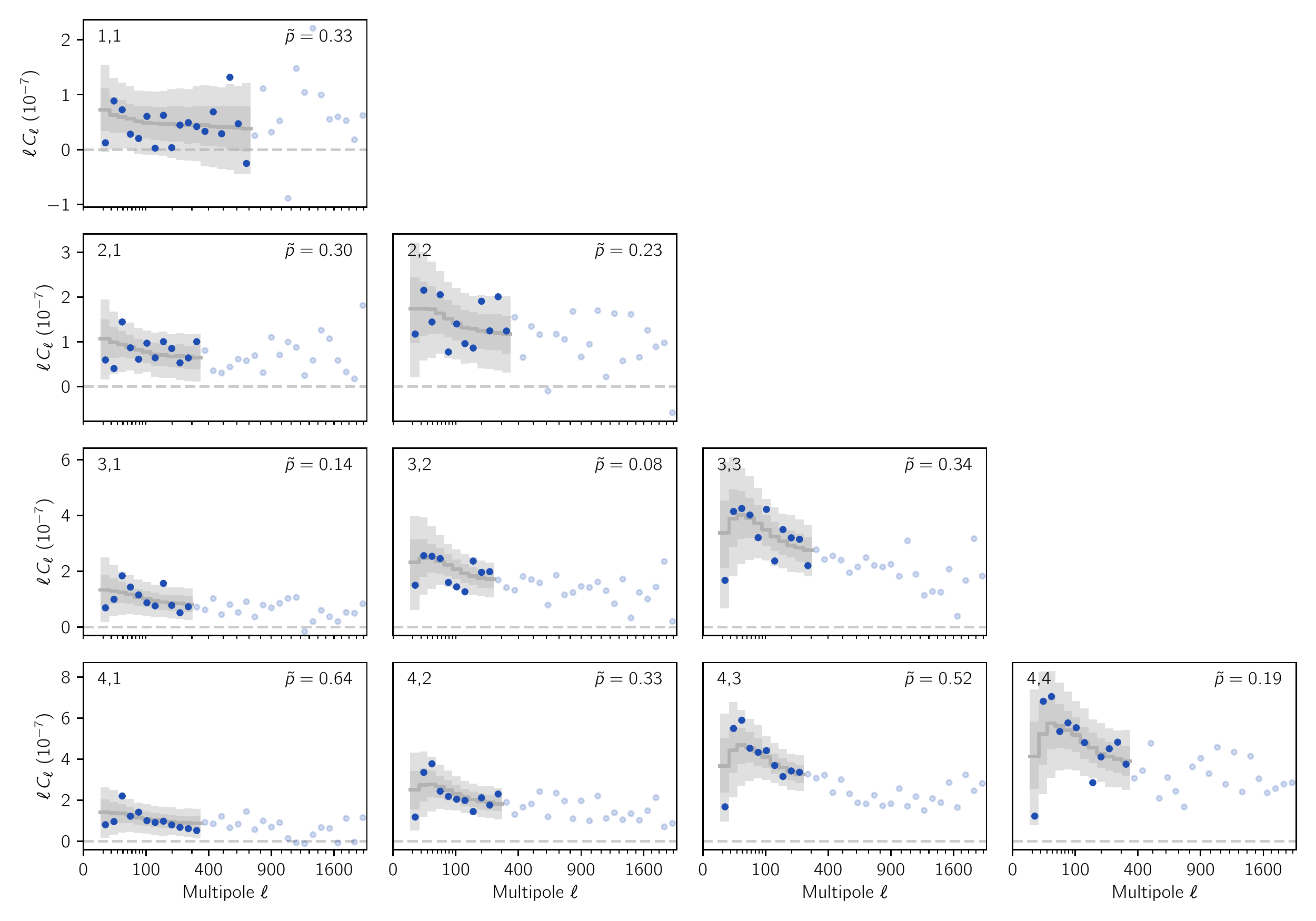}
    \caption{Goodness-of-fit test for the fiducial run using the posterior predictive distribution (PPD) methodology of \citet*{y3-inttensions}. The data is shown by the blue circles, which are filled for data points within fiducial scale cuts. The gray line shows the mean of the PPD realizations, whereas the gray bands show the \SI{1}{$\sigma$} and \SI{2}{$\sigma$} percentiles of the PPD. The calibrated $p$-value for each panel is shown in the upper right corner.}
    \label{fig:ppd_gof}
\end{figure*}

\begin{figure*}
    \centering
    Bin 1 \vs no bin 1
    \includegraphics[scale=0.65]{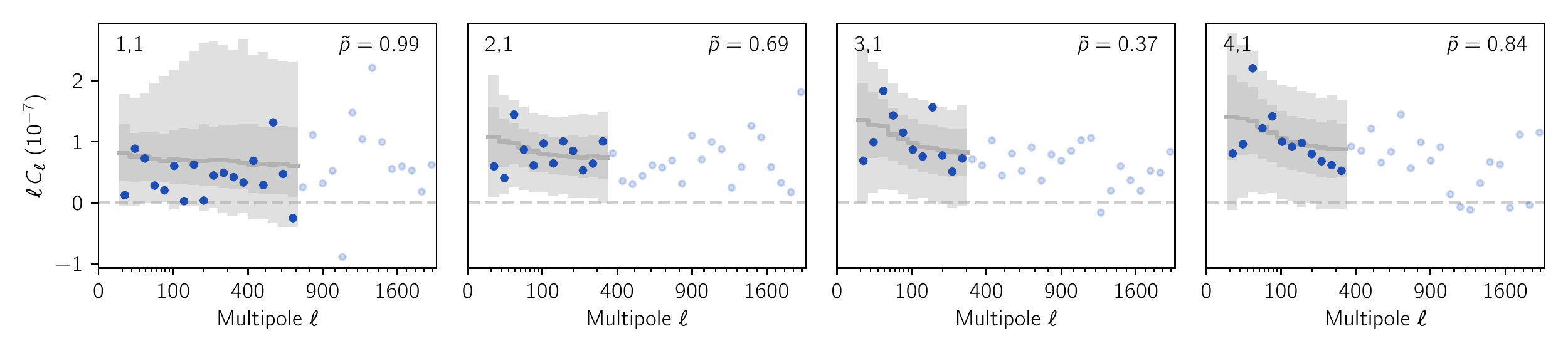}\\
    Bin 2 \vs no bin 2
    \includegraphics[scale=0.65]{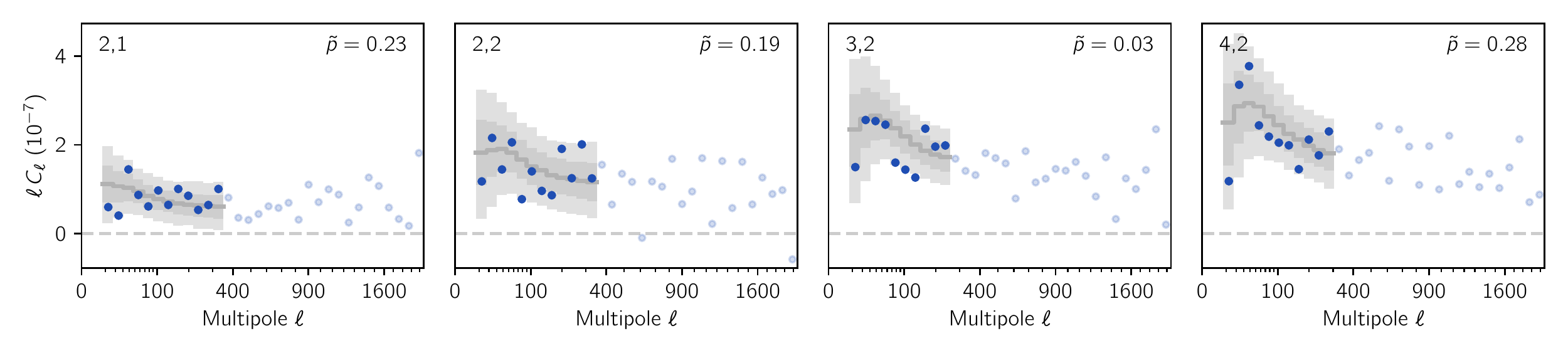}\\
    Bin 3 \vs no bin 3
    \includegraphics[scale=0.65]{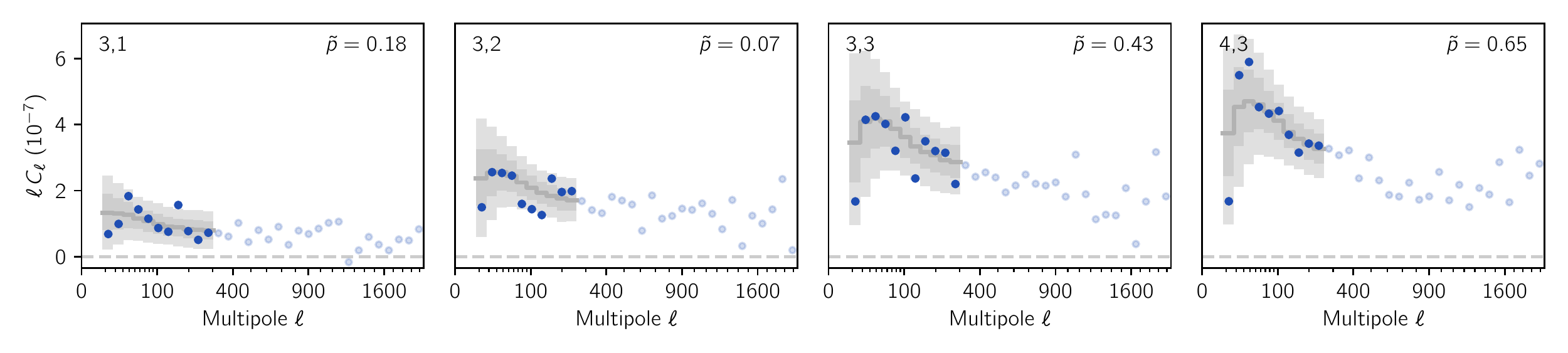}\\
    Bin 4 \vs no bin 4
    \includegraphics[scale=0.65]{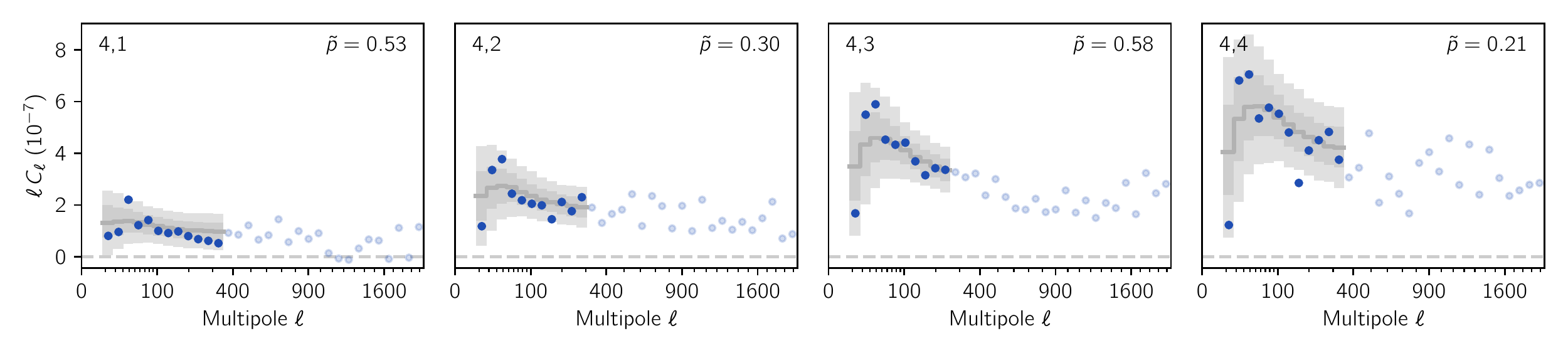}
    \caption{Internal consistency of the four redshift bins (removing one at a time) with the PPD in gray and data in blue. See \cref{fig:ppd_gof} for details.}
    \label{fig:ppd_nozbin}
\end{figure*}

We apply the methodology developed of \citet*{y3-inttensions} based on the posterior predictive distribution (PPD) to test the internal consistency of our data. In a nutshell, the method uses a parameter posterior sample and compares simulated realizations of the data vector drawn from the likelihood at these parameter values to the observed data vector. The test is subsequently calibrated using simulated data vectors, to correct for posterior volume effects, as detailed in \citet*{y3-inttensions}.

We first perform a goodness-of-fit test, where the posterior sample comes from the fiducial run, and simulated realizations are independent of the observed data, and find a calibrated $p$-value of 11.6\%. The PPD samples are shown in gray in \cref{fig:ppd_gof} along with the observed data in blue.

We then perform consistency tests of the type \textit{A \vs B}, \ie where we divide the data in two disjoint parts $A$ and $B$, use $B$ to obtain a posterior sample, and generate from those samples realizations of $A$ to be compared to the real data, in a way that accounts for the correlation between $A$ and $B$. Specifically, we split the data according to redshift bins and scales, using the same splits as in \cref{item:ictests_z} and \cref{item:ictests_scales} of the previous section. We illustrate the redshift consistency test in \cref{fig:ppd_nozbin} and summarize the results in \cref{tab:ppd}, finding no indication of inconsistency.

\section{Full posterior distribution}
\label{app:full_posterior}

\begin{figure*}
    \centering
    \includegraphics[scale=0.65]{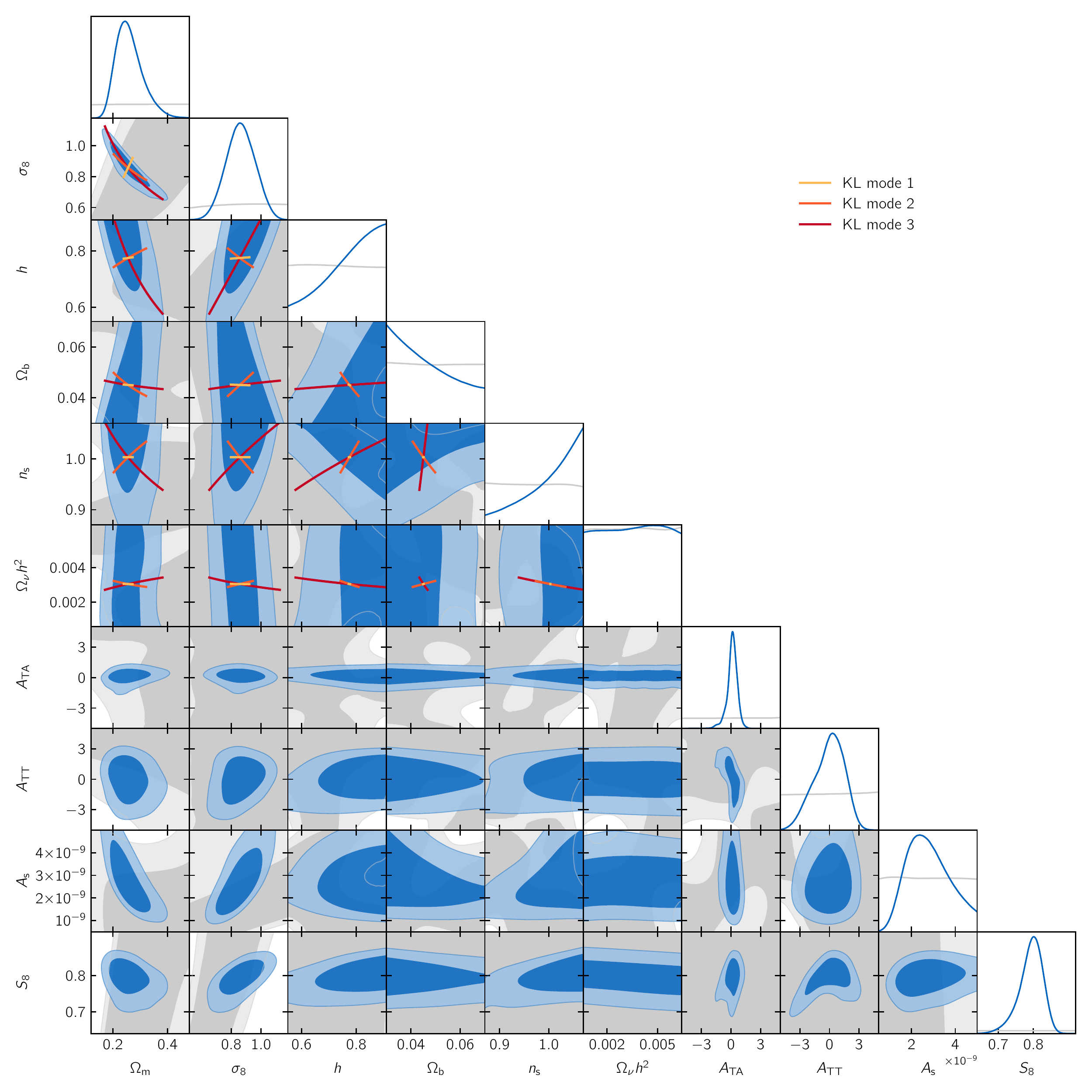}
    \caption{Posterior (in blue) and prior (in gray) distributions for the fiducial \lcdm constraints from DES~Y3 shear power spectra (without shear ratios) presented in \cref{sec:res_lcdm}, showing cosmological and intrinsic alignment parameters (note that the ranges are adjusted to the posterior for readability). Although we sample over $\As$ with a flat prior, we apply the Karhunen–Loève (KL) decomposition \citep{2019PhRvD..99d3506R,2020PhRvD.101j3527R,2021arXiv210503324R} in the space of $\qty{\log\Om,\log\sigma_8,\log h,\log\Ob,\log\ns,\log\Onu h^2}$. The best constrained directions in this parameter space, corresponding to the first three modes of the KL decomposition, are represented in yellow, orange and red.}
    \label{fig:full_posterior}
\end{figure*}

\Cref{fig:full_posterior} shows the prior and posterior distributions for the fiducial constraints presented in \cref{sec:res_lcdm} (without shear ratios). We also perform a Karhunen–Loève (KL) decomposition \citep{2019PhRvD..99d3506R,2020PhRvD.101j3527R,2021arXiv210503324R,2021arXiv211205737D} in order to determine the directions, in parameter space, that are best constrained by the data, as quantified by the improvement between the prior and the posterior. We use the \texttt{tensiometer}\footnote{\url{https://tensiometer.readthedocs.io}} package and work in the space of $\qty{\log\Om,\log\sigma_8,\log h,\log\Ob,\log\ns,\log\Onu h^2}$ in order to express the KL modes as power laws in the original parameters. We find that the three first KL modes are the following (the improvements are in parentheses):
\begin{align}
\qty(\frac{\Omega_{\rm m}}{0.255})^{0.521}
\qty(\frac{\sigma_8}{0.857})
          & = 1.000 \pm 0.116, && (978.7\%)\\
\qty(\frac{\Omega_{\rm m}}{0.255})
\qty(\frac{\sigma_8}{0.857})^{-1.219}
\qty(\frac{n_{\rm s}}{1.003})^{2.651}
          & = 1.000 \pm 0.868, && (202.5\%)\\
\qty(\frac{\Omega_{\rm m}}{0.255})^{-0.149}
\qty(\frac{h}{0.774})
\qty(\frac{n_{\rm s}}{1.003})^{1.681}
          & = 1.000 \pm 0.426. && (77.3 \%)
\end{align}
The first mode nearly matches the $S_8$ parameter, while subsequent modes, with much weaker improvements, include the Hubble constant $h$ and the tilt of the primordial power spectrum $\ns$.

\section*{Affiliations}

$^{1}$ Department of Physics and Astronomy, University of Pennsylvania, Philadelphia, PA 19104, USA\\
$^{2}$ Department of Physics, ETH Zurich, Wolfgang-Pauli-Strasse 16, CH-8093 Zurich, Switzerland\\
$^{3}$ Department of Astronomy, University of California, Berkeley,  501 Campbell Hall, Berkeley, CA 94720, USA\\
$^{4}$ Department of Astronomy/Steward Observatory, University of Arizona, 933 North Cherry Avenue, Tucson, AZ 85721-0065, USA\\
$^{5}$ ICTP South American Institute for Fundamental Research\\ Instituto de F\'{\i}sica Te\'orica, Universidade Estadual Paulista, S\~ao Paulo, Brazil\\
$^{6}$ Laborat\'orio Interinstitucional de e-Astronomia - LIneA, Rua Gal. Jos\'e Cristino 77, Rio de Janeiro, RJ - 20921-400, Brazil\\
$^{7}$ Kavli Institute for Particle Astrophysics \& Cosmology, P. O. Box 2450, Stanford University, Stanford, CA 94305, USA\\
$^{8}$ Instituto de F\'{i}sica Te\'orica, Universidade Estadual Paulista, S\~ao Paulo, Brazil\\
$^{9}$ California Institute of Technology, 1200 East California Blvd, MC 249-17, Pasadena, CA 91125, USA\\
$^{10}$ Kavli Institute for Cosmological Physics, University of Chicago, Chicago, IL 60637, USA\\
$^{11}$ Department of Physics, Northeastern University, Boston, MA 02115, USA\\
$^{12}$ Laboratory of Astrophysics, \'Ecole Polytechnique F\'ed\'erale de Lausanne (EPFL), Observatoire de Sauverny, 1290 Versoix, Switzerland\\
$^{13}$ Department of Astronomy and Astrophysics, University of Chicago, Chicago, IL 60637, USA\\
$^{14}$ Institut d'Estudis Espacials de Catalunya (IEEC), 08034 Barcelona, Spain\\
$^{15}$ Institute of Space Sciences (ICE, CSIC),  Campus UAB, Carrer de Can Magrans, s/n,  08193 Barcelona, Spain\\
$^{16}$ Department of Physics \& Astronomy, University College London, Gower Street, London, WC1E 6BT, UK\\
$^{17}$ Laboratoire de Physique de l'Ecole Normale Sup\'erieure, ENS, Universit\'e PSL, CNRS, Sorbonne Universit\'e, Universit\'e de Paris, Paris, France\\
$^{18}$ Department of Physics, Carnegie Mellon University, Pittsburgh, Pennsylvania 15312, USA\\
$^{19}$ Argonne National Laboratory, 9700 South Cass Avenue, Lemont, IL 60439, USA\\
$^{20}$ Department of Physics, University of Michigan, Ann Arbor, MI 48109, USA\\
$^{21}$ Institute for Astronomy, University of Hawai'i, 2680 Woodlawn Drive, Honolulu, HI 96822, USA\\
$^{22}$ Physics Department, 2320 Chamberlin Hall, University of Wisconsin-Madison, 1150 University Avenue Madison, WI  53706-1390\\
$^{23}$ Instituto de Astrofisica de Canarias, E-38205 La Laguna, Tenerife, Spain\\
$^{24}$ Universidad de La Laguna, Dpto. Astrofísica, E-38206 La Laguna, Tenerife, Spain\\
$^{25}$ Center for Astrophysical Surveys, National Center for Supercomputing Applications, 1205 West Clark St., Urbana, IL 61801, USA\\
$^{26}$ Department of Astronomy, University of Illinois at Urbana-Champaign, 1002 W. Green Street, Urbana, IL 61801, USA\\
$^{27}$ Physics Department, William Jewell College, Liberty, MO, 64068\\
$^{28}$ Department of Physics, Duke University Durham, NC 27708, USA\\
$^{29}$ Jodrell Bank Center for Astrophysics, School of Physics and Astronomy, University of Manchester, Oxford Road, Manchester, M13 9PL, UK\\
$^{30}$ Lawrence Berkeley National Laboratory, 1 Cyclotron Road, Berkeley, CA 94720, USA\\
$^{31}$ NSF AI Planning Institute for Physics of the Future, Carnegie Mellon University, Pittsburgh, PA 15213, USA\\
$^{32}$ Fermi National Accelerator Laboratory, P. O. Box 500, Batavia, IL 60510, USA\\
$^{33}$ Jet Propulsion Laboratory, California Institute of Technology, 4800 Oak Grove Dr., Pasadena, CA 91109, USA\\
$^{34}$ Center for Cosmology and Astro-Particle Physics, The Ohio State University, Columbus, OH 43210, USA\\
$^{35}$ Department of Physics, The Ohio State University, Columbus, OH 43210, USA\\
$^{36}$ Santa Cruz Institute for Particle Physics, Santa Cruz, CA 95064, USA\\
$^{37}$ Kavli Institute for Cosmology, University of Cambridge, Madingley Road, Cambridge CB3 0HA, UK\\
$^{38}$ Institut de F\'{\i}sica d'Altes Energies (IFAE), The Barcelona Institute of Science and Technology, Campus UAB, 08193 Bellaterra (Barcelona) Spain\\
$^{39}$ University Observatory, Faculty of Physics, Ludwig-Maximilians-Universit\"at, Scheinerstr. 1, 81679 Munich, Germany\\
$^{40}$ Department of Physics, University of Oxford, Denys Wilkinson Building, Keble Road, Oxford OX1 3RH, UK\\
$^{41}$ School of Physics and Astronomy, Cardiff University, CF24 3AA, UK\\
$^{42}$ Department of Astronomy, University of Geneva, ch. d'\'Ecogia 16, CH-1290 Versoix, Switzerland\\
$^{43}$ Department of Physics, University of Arizona, Tucson, AZ 85721, USA\\
$^{44}$ Department of Physics and Astronomy, Pevensey Building, University of Sussex, Brighton, BN1 9QH, UK\\
$^{45}$ Instituto de Astrof\'{\i}sica e Ci\^{e}ncias do Espa\c{c}o, Faculdade de Ci\^{e}ncias, Universidade de Lisboa, 1769-016 Lisboa, Portugal\\
$^{46}$ Department of Applied Mathematics and Theoretical Physics, University of Cambridge, Cambridge CB3 0WA, UK\\
$^{47}$ Perimeter Institute for Theoretical Physics, 31 Caroline St. North, Waterloo, ON N2L 2Y5, Canada\\
$^{48}$ Department of Physics, Stanford University, 382 Via Pueblo Mall, Stanford, CA 94305, USA\\
$^{49}$ SLAC National Accelerator Laboratory, Menlo Park, CA 94025, USA\\
$^{50}$ Instituto de F\'isica Gleb Wataghin, Universidade Estadual de Campinas, 13083-859, Campinas, SP, Brazil\\
$^{51}$ Kavli Institute for the Physics and Mathematics of the Universe (WPI), UTIAS, The University of Tokyo, Kashiwa, Chiba 277-8583, Japan\\
$^{52}$ Laboratoire de physique des 2 infinis Ir\`ene Joliot-Curie, CNRS Universit\'e Paris-Saclay, B\^at. 100, Facult\'e des sciences, F-91405 Orsay Cedex, France\\
$^{53}$ Centro de Investigaciones Energ\'eticas, Medioambientales y Tecnol\'ogicas (CIEMAT), Madrid, Spain\\
$^{54}$ Brookhaven National Laboratory, Bldg 510, Upton, NY 11973, USA\\
$^{55}$ D\'{e}partement de Physique Th\'{e}orique and Center for Astroparticle Physics, Universit\'{e} de Gen\`{e}ve, 24 quai Ernest Ansermet, CH-1211 Geneva, Switzerland\\
$^{56}$ Excellence Cluster Origins, Boltzmannstr.\ 2, 85748 Garching, Germany\\
$^{57}$ Max Planck Institute for Extraterrestrial Physics, Giessenbachstrasse, 85748 Garching, Germany\\
$^{58}$ Universit\"ats-Sternwarte, Fakult\"at f\"ur Physik, Ludwig-Maximilians Universit\"at M\"unchen, Scheinerstr. 1, 81679 M\"unchen, Germany\\
$^{59}$ Institute for Astronomy, University of Edinburgh, Edinburgh EH9 3HJ, UK\\
$^{60}$ Cerro Tololo Inter-American Observatory, NSF's National Optical-Infrared Astronomy Research Laboratory, Casilla 603, La Serena, Chile\\
$^{61}$ Institute of Cosmology and Gravitation, University of Portsmouth, Portsmouth, PO1 3FX, UK\\
$^{62}$ CNRS, UMR 7095, Institut d'Astrophysique de Paris, F-75014, Paris, France\\
$^{63}$ Sorbonne Universit\'es, UPMC Univ Paris 06, UMR 7095, Institut d'Astrophysique de Paris, F-75014, Paris, France\\
$^{64}$ Astronomy Unit, Department of Physics, University of Trieste, via Tiepolo 11, I-34131 Trieste, Italy\\
$^{65}$ INAF-Osservatorio Astronomico di Trieste, via G. B. Tiepolo 11, I-34143 Trieste, Italy\\
$^{66}$ Institute for Fundamental Physics of the Universe, Via Beirut 2, 34014 Trieste, Italy\\
$^{67}$ Observat\'orio Nacional, Rua Gal. Jos\'e Cristino 77, Rio de Janeiro, RJ - 20921-400, Brazil\\
$^{68}$ Hamburger Sternwarte, Universit\"{a}t Hamburg, Gojenbergsweg 112, 21029 Hamburg, Germany\\
$^{69}$ Department of Physics, IIT Hyderabad, Kandi, Telangana 502285, India\\
$^{70}$ Institute of Theoretical Astrophysics, University of Oslo. P.O. Box 1029 Blindern, NO-0315 Oslo, Norway\\
$^{71}$ Instituto de Fisica Teorica UAM/CSIC, Universidad Autonoma de Madrid, 28049 Madrid, Spain\\
$^{72}$ Department of Astronomy, University of Michigan, Ann Arbor, MI 48109, USA\\
$^{73}$ Institute of Astronomy, University of Cambridge, Madingley Road, Cambridge CB3 0HA, UK\\
$^{74}$ School of Mathematics and Physics, University of Queensland,  Brisbane, QLD 4072, Australia\\
$^{75}$ Center for Astrophysics $\vert$ Harvard \& Smithsonian, 60 Garden Street, Cambridge, MA 02138, USA\\
$^{76}$ Australian Astronomical Optics, Macquarie University, North Ryde, NSW 2113, Australia\\
$^{77}$ Lowell Observatory, 1400 Mars Hill Rd, Flagstaff, AZ 86001, USA\\
$^{78}$ George P. and Cynthia Woods Mitchell Institute for Fundamental Physics and Astronomy, and Department of Physics and Astronomy, Texas A\&M University, College Station, TX 77843,  USA\\
$^{79}$ Instituci\'o Catalana de Recerca i Estudis Avan\c{c}ats, E-08010 Barcelona, Spain\\
$^{80}$ School of Physics and Astronomy, University of Southampton,  Southampton, SO17 1BJ, UK\\
$^{81}$ Computer Science and Mathematics Division, Oak Ridge National Laboratory, Oak Ridge, TN 37831\\


\bsp	
\label{lastpage}
\end{document}